\documentstyle{mn}

\newif\ifAMStwofonts

\ifoldfss
  \ifCUPmtlplainloaded \else
    \NewTextAlphabet{textbfit} {cmbxti10} {}
    \NewTextAlphabet{textbfss} {cmssbx10} {}
    \NewMathAlphabet{mathbfit} {cmbxti10} {} 
    \NewMathAlphabet{mathbfss} {cmssbx10} {} 
  \fi
  \ifAMStwofonts
    \ifCUPmtlplainloaded \else
      \NewSymbolFont{upmath} {eurm10}
      \NewSymbolFont{AMSa} {msam10}
      \NewMathSymbol{\upi}     {0}{upmath}{19}
      \NewMathSymbol{\umu}     {0}{upmath}{16}
      \NewMathSymbol{\upartial}{0}{upmath}{40}
      \NewMathSymbol{\leqslant}{3}{AMSa}{36}
      \NewMathSymbol{\geqslant}{3}{AMSa}{3E}

      \let\leq=\leqslant \let\le=\leqslant
      \let\geq=\geqslant 
    \fi
  \fi
\fi 

\ifnfssone
  \newmathalphabet{\mathit}
  \addtoversion{normal}{\mathit}{cmr}{m}{it}
  \addtoversion{bold}{\mathit}{cmr}{bx}{it}
  \newmathalphabet{\mathbfit} 
  \addtoversion{normal}{\mathbfit}{cmr}{bx}{it}
  \addtoversion{bold}{\mathbfit}{cmr}{bx}{it}
  \newmathalphabet{\mathbfss} 
  \addtoversion{normal}{\mathbfss}{cmss}{bx}{n}
  \addtoversion{bold}{\mathbfss}{cmss}{bx}{n}
  \ifAMStwofonts
    \ifCUPmtlplainloaded \else
      \UseAMStwoboldmath
      \makeatletter
      \new@mathgroup\upmath@group
      \define@mathgroup\mv@normal\upmath@group{eur}{m}{n}
      \define@mathgroup\mv@bold\upmath@group{eur}{b}{n}
      \edef\UPM{\hexnumber\upmath@group}
      \new@mathgroup\amsa@group
      \define@mathgroup\mv@normal\amsa@group{msa}{m}{n}
      \define@mathgroup\mv@bold\amsa@group{msa}{m}{n}
      \edef\AMSa{\hexnumber\amsa@group}
      \makeatother
      \mathchardef\upi="0\UPM19
      \mathchardef\umu="0\UPM16
      \mathchardef\upartial="0\UPM40
      \mathchardef\leqslant="3\AMSa36
      \mathchardef\geqslant="3\AMSa3E

      \let\leq=\leqslant \let\le=\leqslant
      \let\geq=\geqslant 
    \fi
  \fi
\fi 

\ifnfsstwo
  \DeclareMathAlphabet{\mathbfit}{OT1}{cmr}{bx}{it}
  \SetMathAlphabet\mathbfit{bold}{OT1}{cmr}{bx}{it}
  \DeclareMathAlphabet{\mathbfss}{OT1}{cmss}{bx}{n}
  \SetMathAlphabet\mathbfss{bold}{OT1}{cmss}{bx}{n}
  \ifAMStwofonts
    \ifCUPmtlplainloaded \else
      \DeclareSymbolFont{UPM}{U}{eur}{m}{n}
      \SetSymbolFont{UPM}{bold}{U}{eur}{b}{n}
      \DeclareSymbolFont{AMSa}{U}{msa}{m}{n}
      \DeclareMathSymbol{\upi}{0}{UPM}{"19}
      \DeclareMathSymbol{\umu}{0}{UPM}{"16}
      \DeclareMathSymbol{\upartial}{0}{UPM}{"40}
      \DeclareMathSymbol{\leqslant}{3}{AMSa}{"36}
      \DeclareMathSymbol{\geqslant}{3}{AMSa}{"3E}

      \let\leq=\leqslant \let\le=\leqslant
      \let\geq=\geqslant 
    \fi
  \fi
\fi 

\ifCUPmtlplainloaded \else
  \ifAMStwofonts \else 
    \def\upi{\pi}
    \def\umu{\mu}
    \def\upartial{\partial}
  \fi
\fi

\title{Stellar population gradients in Seyfert 2 galaxies. Northern sample}
\author[D. Raimann et al.]
  {D.~Raimann$^1$, T.~Storchi-Bergmann$^1$, R.M.~Gonz\' alez Delgado$^2$, R. Cid Fernandes$^3$,
  \newauthor T. Heckman$^4$, C. Leitherer$^5$ and H. Schmitt$^{6,7}$
\thanks{E-mail: raimann@if.ufrgs.br; thaisa@if.ufrgs.br; rosa@iaa.es; cid@fsc.ufsc.br;
 heckman@pha.jhu.edu; leitherer@stsci.edu; hschmitt@nrao.edu} \\
  $^1$ Universidade Federal do Rio Grande do Sul, IF, CP15051, Porto Alegre 91501-970, RS, Brazil\\
  $^2$ Instituto de Astrof\'{\i}sica Andaluc\'{\i}a (CSIC), Apto.3004, 18080, Granada, Spain \\
  $^3$ Departamento de F\'{\i}sica, CFM - UFSC, Campus Universit\' ario - Trindade, CP476, Florian\' opolis 88040-900, SC, Brazil \\
  $^4$ Department of Physics \& Astronomy, Johns Hopkins University, 3400 N. Charles St., Baltimore, MD, 21218 \\
  $^5$ Space Telescope Science Institute, 3700 San Martin Drive, Baltimore, MD, 21218 \\
  $^6$ National Radio Astronomy Observatory, PO Box 0, Socorro, NM 87801 \\
  $^7$ Jansky Fellow}

\date{Accepted
      Received
      in original form}

\pagerange{\pageref{firstpage}--\pageref{lastpage}}
\pubyear{2000}

\begin{document}

\maketitle

\title{Stellar population gradients in Seyfert 2 galaxies}

\label{firstpage}

\begin{abstract}
We use high signal-to-noise ratio long-slit spectra in the $\lambda$$\lambda$3600-4700 \AA\
range of the twenty brightest northern Seyfert 2 galaxies to study the variation of the stellar
population properties as a function of distance from the nucleus. In order to characterize
the stellar population and other continuum sources (e.g. featureless continuum FC) we have
measured equivalent widths Ws of six absorption features, four continuum colours and their
radial variations, and performed spectral population synthesis as a function of distance
from the nucleus. About half the sample has Ca{\sevensize II}K and G-band W values
smaller at the nucleus than at 1\,kpc from it, due to a younger population and/or FC.
The stellar population synthesis shows that, while at the nucleus, 75 per cent of the
galaxies present contribution $>$20 per cent of ages $\le$100\,Myr and/or of a FC,
this proportion decreases to 45 per cent at 3\,kpc. In particular, 55 per cent of the
galaxies have contribution $>$10 per cent of the 3\,Myr/FC component (a degenerate component
in which one cannot separate what is due to a FC or to a 3\,Myr stellar population)
at the nucleus, but only 25 per cent of them have this contribution at 3\,kpc.
As reference, the stellar population of 10 non-Seyfert galaxies, spanning the Hubble
types of the Seyfert (from S0 to Sc) was also studied. A comparison between the stellar
population of the Seyferts and that of the non-Seyferts shows systematic differences: the
contribution of ages younger than 1\,Gyr is in most cases larger in the Seyfert galaxies
than in non-Seyferts, not only at the nucleus but up to 1\,kpc from it.

\end{abstract}


\section{Introduction}

The relation between recent star-formation and nuclear activity in galaxies has been
the subject of a number of recent studies. Cid Fernandes \& Terlevich (1995) were the first
to argue that the blue continuum in Seyfert 2 galaxies was rather mostly due to the 
contribution of young stars than to a scattered featureless continuum (FC), if the
hypotheses of Unified Models for Seyfert galaxies were correct (Antonucci 1993). 
Heckman et al. (1995) reached similar conclusions using a different method.
Heckman et al. (1997) and Gonz\'alez Delgado et al. (1998) have proved this to be
the case for a few Seyfert 2 galaxies with available UV spectra with enough 
signal-to-noise ratio to show the wind lines of young O and B stars. 
Gonz\'alez Delgado et al. (1998) have also shown that, in these cases, high-order 
Balmer lines in absorption, characteristic of young and intermediate age stars 
($\approx$ 100\,Myr), were also present in the spectra. As optical spectra are more
easily available, the high-order Balmer lines could thus be used as unambiguous 
signatures of the presence of young to intermediate age stars.

In Cid Fernandes, Storchi-Bergmann \& Schmitt (1998), Storchi-Bergmann, Cid
Fernandes \& Schmitt (1998) and Schmitt, Storchi-Bergmann \& Cid Fernandes (1999), 
we used long-slit spectra of about 40 AGNs to measure absorption line and
continuum spectral features within a few kiloparsecs from the nucleus, to characterize 
the stellar population. For the 20 Seyfert 2 galaxies of the sample, we have then 
performed spectral synthesis of the nuclear spectra, having concluded that the main 
common characteristic was an excess of an intermediate age stellar population
($\approx$100\,Myr) when compared to early-type non-Seyfert galaxies.

Storchi-Bergmann et al. (2000) used stellar population templates to
investigate the effect of combining an old bulge population (present in
most Seyferts) with a younger one, in particular to investigate the detectability of
the high-order Balmer lines as signatures of the young and intermediate age stars.
By comparing the models with the data, we concluded that unambiguous signatures of
such stars were present in only 30 per cent of the sample. We have also concluded that
is not possible to distinguish a stellar population younger than 10\,Myr from a
power-law continuum, when these components contribute with less than 40 per cent of
the light at 4020\AA.

Combining the above results with the stellar population analysis performed by
Gonz\'alez Delgado et al. (2001) for a northern sample of 20 Seyfert 2 galaxies, we
conclude that 40 per cent of nearby Seyfert 2 galaxies present unambiguous signatures
of recent episodes of star-formation in the nuclear region. In another 30 per cent, a
blue component is also necessary to reproduce the near-UV continuum, but its faintness
precludes a firm identification, and this component can either be due to a FC  or
a stellar population younger than 10\,Myr, or both. Joguet et al. (2001),
studying the circumnuclear stellar population of the host galaxies of 79 Seyfert 2 nuclei
also found high-order Balmer lines seen in absorption in many cases, concluding
that young stellar populations are present in around 50 per cent of Seyfert 2 nuclei,
a rate similar to that found by us.

In Cid Fernandes et al. (2001b) and Storchi-Bergmann et al. (2001) we have looked for
relations between the age of the stellar population in the nuclear region
and the host galaxy properties and environment. We have found that the AGNs with
largest contributions of recent star-formation occur in galaxies with
the largest IR luminosities; these galaxies are frequently in interactions and present
distorted inner morphologies. An evolutionary scenario was suggested, in which the
interactions trigger both the nuclear activity and bursts of star formation. The bursts
then fade before the nuclear activity, which is thus found also in galaxies with
old stellar populations.

In the near-IR, Oliva et al. (1999) studied stellar absorption lines in 13 obscured AGNs
and 8 genuine Seyfert 1's and found that the presence of powerful starbursts in obscured
AGNs is relatively common while they are not present in Seyfert 1's.

In more luminous objects, Aretxaga et al. (2000) found young stellar populations in the
nuclear regions of some nearby radio-galaxies, as well as Tadhunter et al. (2002) for
intermediate redshift ones. Canalizo \& Stockton (2001) studied a sample
of low-redshift QSOs that may be in a transitionary stage between ULIGs and QSOs
and found that every object shows strong recent star-forming activity (younger than
300\,Myr) and in many cases this activity is closely related to tidal interactions.

A number of questions are nevertheless still open regarding the hosts of AGNs,
and one of fundamental nature is if there
are any systematic differences between the hosts of AGNs and non-active galaxies of the
same Hubble type. In other words: does the onset of nuclear activity relate to
large-scale properties of the host galaxies? In the evolutionary scenario
described above, for example, if interactions are responsible for triggering the activity
and circumnuclear bursts of star formation, wouldn't the
more external regions also be affected?

In order to answer the above questions we now extend
the study of the stellar population to the extranuclear regions of Seyfert galaxies.
As a first step towards this goal, in Raimann et al. (2001, hereafter R01) we have
analysed the nuclear and extranuclear stellar population of 3 early-type (Hubble
type S0) non-Seyfert, 2 LINER and 3 Seyfert 2 galaxies as a function of distance from the
nucleus.
While we found a common behaviour for non-Seyfert galaxies and LINERs,
this is not true for the 3 Seyferts, and a larger sample is needed
to characterize the stellar population gradients in Seyferts.
We have thus decided to extend this study
to the samples of Storchi-Bergmann et al. (2000, called southern sample) and Gonz\' alez
Delgado et al. (2001, called northern sample), which comprise 25 Seyfert 2 galaxies with
$z<0.02$ plus 10 with $0.02<z<0.05$. This sample can be considered as representative of
nearby Seyfert 2 galaxies. The common selection criterion for all the galaxies is a
lower limit for the luminosity of the central source, such that the luminosity in
the [OIII]$\lambda$5007 line is L$_{[OIII]}>10^{40}$ ergs s$^{-1}$. As it was not selected
by any property related to the stellar population, it is suitable to explore this property.

As the spectra of the northern and southern samples cover distinct wavelength ranges, we
have performed the two studies separately. In this work we analyse the northern sample
and in a future work the southern one. We first quantify the radial
variations of the equivalent widths and continuum fluxes as a function of distance
from the nucleus, used to measure the dilution of the bulge
absorption lines by an FC or young stellar population at the nucleus.
Then we use these measurements to perform spectral synthesis and derive
the stellar population properties and gradients.

This paper is organized as follows. In Section 2 we describe the sample galaxies and the
observations. In Section 3 we discuss the spatial variation of the equivalent widths and
continuum. We present the spectral synthesis results in Section 4 and, in Section 5,
the discussion and in Section 6 the conclusions of this work.

\section{Sample and Observations}

\begin{table}
\caption{The Seyfert 2 galaxies sample.}
\label{Obj1}
\begin{tabular}{lcccc} \hline
Name    & Type          & v (km/s) & 1 arcsec (pc) & E(B-V)$_G$ \\ \hline
Mrk1    & S             &  4780    & 310           & 0.060 \\
Mrk3    & S0:           &  4050    & 260           & 0.188 \\
Mrk34   & S:            &  15150   & 980           & 0.009 \\
Mrk78   & SB            &  11145   & 720           & 0.035 \\
Mrk273  & Ring. Gal pec &  11334   & 730           & 0.008 \\
Mrk348  & SA(s)0/a:     &  4540    & 290           & 0.067 \\
Mrk463E & S pec         &  14904   & 960           & 0.030 \\
Mrk477  & S             &  11340   & 730           & 0.011 \\
Mrk533  & SA(r)bc pec   &  8713    & 563           & 0.059 \\
Mrk573  & SAB(rs)0+:    &  5174    & 335           & 0.023 \\
Mrk1066 & SB(s)0+       &  3605    & 235           & 0.132 \\
Mrk1073 & SB(s)b        &  6991    & 450           & 0.160 \\
NGC1068 & SA(rs)b       &  1136    & 75            & 0.034 \\
NGC1386 & SB(s)0+       &   924    & 60            & 0.012 \\
NGC2110 & SAB0-         &  2284    & 150           & 0.375 \\
NGC5135 & SBab          &  4112    & 270           & 0.060 \\
NGC5929 & Sab:pec       &  2753    & 180           & 0.024 \\
NGC7130 & Sa pec        &  4842    & 310           & 0.029 \\
NGC7212 & S             &  7994    & 515           & 0.072 \\
IC3639  & SBbc          &  3285    & 210           & 0.069 \\ \hline
\end{tabular}
\end{table}

\begin{table}
\caption{The non-Seyfert galaxies sample.}
\label{Obj2}
\begin{tabular}{lcccc} \hline
Name    & Type          & v (km/s) & 1 arcsec (pc) & E(B-V)$_G$ \\ \hline
NGC1232 & SAB(rs)c      & 1500     & 97            & 0.026 \\
NGC1367 & (R')SAB(r'l)a & 1500     & 97            & 0.024 \\
NGC1425 & SA(rs)b       & 1500     & 97            & 0.013 \\
NGC1637 & SAB(rs)c      &  600     & 39            & 0.040 \\
NGC3054 & SAB(r)bc      & 2400     & 155           & 0.074 \\
NGC3223 & SA(r)bc       & 3000     & 194           & 0.108 \\
NGC3358 & (R'2)SAB(l)ab & 3000     & 194           & 0.055 \\
NGC7049$^a$ & SA(s)0        & 2231     & 116           & 0.007 \\ \hline
\end{tabular}

$^a$ Observed by Raimann et al. (2001).
\end{table}

Detailed information about the observations and reductions of the long-slit spectra of the
northern sample can be found in Gonz\' alez Delgado et al. (2001).
It comprises the brightest 20 Seyfert 2 galaxies from the compilation of
Whittle (1992a,b). The radial velocities range from 1000\,km/s to 15000\,km/s but only
five objects have radial velocities $>$ 9000\,km/s. The value mean is 4811\,km/s, what gives
a mean spatial scale of 310\,pc/arcsec. In Table \ref{Obj1}
we list the morphological type, radial velocity, spatial scale (using H$_{0}$ = 75 km
s$^{-1}$ Mpc$^{-1}$) and foreground galactic E(B-V)$_G$ values. The data were extracted from
the NASA/IPAC Extragalactic Database (NED)\footnote{The NASA/IPAC Extragalactic Database
(NED) is operated by the Jet Propulsion Laboratory, California Institute of Technology,
under contract with the National Aeronautics and Space Administration.}.

Long-slit spectra of these galaxies have been obtained using the Richey-Chr\' etien
spectrograph at the 4m Mayall telescope of Kitt Peak National Observatory in February and
Ocotber 1996. The long-slit spectra used in this work cover the wavelength range 3600 --
4700 \AA, at a resolution of $\approx$ 3 \AA. Each pixel at the detector corresponds to 0.7
arcsec in the sky. A slit width corresponding to 1.5 arcsec on the sky was oriented along the
parallactic angle except for NGC1068 on which it was oriented at P.A.=123$^{\circ}$. For two
galaxies, NGC5929 and Mrk477, we have two position angles, P.A.=60,80 and P.A.=44,158 respectively.

One-dimensional spectra were extracted in windows of 2.1 arcsec in the bright nuclear regions
and progressively larger windows towards the fainter outer regions. The spectra were dereddened
according to foreground galactic E(B-V)$_G$. The spatial coverage ranged between 3 (for Mrk34)
and 50 arcsec (NGC1068) from the nucleus. The signal-to-noise (S/N) ratio of the extracted spectra
ranges between 10 and 40.

In order to compare the stellar population of these active galaxies with that of
non-active ones of the same Hubble type, we have also obtained long-slit spectra of
nearby non-Seyfert galaxies with morphological types between Sa and Sc. We used the R-C
Cassegrain Spectrograph at the 1.5m telescope of Cerro Tololo Interamerican Observatory
in January 2002 (Table \ref{Obj2}). Due to the proximity of these galaxies the nuclear
spectrum was extracted in a window of 3 arcsec, in order to allow a similar sampling at
the galaxy to that of the Seyferts. Extranuclear spectra were extracted up to about 8
arcsec from the nucleus. The spectra were dereddened according to foreground galactic
E(B-V)$_G$. In addition, to represent the S0 Hubble type, we use the data from the S0
non-Seyfert galaxy NGC7049 (see bottom of Table \ref{Obj2}), already presented in R01,
which is representative of the 3 S0 galaxies studied in that paper.

Because the number of non-Seyfert galaxies is still small, we use as additional reference the study
of Bica (1988, hereafter B88), in which he performed stellar population synthesis for about 100
non-Seyfert galaxies, using integrated spectra from regions of about 1\,kpc $\times$
1\,kpc. B88 grouped the galaxies spectra in six templates, called S1 to S6.
The templates S1 and S2 are dominated by galaxies with Hubble types Sa, the templates S3 and S4
by Sb's and the templates S5 and S6 by Sc's.

\section{Radial variation of the equivalent widths and continuum colours}

The variation of the equivalent widths (Ws) of absorption lines and the 
continuum colours as a function of distance from the nucleus allows the study of stellar
population gradients in the galaxy. Usually, in non-active galaxies, the Ws increase
from the external regions towards the bulge, where they remain approximately constant.
The presence of a burst of star-formation and/or a featureless AGN continuum
will produce a ``dilution'' of the absorption lines, whose Ws will then decrease at the
nucleus when compared to the values at adjacent locations (Cid Fernandes et al. 1998).

In order to study the run of the stellar absorption features with distance from the nucleus
we determined a pseudo-continuum at selected pivot-points and measured the Ws of six
absorption features as described in R01. Due to the shorter
spectral range of the present work the pivot points used here for the continuum were
3660, 3780, 4020 and 4510 \AA\ and the absorption features for which we measured the Ws
were: W$_{wlb}$ (a blend of weak
lines in the near-UV, within the spectral window $\lambda\lambda$3810-3822 \AA), H9
(a blend of absorption lines which includes H9, $\lambda\lambda$3822-3858 \AA),
Ca{\sevensize II}K ($\lambda\lambda$3908-3952 \AA), Ca{\sevensize II}H+H$\epsilon$
($\lambda\lambda$3952-3988 \AA), CN-band ($\lambda\lambda$4150-4214 \AA) and G-band
($\lambda\lambda$4284-4318 \AA).

\begin{figure*}
\vspace{22cm}
\caption{Seyfert 2 galaxies: radial variations of the equivalent widths (Ws), 
continuum colour and surface
 brightness. The first panel shows W$_{wlb}$ (solid line) and W$_{H9}$
(dotted), the second shows W$_{CaIIK}$ (solid) and W$_{CaIIH+H\epsilon}$ (dotted),
the third, W$_{CN-band}$ (solid) and W$_{G-band}$ (dotted). The fourth panel shows
the continuum colour $C_{4510/4020}$. The fifth panel shows the run of the surface brightness
at 4020 \AA\ (in units of 10$^{-15}$ erg cm$^{-2}$ s$^{-1}$ \AA $^{-1}$ arcsec$^{-2}$) along
the slit. The dotted and dashed vertical lines mark distances of 1\,kpc and 3\,kpc from the
nucleus, respectively.}
\label{var1}
\includegraphics{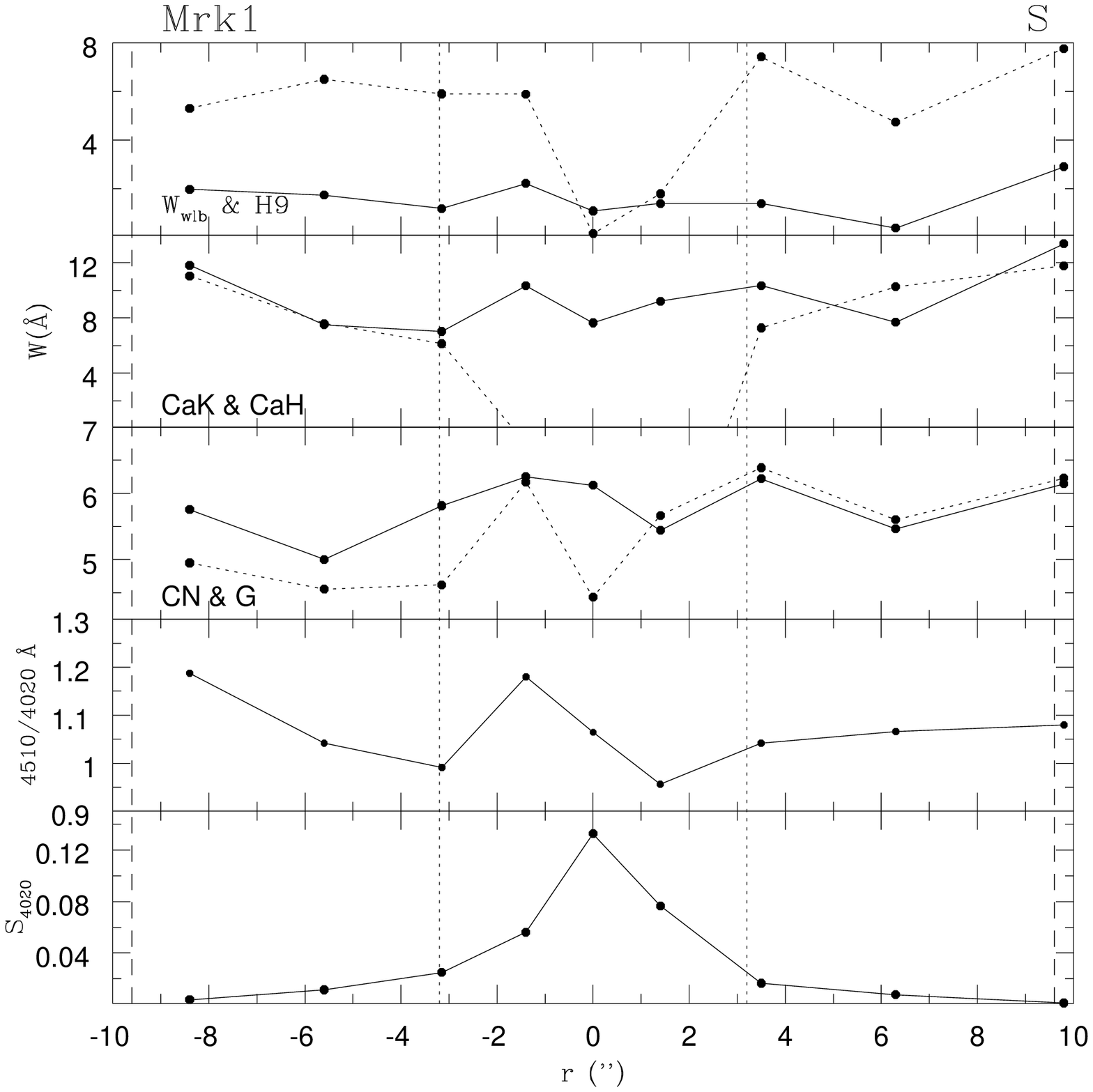}
\includegraphics{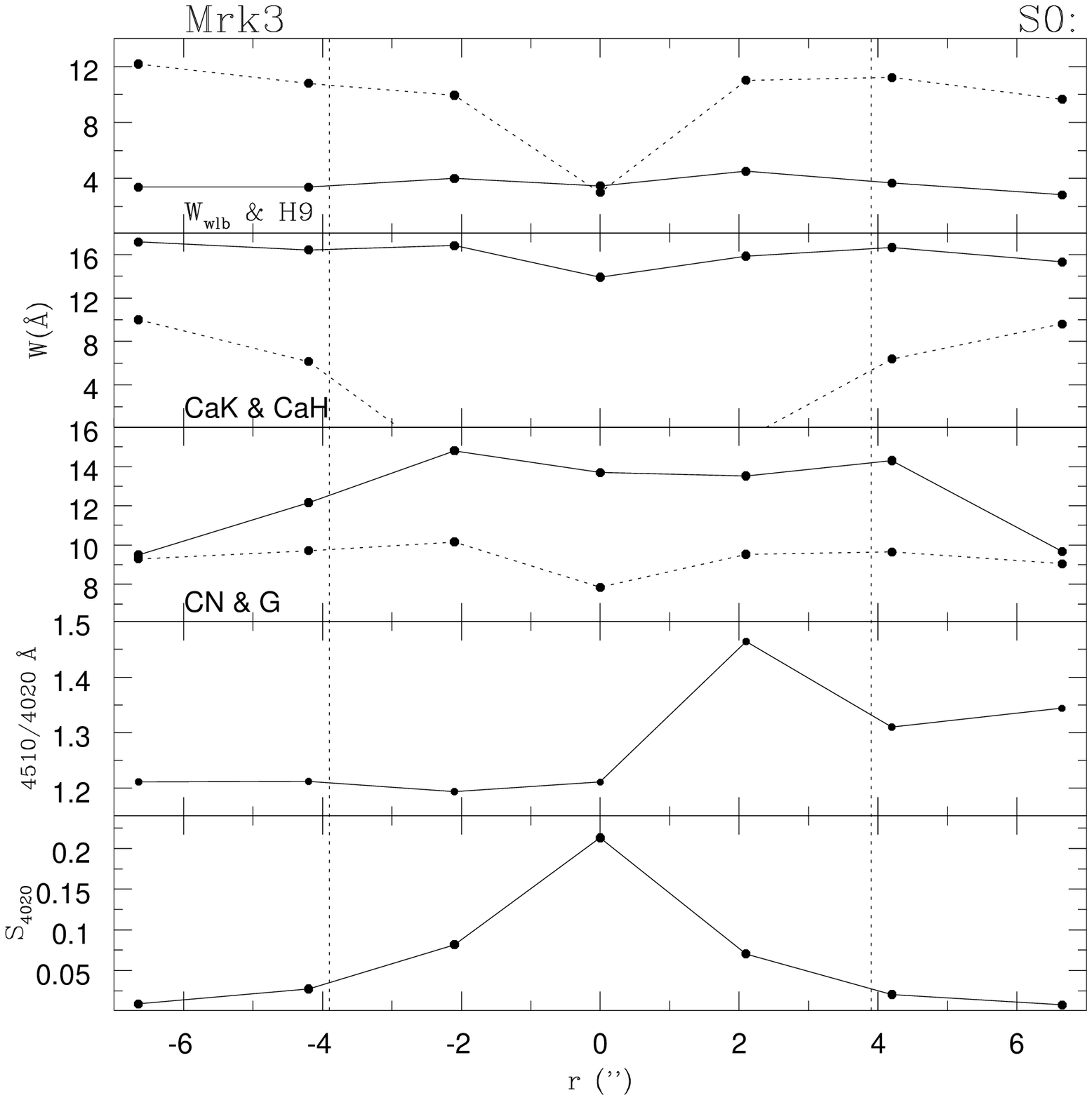}
\includegraphics{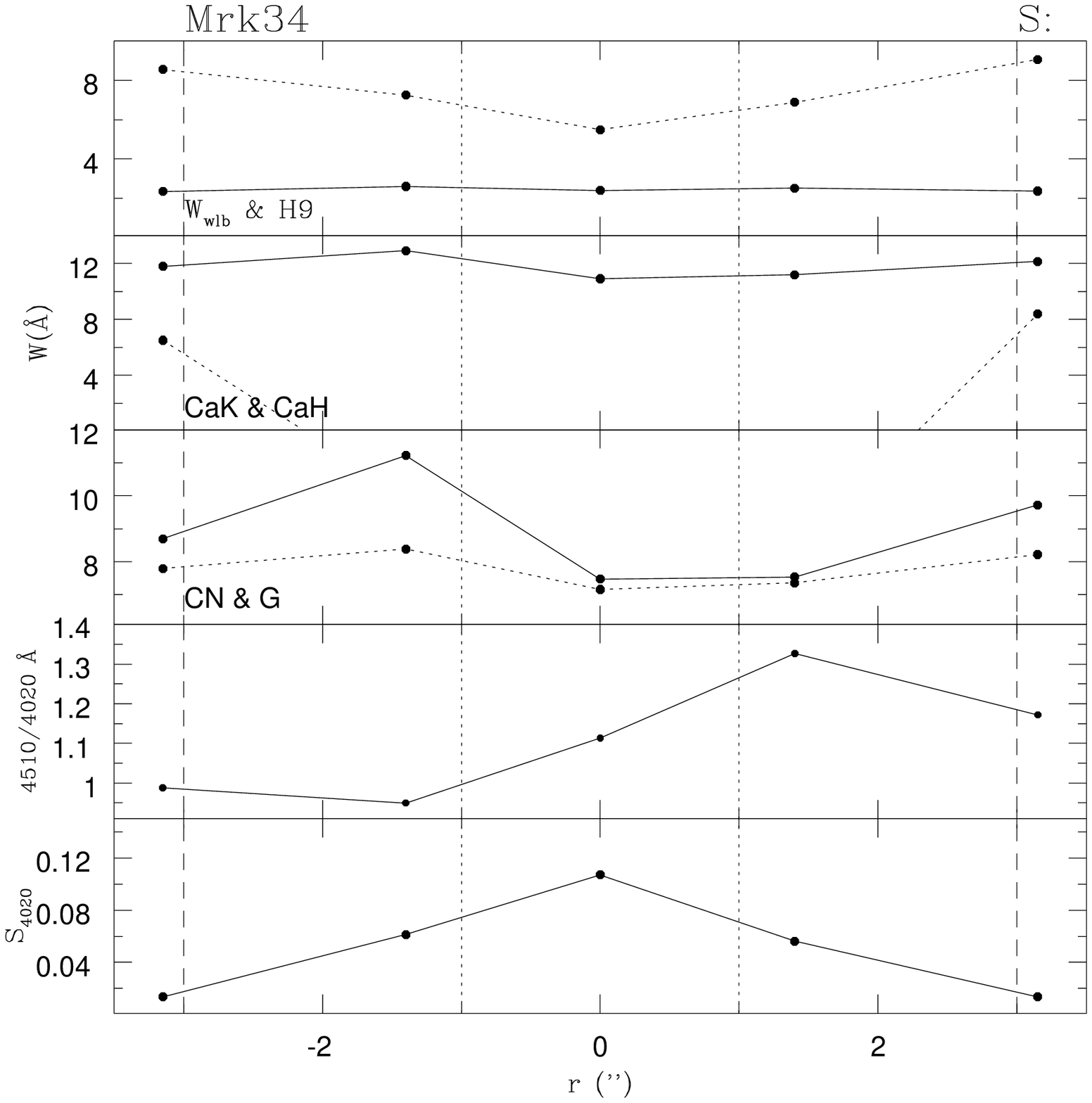}
\includegraphics{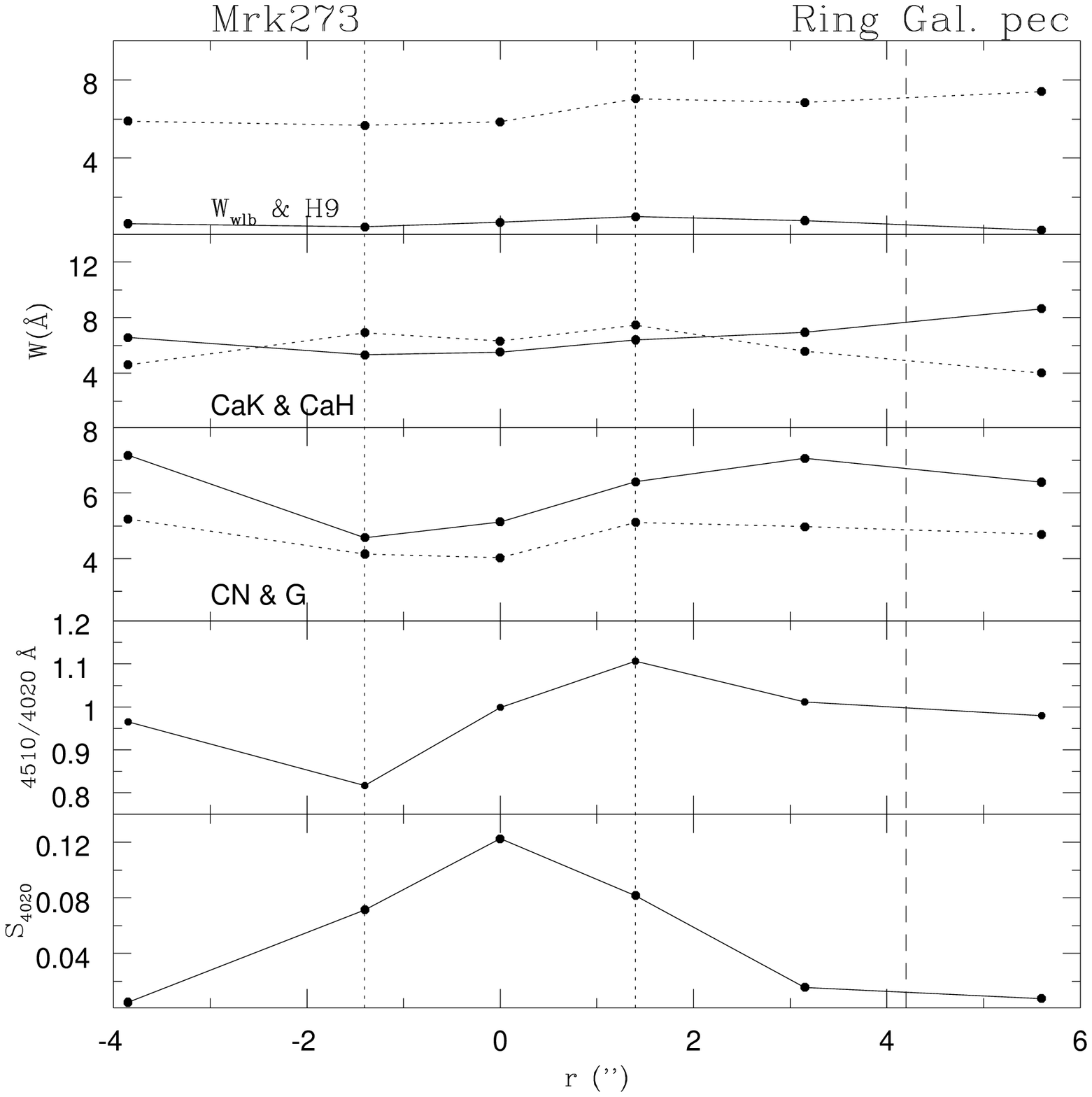}
\includegraphics{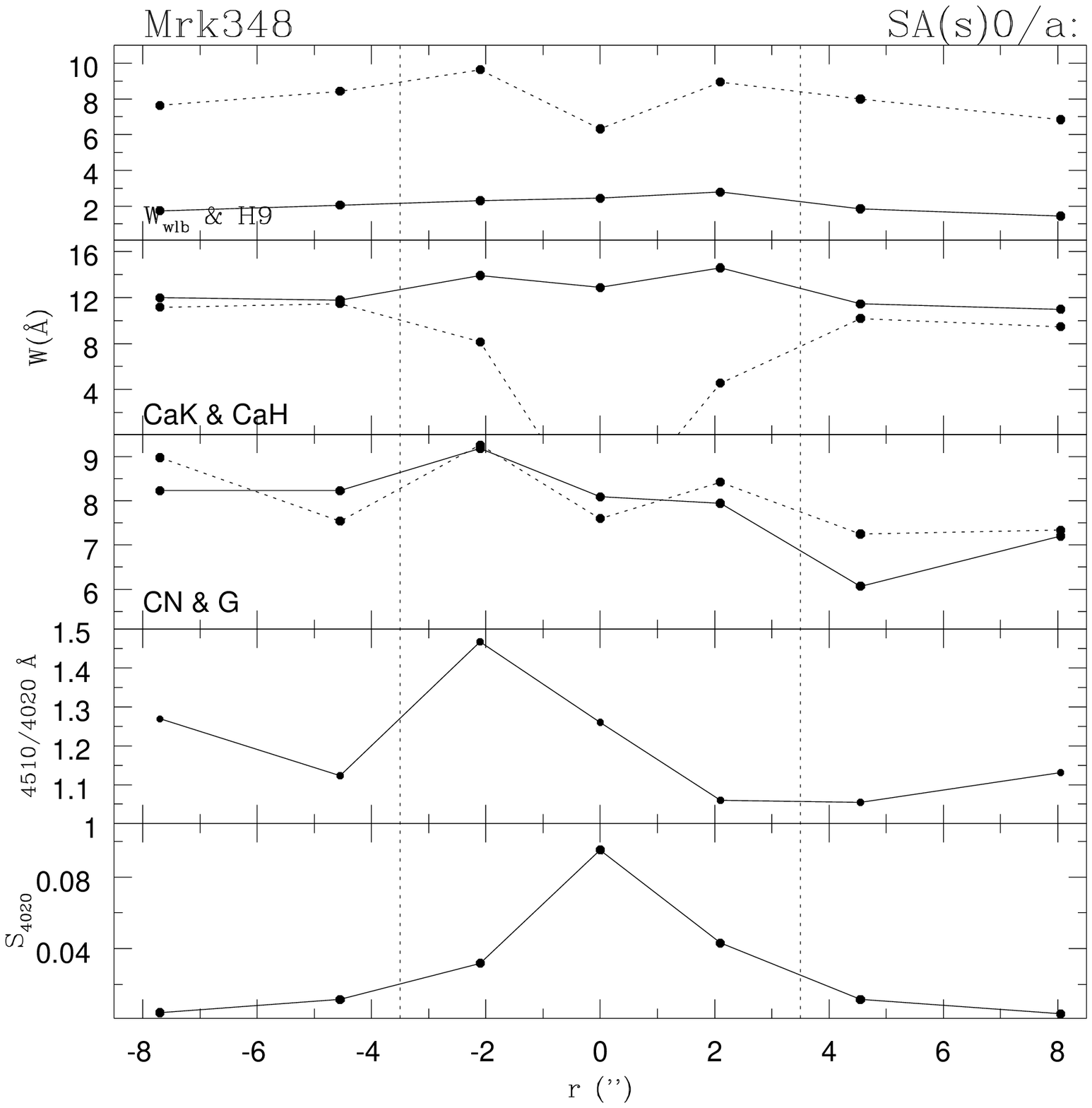}
\includegraphics{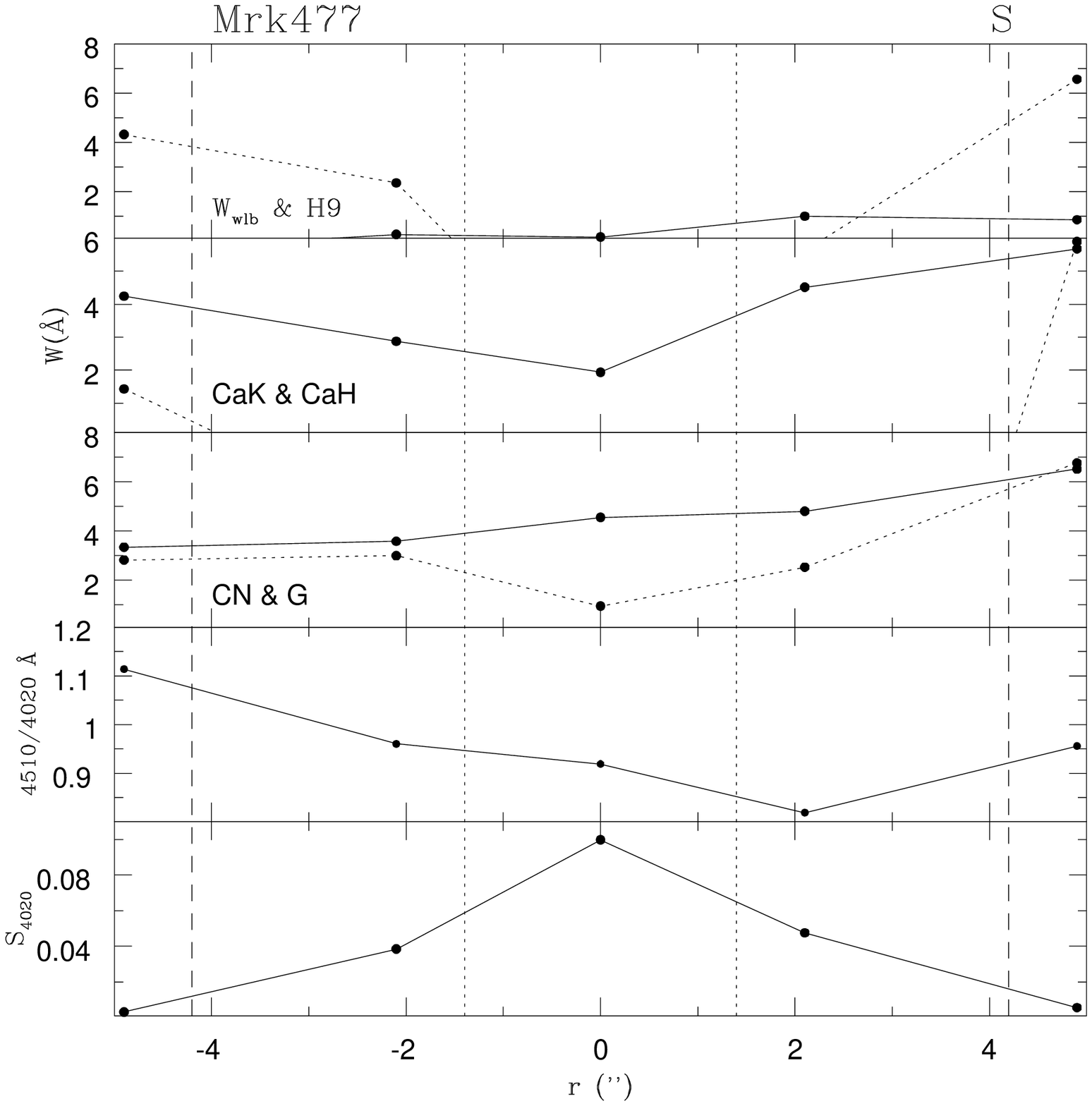}
\end{figure*}

\begin{figure*}
\vspace{22cm}
\caption{Seyfert 2 galaxies: Radial variations of the equivalent widths (Ws), 
continuum colour and surface brightness. Symbols as in Fig. \ref{var1}.}
\label{var2}
\includegraphics{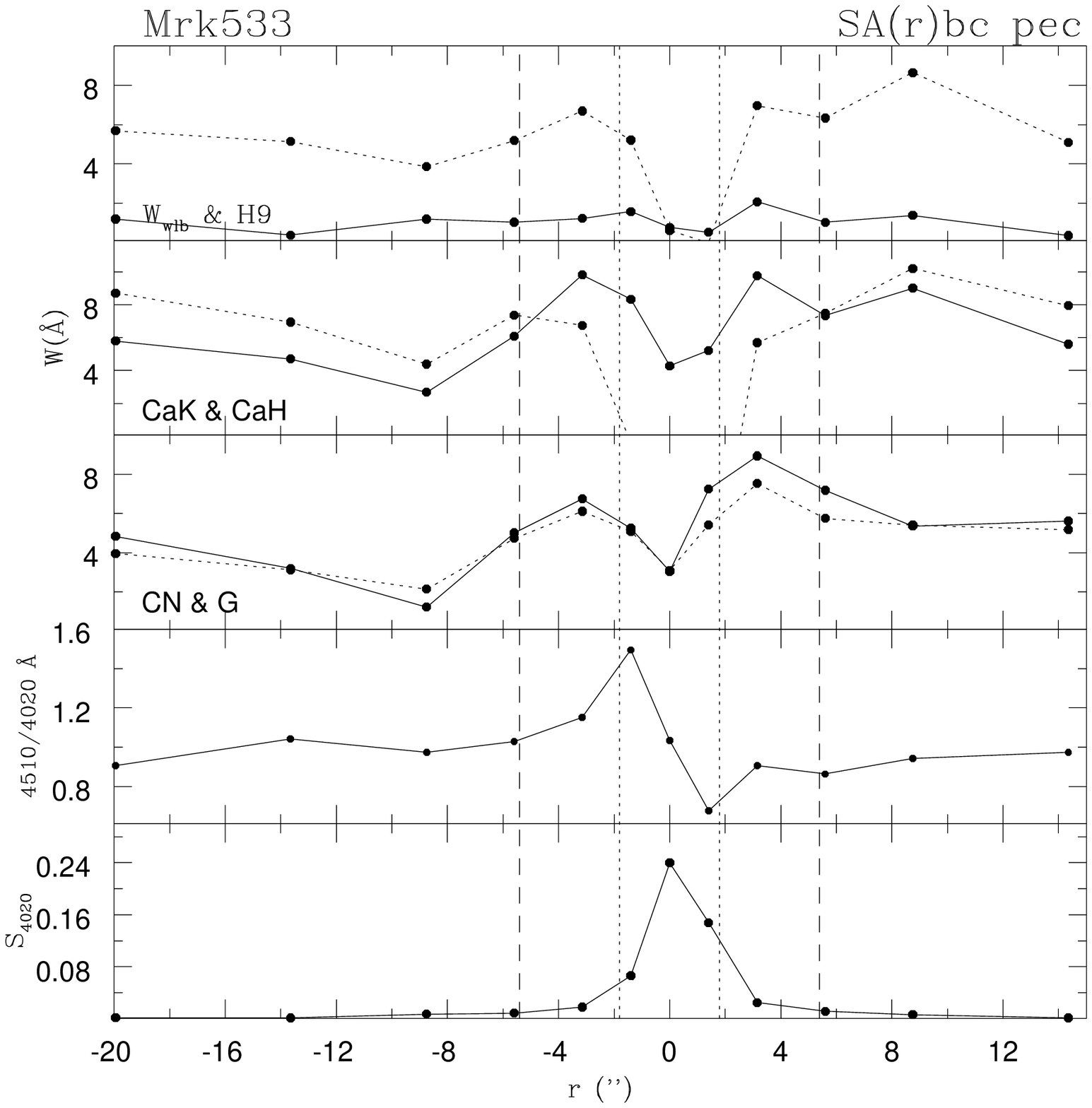}
\includegraphics{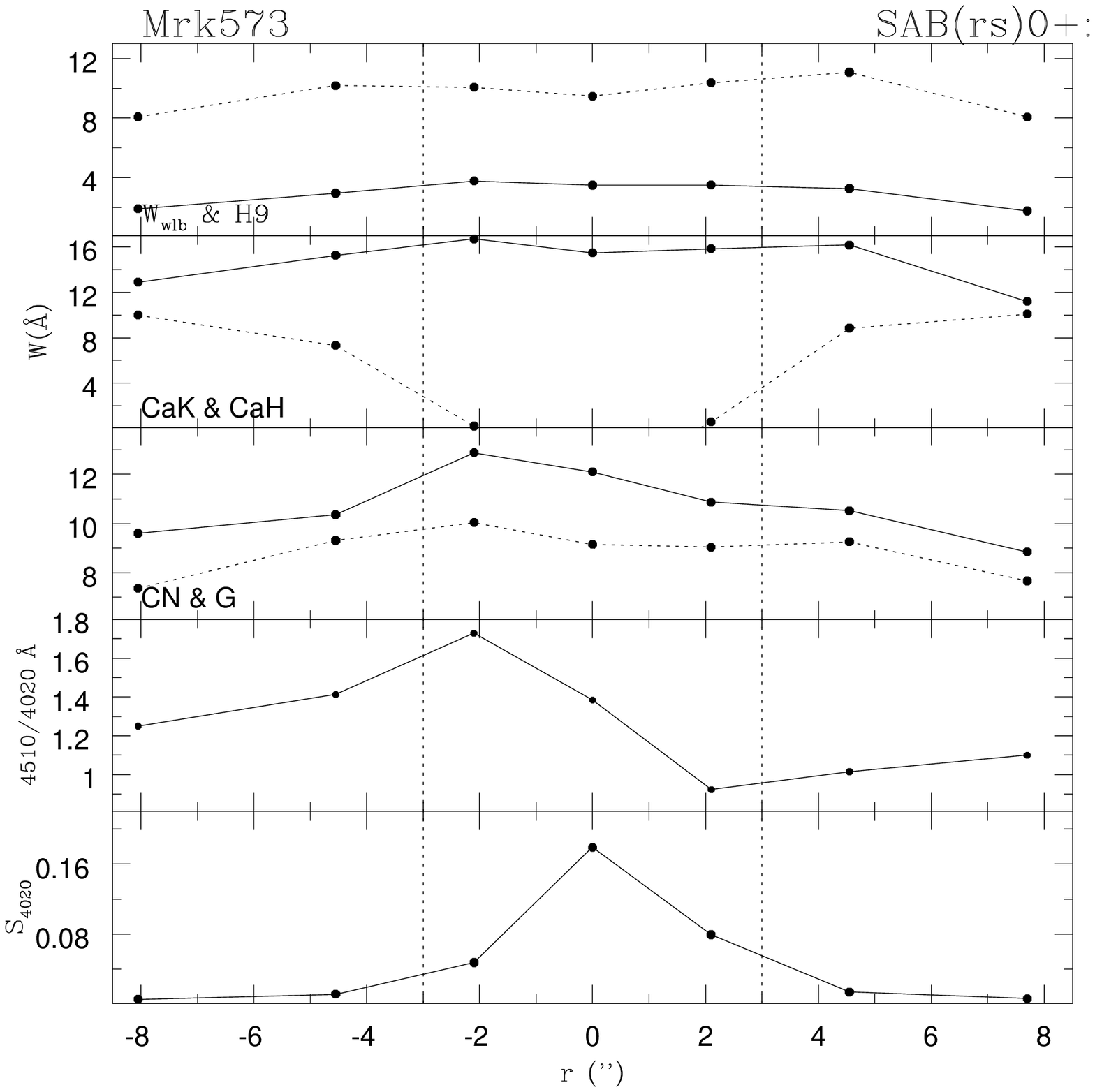}
\includegraphics{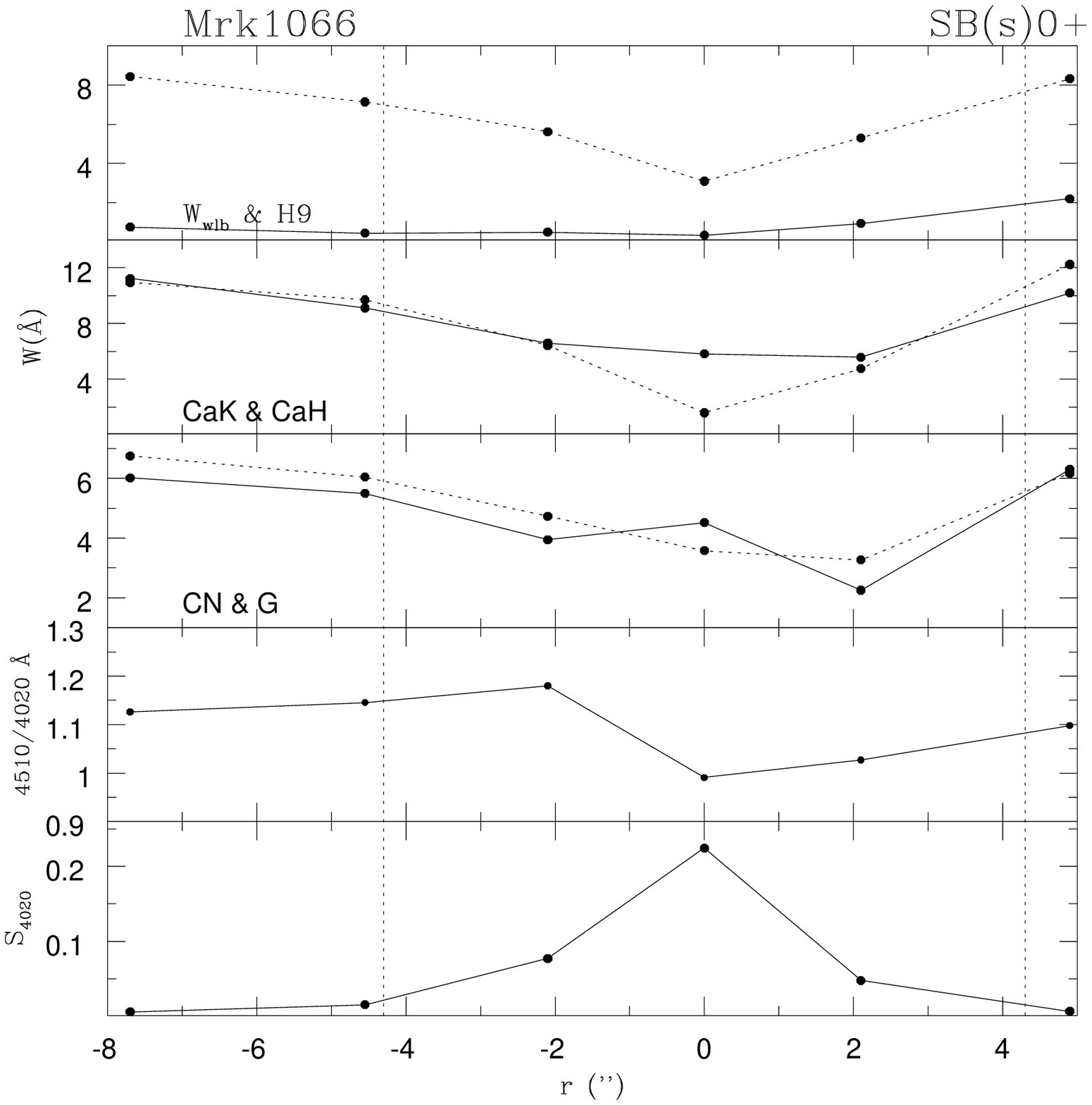}
\includegraphics{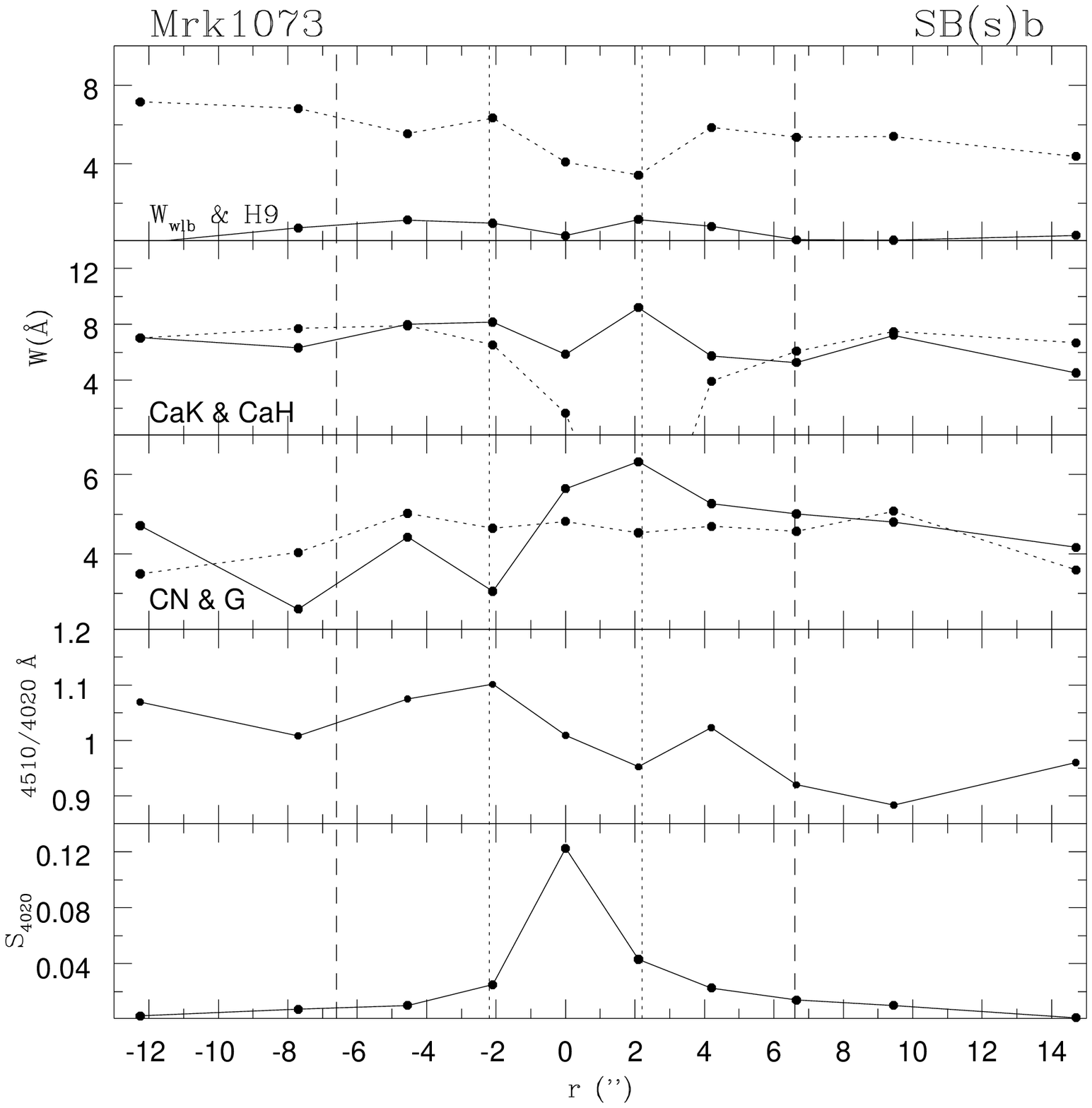}
\includegraphics{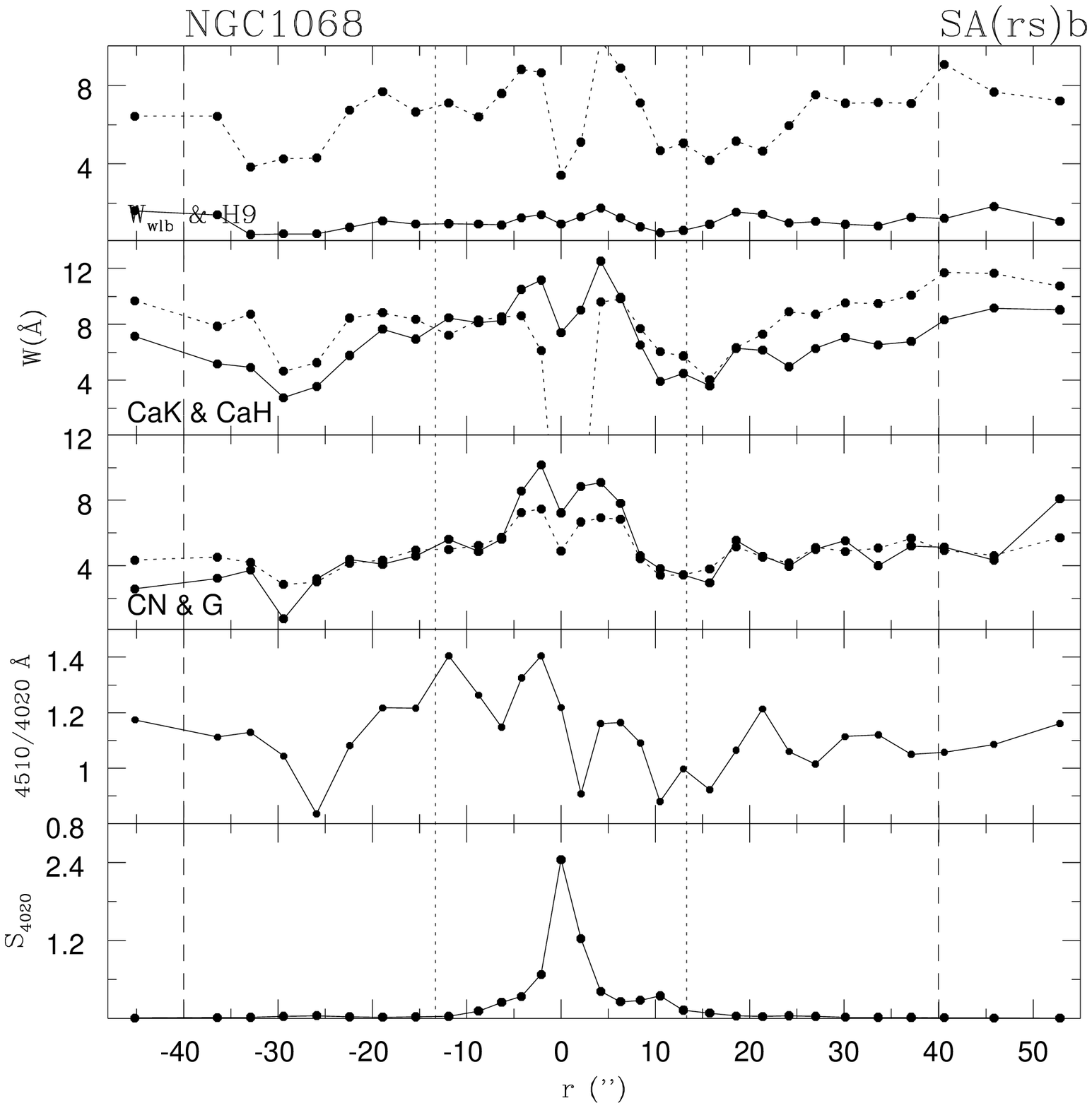}
\includegraphics{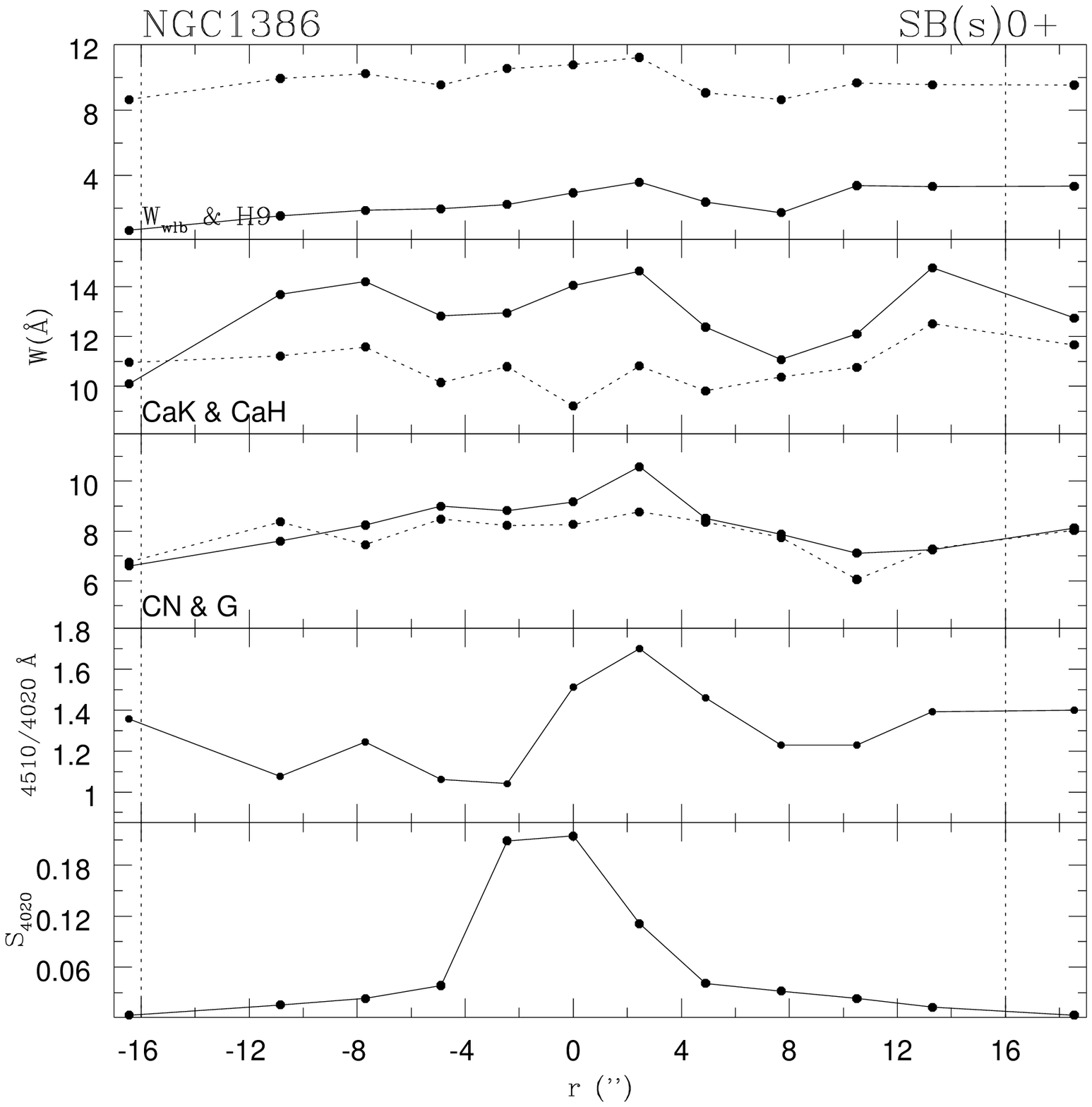}
\end{figure*}

\begin{figure*}
\vspace{22cm}
\caption{Seyfert 2 galaxies: radial variations of the equivalent widths (Ws), 
continuum colour and surface brightness. Symbols as in Fig. \ref{var1}.}
\label{var3}
\includegraphics{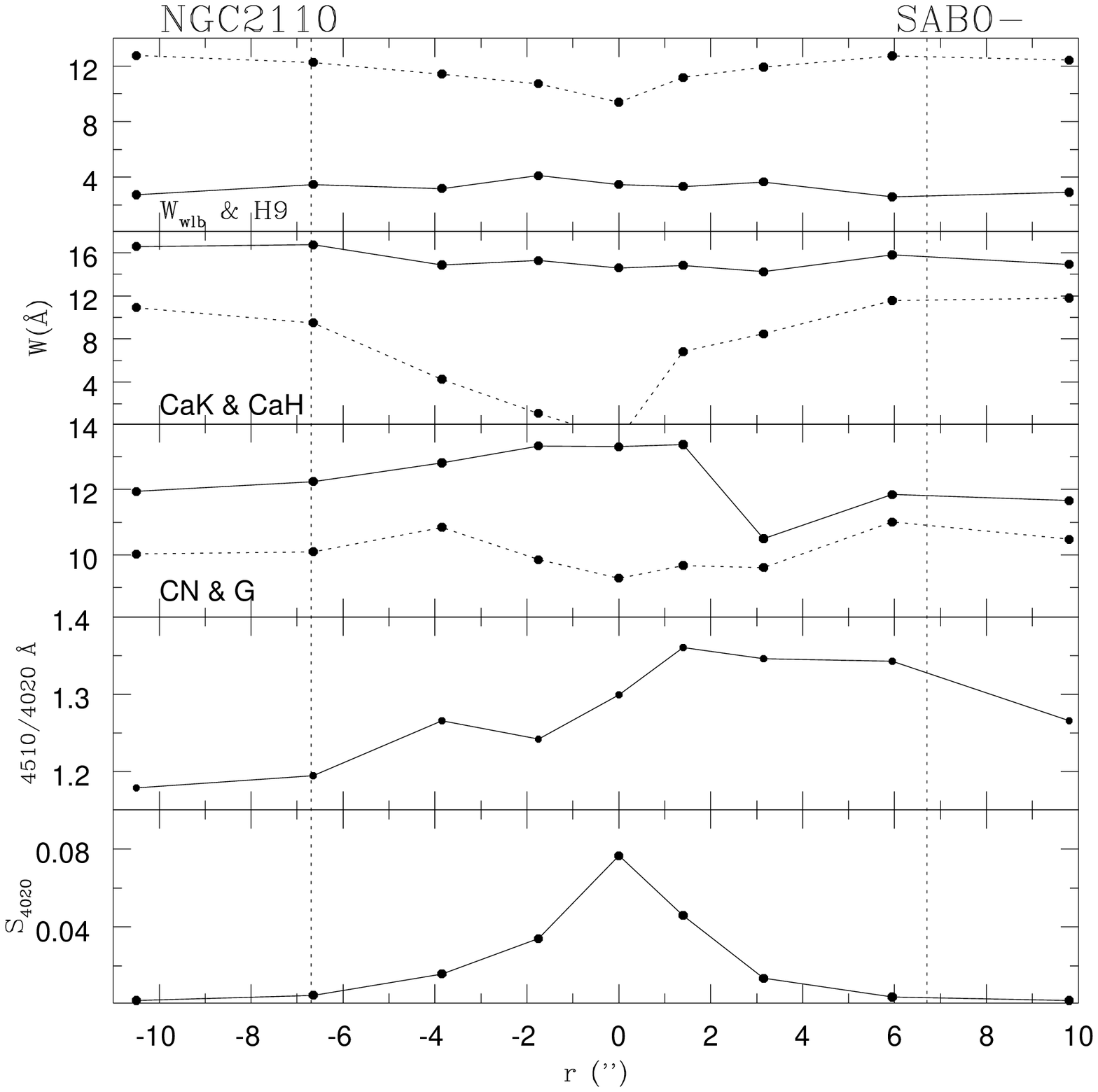}
\includegraphics{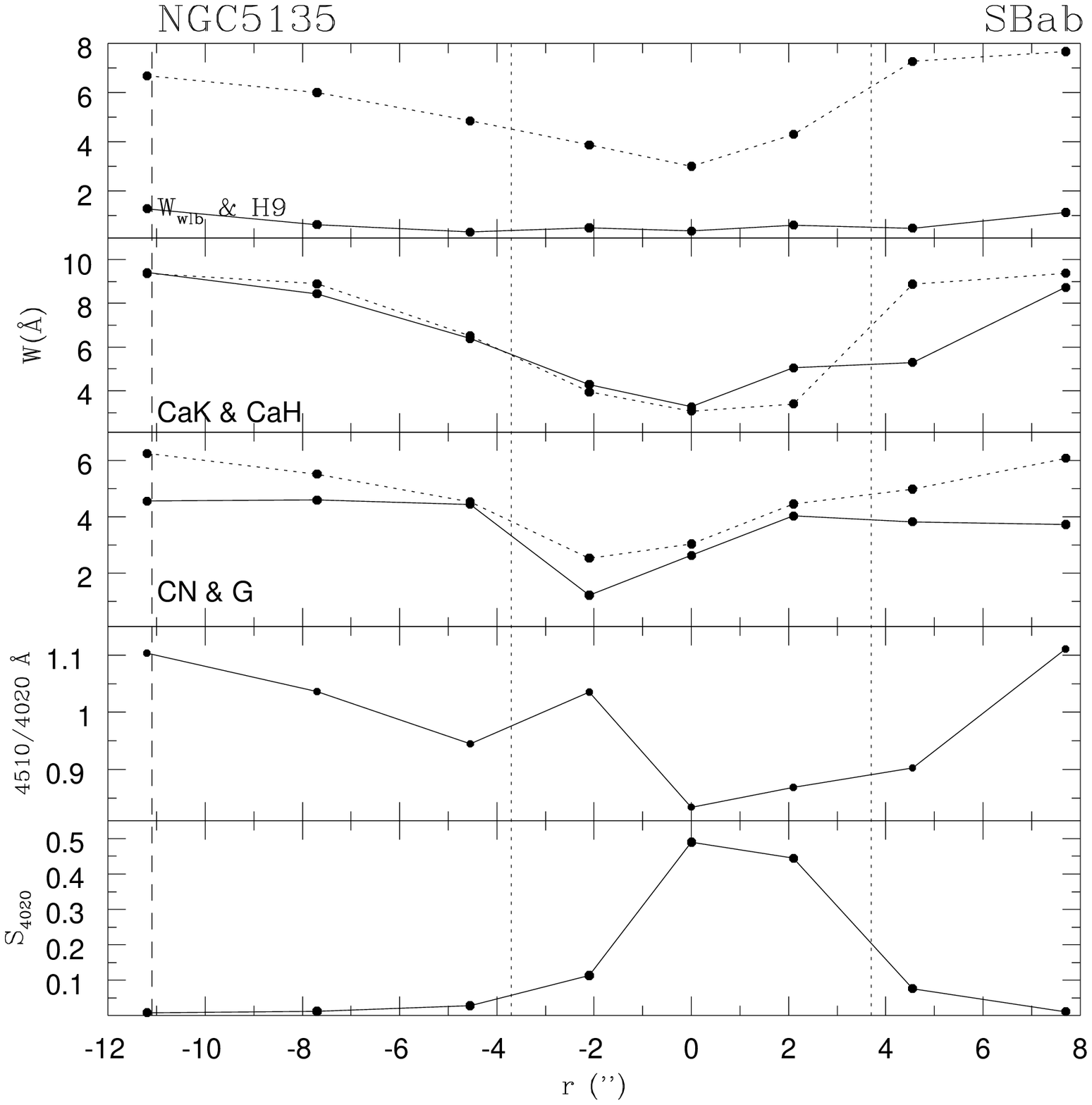}
\includegraphics{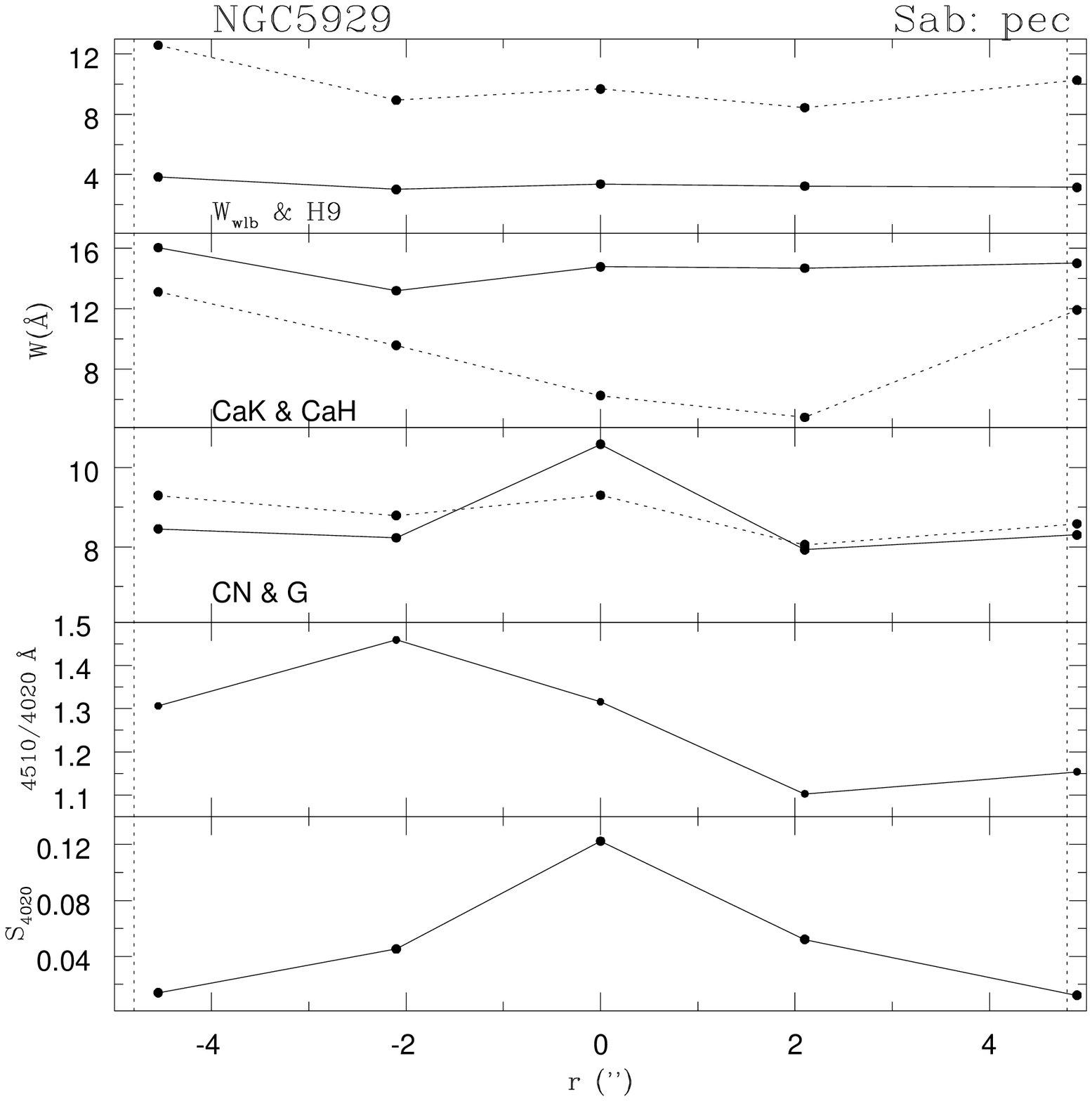}
\includegraphics{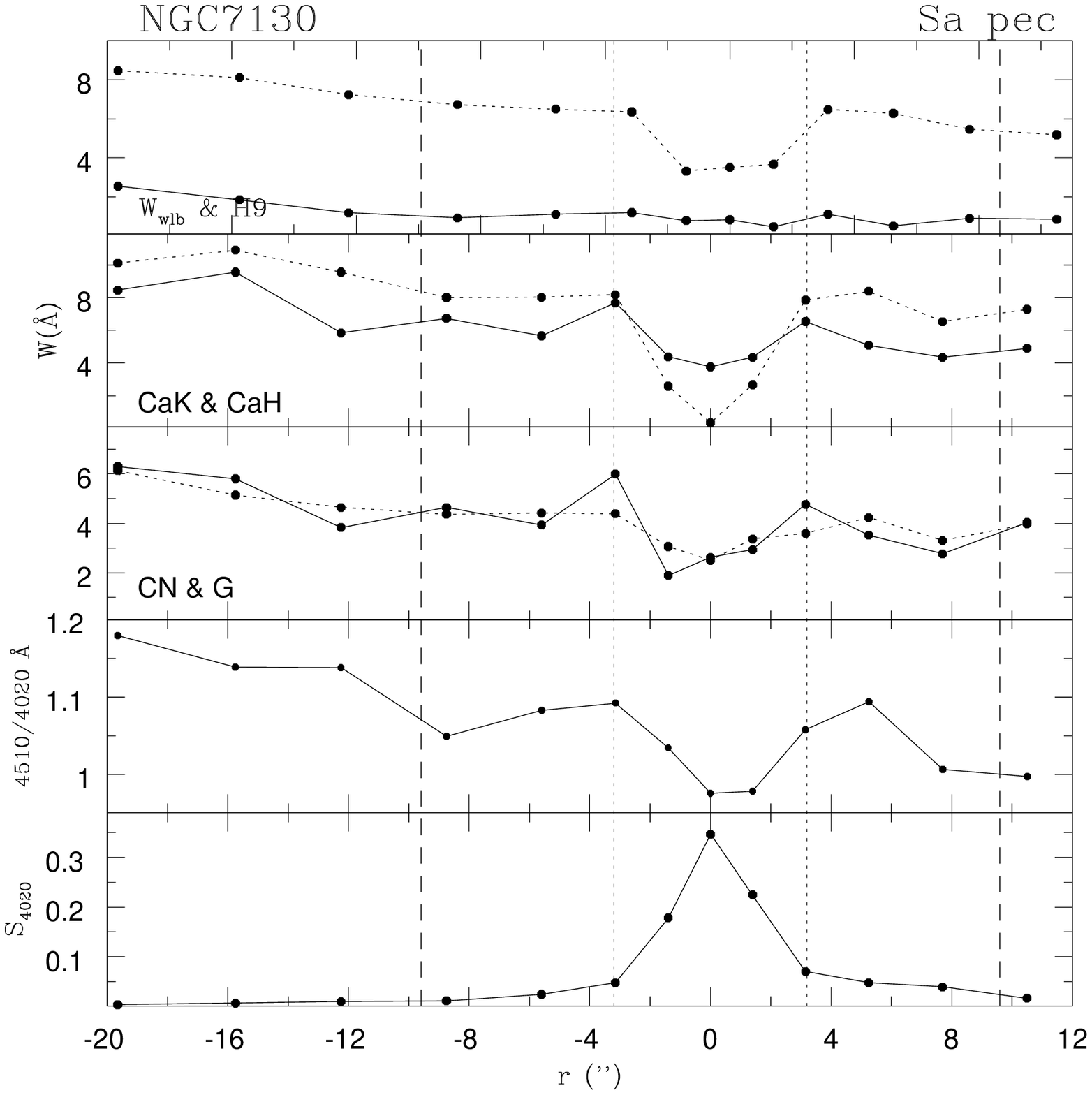}
\includegraphics{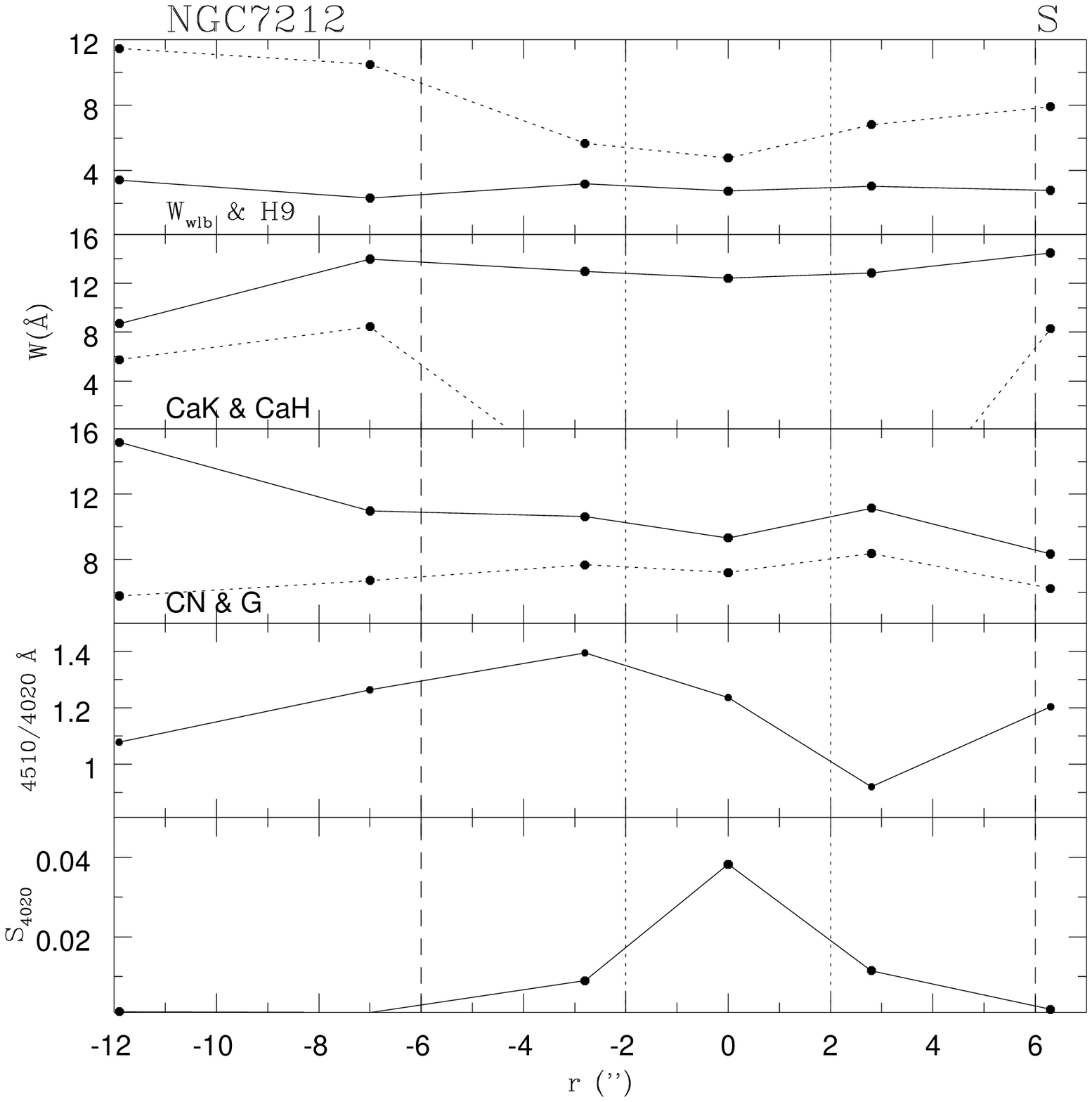}
\includegraphics{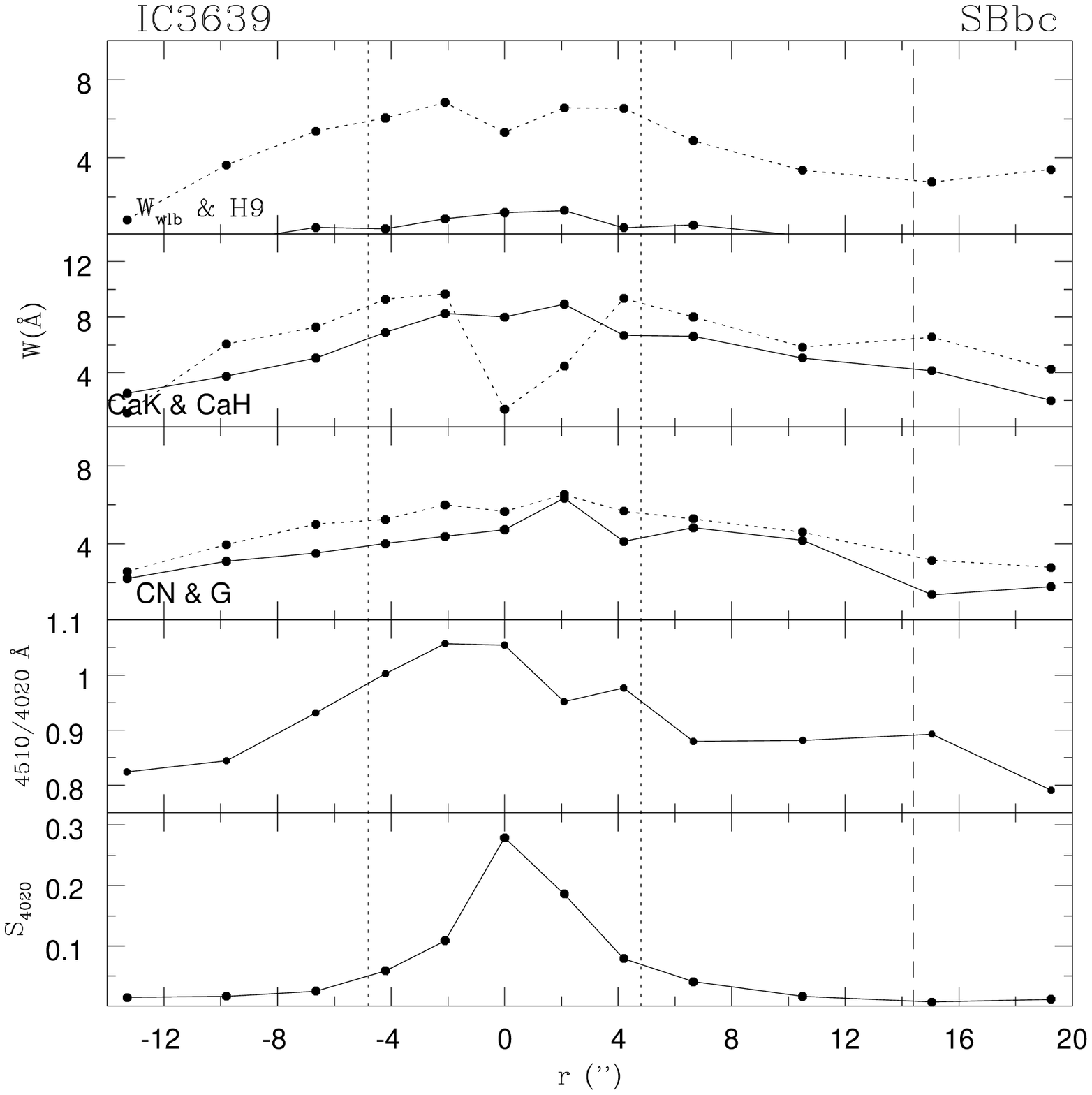}
\end{figure*}

\begin{figure}
\vspace{7cm}
\caption{Non-Seyfert S0 galaxy NGC7049: radial variations of the equivalent widths (Ws),
continuum colour and surface brightness. Symbols as in Fig. \ref{var1}.}
\label{varnormal}
\includegraphics{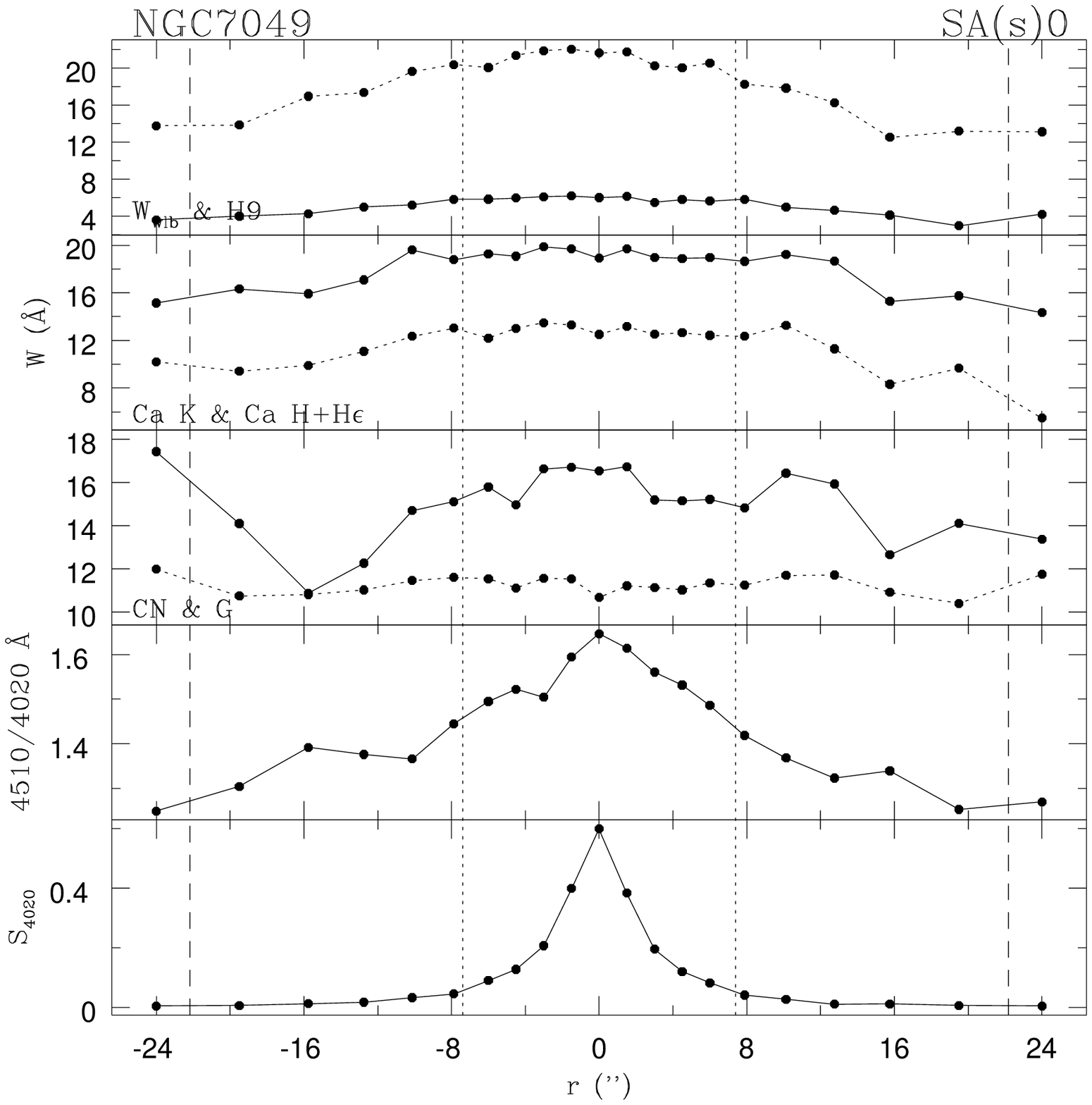}
\end{figure}

In Figs. \ref{var1}-\ref{var3} we present the above Ws, continuum colour F$_{4510}$/F$_{4020}$
(hereafter $C_{4510/4020}$) and surface brightness as a function of the angular distance
from the nucleus for the Seyfert 2 galaxy sample. The typical errors are 0.5\AA\ for W$_{wlb}$,
H9, Ca{\sevensize II}K, Ca{\sevensize II}H+H$\epsilon$ and G-band, 1.0\AA\ for the CN-band and 0.05
for the continuum ratio (Cid Fernandes et al. 1998). The dotted and dashed vertical lines mark
distances at each galaxy of 1\,kpc and 3\,kpc from the nucleus, respectively. The
measurements for Mrk463 are not presented here because this data is already shown in R01.
We also do not show the data for Mrk78 because it has only one
extranuclear spectrum to each  side. We present data for just one position angle of Mrk477 and
NGC5929 because the other p.a.'s show similar behavious.
Measurements of Ca{\sevensize II}K and G-band Ws for some of these objects were showed
in Gonz\' alez Delgado et al. (2001) but they are different from those showed here
due to a different pseudo-continuum adopted. As five galaxies of the present sample are also in
the sample of Cid Fernandes et al. (1998), we can compare our measurements with theirs in order
to check the robustness of on method. We point out that the measurements in the above paper were
obtained with a different instrument and along different position angles, although using the
same technique. Nevertheless, a careful comparison shows that the W values for the same feature
are remarkably similar, usually within the measurement errors.

In Fig. \ref{varnormal} we show the same quantities for the non-Seyfert S0 galaxy NGC7049, first
presented in R01. For the non-Seyfert Sa to Sc galaxies, it was only possible to obtain
signal-to-noise ratios high enough for reliable measurements for the nucleus and usually a
couple of extractions to each side of the nucleus. The corresponding values are summarized
below.

The range of nuclear Ws obtained for Seyfert 2 galaxies is 0$<$W$_{wlb}<$3.5\AA, 0$<$W$_{H9}<$11\AA,
2$<$W$_{CaIIK}<$15.5\AA, 0$<$W$_{CaIIH+H\epsilon}<$9\AA, 2.5$<$W$_{CN-band}<$14\AA\ and
1$<$W$_{G-band}<$9.5\AA. While the nuclear values for the S0 are W$_{wlb}\approx$6\AA,
W$_{H9}\approx$20\AA, 18$<$W$_{CaIIK}<$20\AA, 12$<$W$_{CaIIH+H\epsilon}<$14\AA,
17$<$W$_{CN-band}<$19\AA\ and W$_{G-band}\approx$11\AA, those of the Sa to Sc are
1$<$W$_{wlb}<$4\AA, 6$<$W$_{H9}<$15\AA, 7$<$W$_{CaIIK}<$18\AA,
8$<$W$_{CaIIH+H\epsilon}<$13\AA, 3$<$W$_{CN-band}<$13\AA\ and 4$<$W$_{G-band}<$10\AA.
The non-Seyfert Sa to Sc galaxies thus present nuclear W values considerably smaller than that
of the non-Seyfert S0, but the radial variations show a similar trend.

Almost all Seyfert 2s have some gas emission in the nuclear lines H9 and Ca{\sevensize II}H+H$\epsilon$,
and sometimes this emission is present in the extranuclear spectra too. Mrk463E and Mrk477 are
examples of objects dominated by extended gas emission. Therefore, almost all Seyfert 2 galaxies have
these lines with smaller Ws in and near the nucleus.

In non-Seyfert S0 galaxies most Ws have the largest values at the nucleus, decreasing outwards (e.g.,
Fig. \ref{varnormal}). Two Seyferts, Mrk348 and IC3639, present a similar trend, with Ws
increasing towards the nucleus, although at the nucleus proper they show a sudden decrease.
If we focus our attention in the behaviour of W$_{CaIIK}$ and G-band, the lines with the smallest
measurement errors, we realize that about half the sample (galaxies listed in Table \ref{varws})
have these nuclear Ws smaller than the extranuclear ones. This behaviour is probably due to
dilution of the bulge stellar population by a blue continuum which can either be due to a FC
component or a young stellar population, or both.

We estimate the fraction of the continuum associated with the above blue component $f(\lambda)$ 
at the wavelength of an absorption line as:

\begin{equation}
f(\lambda)=\frac{W_{off-nuc}-W_{nuc}}{W_{off-nuc}}
\end{equation}

\noindent where $W_{nuc}$ and $W_{off-nuc}$ are the Ws at the nucleus and outside,
respectively (Cid Fernandes et al. 1998). 

\begin{table}
\caption{Dilution percentage of the nuclear Ca{\sevensize II}K and G-band in relation to
1 kpc from the nucleus.}
\label{varws}
\begin{center}
\begin{tabular}{lcc} \hline
Name        & \hspace{2cm} f($\lambda$) [per cent] & \\
            & Ca{\sevensize II}K & G-band \\ \hline
Mrk1        & 13$\pm$13 & 17$\pm$13 \\
Mrk3        & 10$\pm$6  & 18$\pm$1  \\
Mrk463E     & 55$\pm$6  & 70$\pm$4  \\
Mrk477      & 45$\pm$12 & 66$\pm$2  \\
Mrk533      & 55$\pm$1  & 54$\pm$5  \\
Mrk1066     & 40$\pm$3  & 41$\pm$1  \\
NGC1068$^a$ & 35$\pm$6  & 30$\pm$2  \\
NGC2110     &  5$\pm$3  & 12$\pm$4  \\
NGC5135     & 44$\pm$6  & 36$\pm$3  \\
NGC7130     & 46$\pm$4  & 36$\pm$6  \\
NGC7212     &  3$\pm$1  & 10$\pm$4  \\  \hline
\end{tabular}

$^a$ Relative to 5 arcsec (375\,pc) from the nucleus.
\end{center}
\end{table}

In Table \ref{varws} we show the dilutions $f(\lambda)$ of the nuclear Ca{\sevensize II}K
and G-band. These dilutions are measurements of the contribution of a FC or a younger
population at the nucleus, as compared to that at 1 kpc. The uncertainties listed in the Table
are measurements of the difference between the dilution factors obtained using the two extranuclear
extractions at opposite sides of the nucleus. In the remaining Seyferts there is no obvious
dilution, nor a systematic variation with the distance similar to that observed for the S0 galaxy.

Regarding the continuum, two galaxies (Mrk78 and IC3639) have the nuclear $C_{4510/4020}$ redder than
the extranuclear ones similarly to those of the non-Seyfert early-type galaxies.
Four others have the nuclear continuum  bluer than that of the extranuclear region (Mrk1066,
Mrk477, NGC5135 and NGC7130) and in the remainder (14 galaxies) the $C_{4510/4020}$ behaviour
is not symmetrical relative to the nucleus. This is likely due to a non-uniform dust 
distribution, in agreement with HST images of the inner region of these galaxies, many of which 
were published by Malkan et al. (1998, their Fig. 2).

\subsection{NGC1068}
NGC1068 is considered the prototypical Seyfert 2 and it is the brightest galaxy of the sample,
allowing a more detailed study of its stellar population, as high signal-to-noise ratio spectra
could be extracted up to $\approx$ 50 arcsec (3.7 kpc) from the nucleus, at a spatial sampling
of $\approx$ 200\,pc. 

Neff et al. (1994) showed that this galaxy contains multiple components at UV wavelengths:
the central AGN, a population of very luminous starburst knots, a bright oval inner disk
and a fainter, more circular halo. The knots are mostly located in two star-formation rings.
One at 10 arcsec from the nucleus (750\,pc at the galaxy) -- the nuclear ring -- that
includes the most luminous knot  and other at 28 arcsec from the
nucleus (2.1 kpc) -- the inner ring -- that includes several luminous knots. Our slit
was oriented to cover one luminous knot from each ring, labeled regions C and J by Neff 
and collaborators.

In Fig. \ref{var2} (bottom left panel) we show the Ws, $C_{4510/4020}$ and surface
brightness as a function of the angular distance from the nucleus for NGC1068.
All Ws show a dip at the nucleus relative to the extranuclear values at $\approx$ 5 arcsec,
where the Ws reach the largest values (which are however still smaller than those of the non-Seyfert
early-type galaxy at similar location). Beyond 5 arcsec, the Ws decrease, consistent with the
presence of star-forming regions along the slit. In particular, at 10 arcsec NW on the nuclear
ring, where knot J is located, the Ws decrease to values smaller than those at the nucleus,
while to the other side of the ring the Ws are similar to the nuclear ones, suggesting that
the stellar population is not as young as in knot J. At the inner ring, the smallest Ws occur
at $\approx$ 28 arcsec SE, the location of knot C, and are similar to those of knot J,
indicating that at the knots there are younger star-forming regions than at other locations
along the ring.

The continuum ratio $C_{4510/4020}$ shows a behaviour consistent with that of the Ws,
being bluest close to the regions where the Ws are smallest, except for the nucleus, which has
a continuum redder than that at 2 arcsec NW but bluer than that at 2 arcsec SE.

In order to calculate the dilution of the Ws at the nucleus, we have compared the nuclear
Ws with those from the region at 5 arcsec (375\,pc at the galaxy) instead of using 1 kpc,
because of the nuclear ring at this latter location. All nuclear Ws
are diluted relative to the extranuclear values. H9 and Ca{\sevensize II}H+H$\epsilon$ are
contaminated by gas emission, while Ca{\sevensize II}K and the G-band are diluted by about
30 per cent. This value is consistent with the known contribution of a FC component
(Antonucci, Hurt  \& Miller 1994).

\section{The spectral synthesis}

The spectral synthesis was performed using the probabilistic formalism described in Cid
Fernandes et al. (2001a). We reproduce the observed Ws and continuum ratios (Cs) using a base
of star cluster spectra with different ages and metallicities (Bica \& Alloin 1986). We use 12
components representing the age-metallicity plane plus a 13$^{th}$ component
representing a cannonical AGN continuum F$_{\nu} \propto \nu^{-1.5}$ (Schmitt et al. 1999).

To synthesize the data of this work we have used the Cs $\lambda$3660/4020 and
$\lambda$4510/4020, and the Ws W$_{wlb}$, W$_{H9}$, W$_{CaIIK}$, W$_{CN-band}$ and
W$_{G-band}$. The adopted errors were $\sigma(W_\lambda)=0.5$ \AA\ for W$_{wlb}$,
W$_{H9}$, W$_{CaIIK}$ and W$_{G-band}$, 1.0\AA\ for W$_{CN-band}$ and
$\sigma(C_\lambda)=0.05$ for continuum ratios (Cid Fernandes et al. 1998). In some cases the
synthesis was performed with a smaller number of Ws, due to the contamination by emission lines. 

It is important to keep in mind that in this spectral range is not possible to discriminate
between the FC and 3\,Myr  components for flux contributions smaller than 40 per cent at 4020\AA,
because they have very similar continua (Storchi-Bergmann et al. 2000). Therefore, in the
description and discussion of the synthesis results we combine the 3\,Myr
and FC components in one, the 3\,Myr/FC component.

According to Cid Fernandes et al. (2001a), this method of spectral synthesis has also difficulty
to accurately
determine the contributions of all 12 components of Bica's base for modest signal-to-noise
ratio and/or reduced number of observables. These constraints act primarily in the sense of
spreading a strong contribution in one component preferentially among base elements of same age.
Thus, in order to produce more robust results we have grouped components of same age.

Due to the fact that the spectral resolution of the Bica \& Alloin's (1986) templates is
lower than that of the present galaxy spectra we need to estimate the effect of this difference
in the equivalent width measurements, and in the synthesis results. The equivalent widths are
weaker in the lower resolution case. Therefore, when we use a base with lower resolution than 
the data, the contribution of the older and/or more metallic populations are overestimated
and the contribution of younger and/or less metallic are underestimated. These effects are
nevertheless small in our sample, as explained below.

Smoothing a few spectra down to the resolution of the Bica \& Alloin's templates we have 
estimated that the differences in the equivalent widths are, in most cases, within the 
measurement errors (see above). Performing the synthesis with the smoothed spectra,
we realized that, on average, the percent contribution to the light at 4020\AA\ 
of the oldest components 10\,Gyr or 1\,Gyr becomes 5 per cent smaller while the contribution
of the youngest component 3\,Myr/FC becomes 5 per cent larger, while E(B-V) becomes on 
average 0.06 larger. These differences do not affect the conclusions of our work, as can be 
seen in the following sections, and we thus have opted for keeping the full resolution
of our data in the synthesis.

\subsection{Synthesis results for the non-Seyfert galaxies}

\begin{figure}
\vspace{9.5cm}
\caption{Results of stellar population synthesis using a base of star clusters plus a power
law component F$_{\nu} \propto \nu^{-1.5}$ for the non-Seyfert galaxy NGC7049. The dots represent
percent contribution to the flux at 4020 \AA\ from different age and metallicity
components as a function of distance from the nucleus. The panels show from bottom to top,
the contribution from: the sum of the four 10\,Gyr  components; the sum of the three 1\,Gyr
 components; the sum of two 100\,Myr  components; the sum of the two 10\,Myr  components;
the sum of 3\,Myr  component and FC (F$_{\nu} \propto \nu^{-1.5}$). The top panel shows
the internal reddening E(B-V)$_i$ obtained from the synthesis.  At the bottom panel we also
show the sum of the two  metal poor components (open dots, see the discussion in the text,
Section 5.1). The vertical bars represent the standard deviation of the results. The vertical
lines spanning the whole figure mark locations at 1 kpc and 3 kpc from the nucleus.}
\label{sin_normal}
\includegraphics{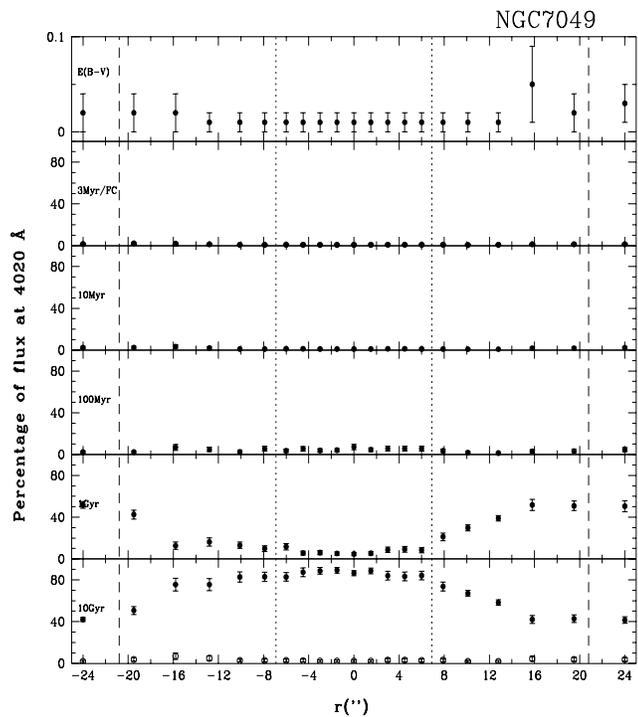}
\end{figure}

In order to compare our results with those for non-Seyfert galaxies we performed again the
spectral synthesis for the non-Seyfert galaxy NGC7049, with the new input parameters
specified above in order to verify if there is any change relative to results found
in R01. There are no significant differences: the stellar population is
dominated by the 10\,Gyr  metal rich component at the nucleus, with its contribution
decreasing beyond 1 kpc, while the contribution of the 1\,Gyr  component increases outwards.
In figure \ref{sin_normal} we show the synthesis results for this galaxy.
The results are presented as proportions of the flux contributed by each component at
$\lambda$4020 \AA\ and the internal reddening E(B-V)$_i$ of the galaxy. The panels show from
bottom to top the contribution from: the sum of the four 10\,Gyr  components; the sum of the
three 1\,Gyr  components; the sum of two 100\,Myr  components; the sum of the two 10
Myr  components; the sum of the 3\,Myr  component and FC (F$_{\nu} \propto \nu^{-1.5}$).
The top panel shows the internal reddening E(B-V)$_i$ obtained from the synthesis.  At the
bottom panel we also show separately, with open symbols, the sum of the two  metal poor
components. The vertical bars are the standard deviation of the results. The vertical lines
crossing all panels mark distances of 1\,kpc and 3\,kpc from the nucleus.

\begin{figure}
\vspace{9.5cm}
\caption{Results of stellar population synthesis for the non-Seyfert Sa-Sc galaxies
ploted altogether. In order to group the results, we had first to transform the angular
distances to linear distances at
each galaxy. The galaxy with recent star formation is the Sc NGC1637 (filled circle).}
\label{sin_normaltodos}
\includegraphics{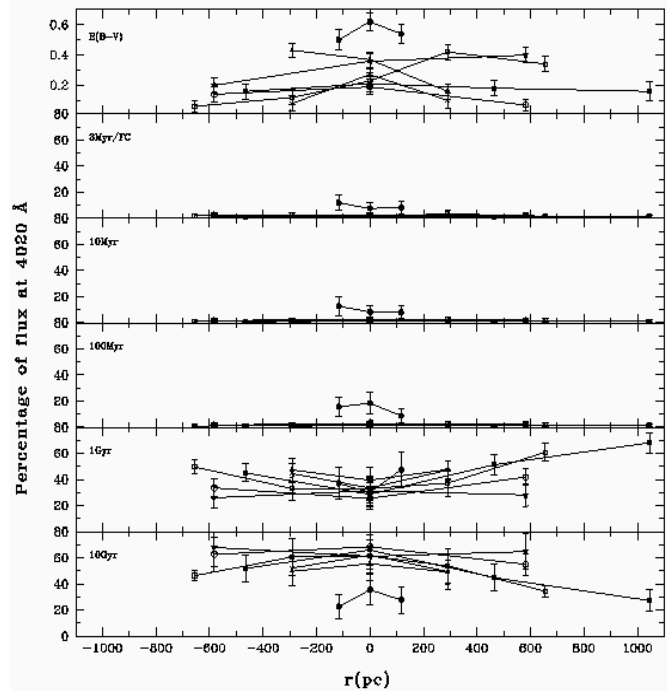}
\end{figure}

The results of the synthesis for the Sa-Sc non-Seyfert galaxies are summarized in Fig.
\ref{sin_normaltodos}, where we show the values for all non-Seyferts together, due to
the small number of points for each individual galaxy. In order to group the results,
we had first to transform the angular distances to linear distances at each galaxy.
For most galaxies, the stellar population in the nucleus
is composed by $\approx$65 per cent of the 10\,Gyr component and $\approx$30 per cent of the
1\,Gyr component. The contribution of the other components are not significant. In six
(of seven) galaxies the  10\,Gyr component decreases while the 1\,Gyr component increases
outwards, similarly to the S0 non-Seyfert galaxies behaviour.

A distinct result is found for the Sc galaxy NGC1637 which has recent star formation:
at the nucleus the component of 10\,Gyr contributes with 35 per cent, that of 1\,Gyr with 30
per cent and those younger than 1\,Gyr contribute with the remaining 35 per cent of the flux
at 4020\AA\ (this galaxy has little information about the stellar population outside the
nucleus -- only up to $\approx$ 100\,pc from the nucleus).

The E(B-V)$_i$ for these galaxies range from 0.1 to 0.6 and it is larger than those for
S0 galaxies. The largest E(B-V)$_i$ is found for NGC1637, the galaxy with recent star
formation.

\subsection{Synthesis results for the Seyfert 2 galaxies}

In order to describe the synthesis results for the Seyfert 2 galaxies, we have grouped the
sample in four classes according to their most important nuclear stellar component and describe
the main characteristics (or variations) in the populations at the nucleus and outside, using
the 1\,kpc and 3\,kpc (when possible) locations as references. In the following discussion, we
consider a population component significant when it contributes with at least 10 per cent of the
total flux at 4020\AA.

\subsubsection{Dominant 10\,Gyr metal rich stellar population}
In this group are Mrk573, NGC5929, NGC2110, Mrk3, NGC7212, Mrk348 and Mrk34, whose
nuclear stellar population is dominated by the 10\,Gyr  metal rich (with solar or
above metalicity) component. The galaxies are sorted above according the importance of the
 rich component, whose contribution ranges from about 60
per cent of the total flux at 4020\AA\ in Mrk573 to 30 per cent in Mrk34. At the nucleus
all these galaxies have a contribution of the 1\,Gyr  component of at least
15 per cent of the total flux. Some of these objects also have significant nuclear
contributions of younger components: Mrk3 with 20 per cent of 3\,Myr/FC, Mrk34 with 10 per
cent of 10\,Myr old and 10 per cent of 3\,Myr/FC and NGC7212 with 15 per cent of 3\,Myr/FC.

\begin{figure*}
\vspace{19cm}
\caption{Results of stellar population synthesis as a function of the distance from the nucleus
for the galaxies with dominant 10\,Gyr  metal rich component at the nucleus. Symbols as in
Fig.\ref{sin_normal}.}
\label{sin1}
\includegraphics{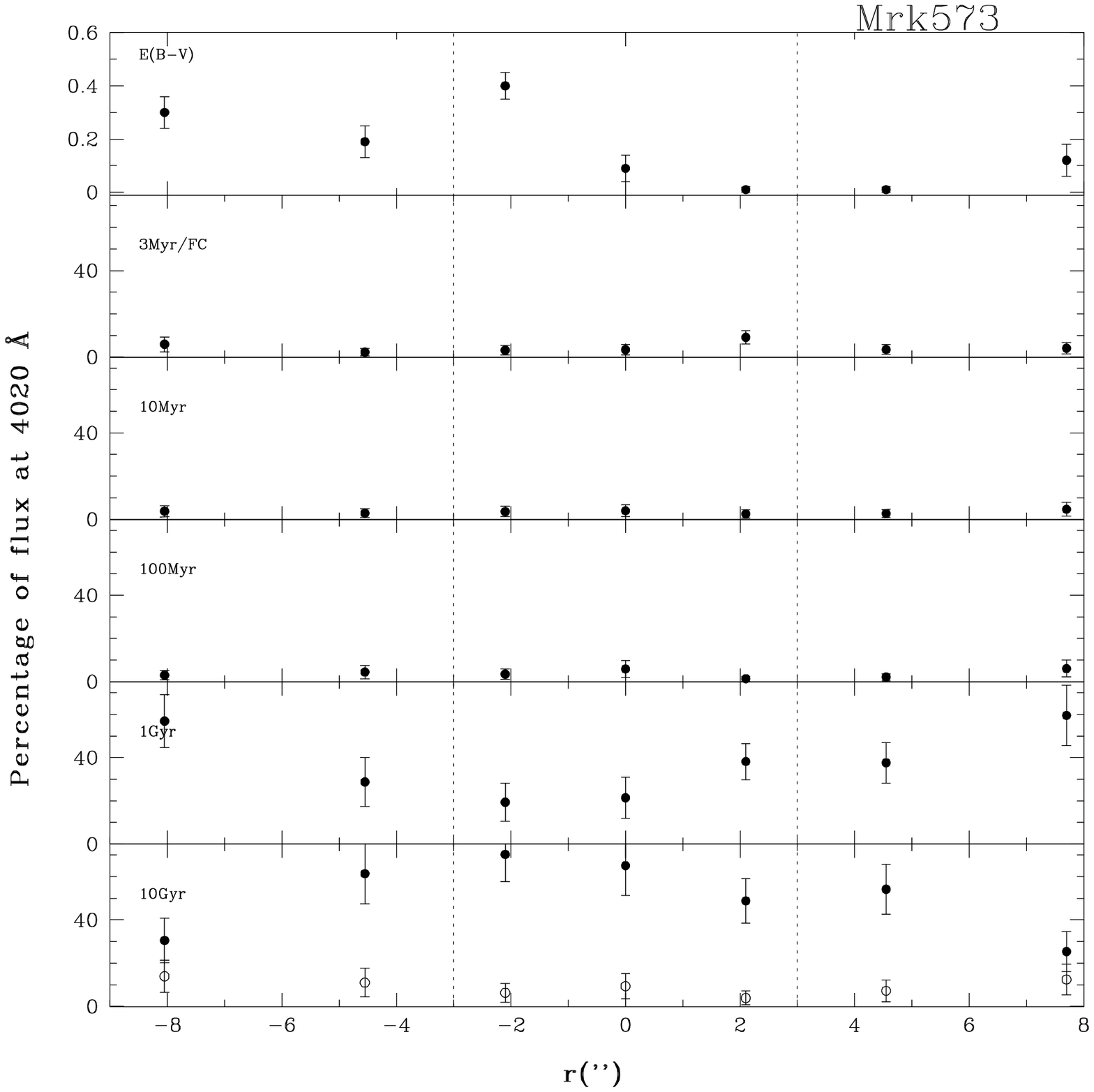}
\includegraphics{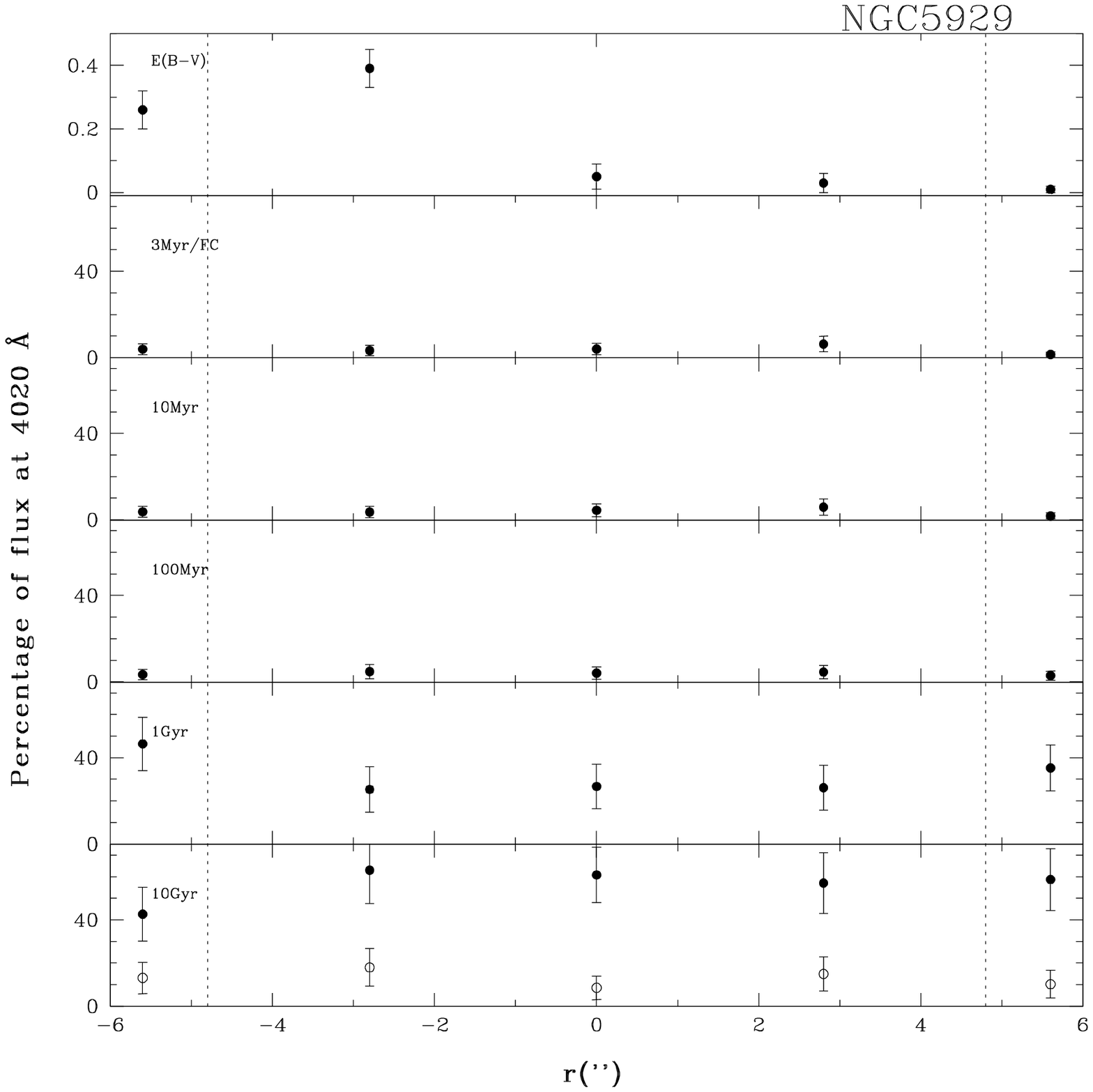}
\includegraphics{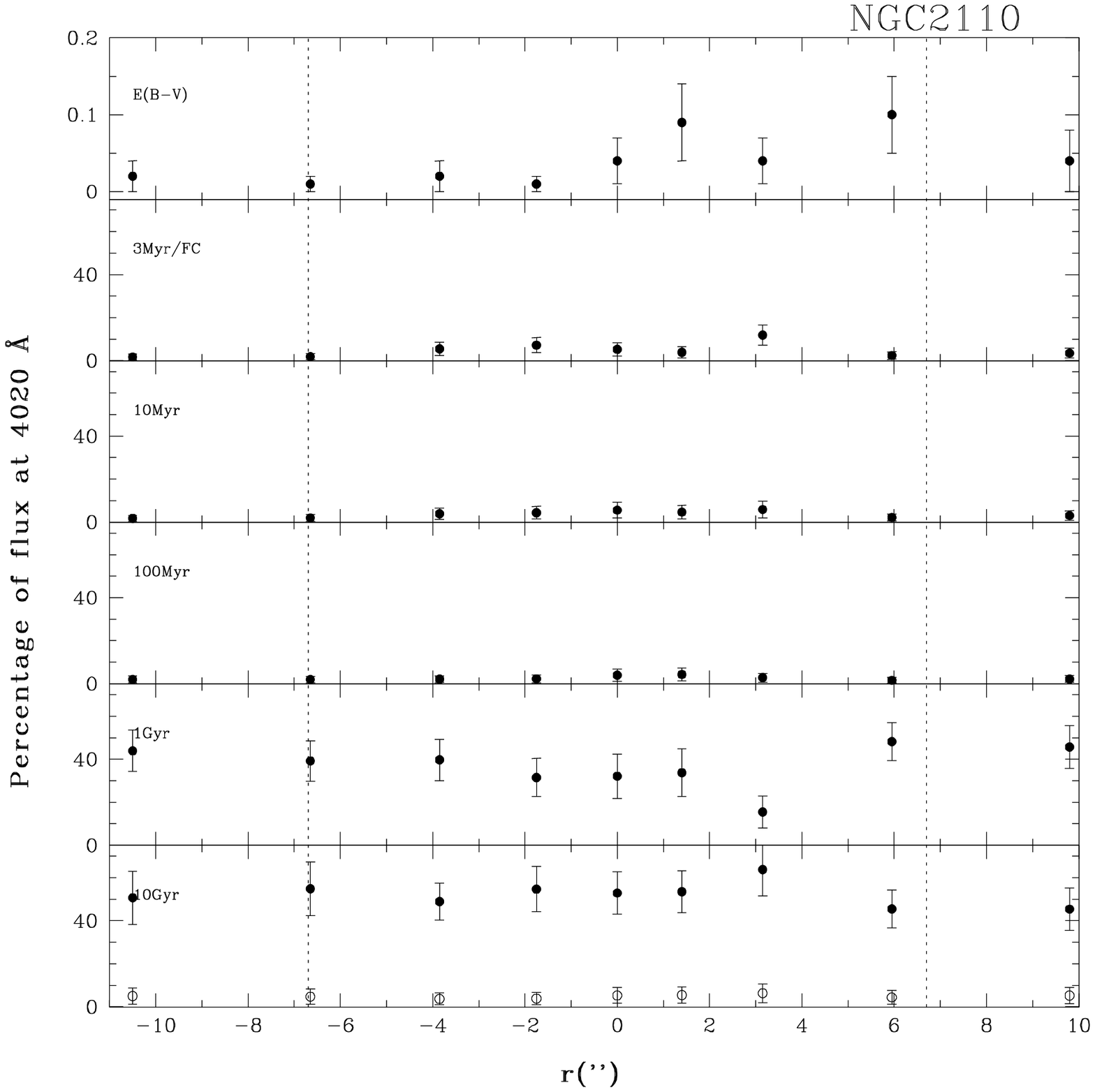}
\includegraphics{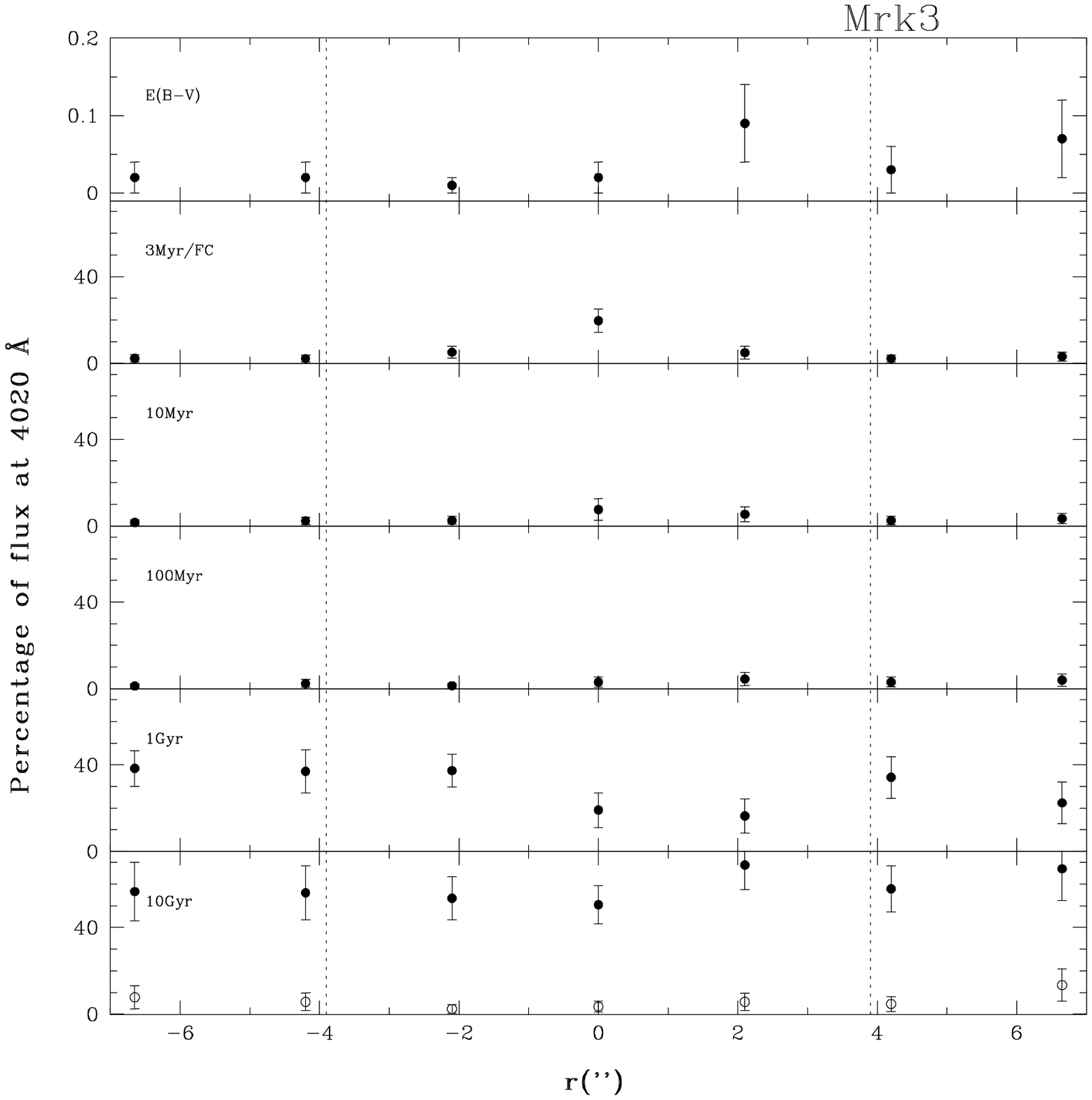}
\end{figure*}

\begin{figure*}
\vspace{19cm}
\caption{Synthesis results: galaxies with dominant 10\,Gyr  metal rich component at the nucleus.}
\label{sin2}
\includegraphics{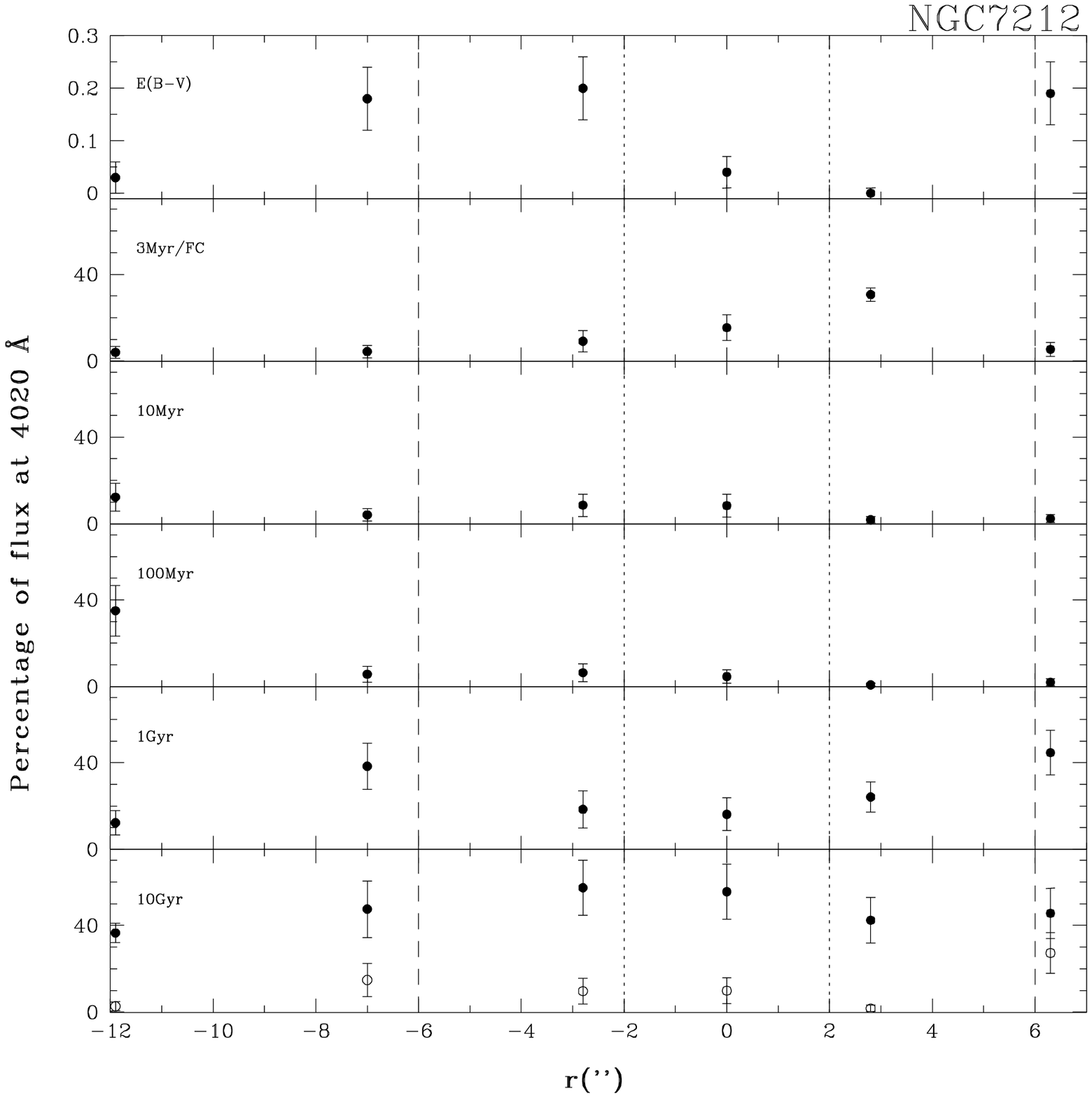}
\includegraphics{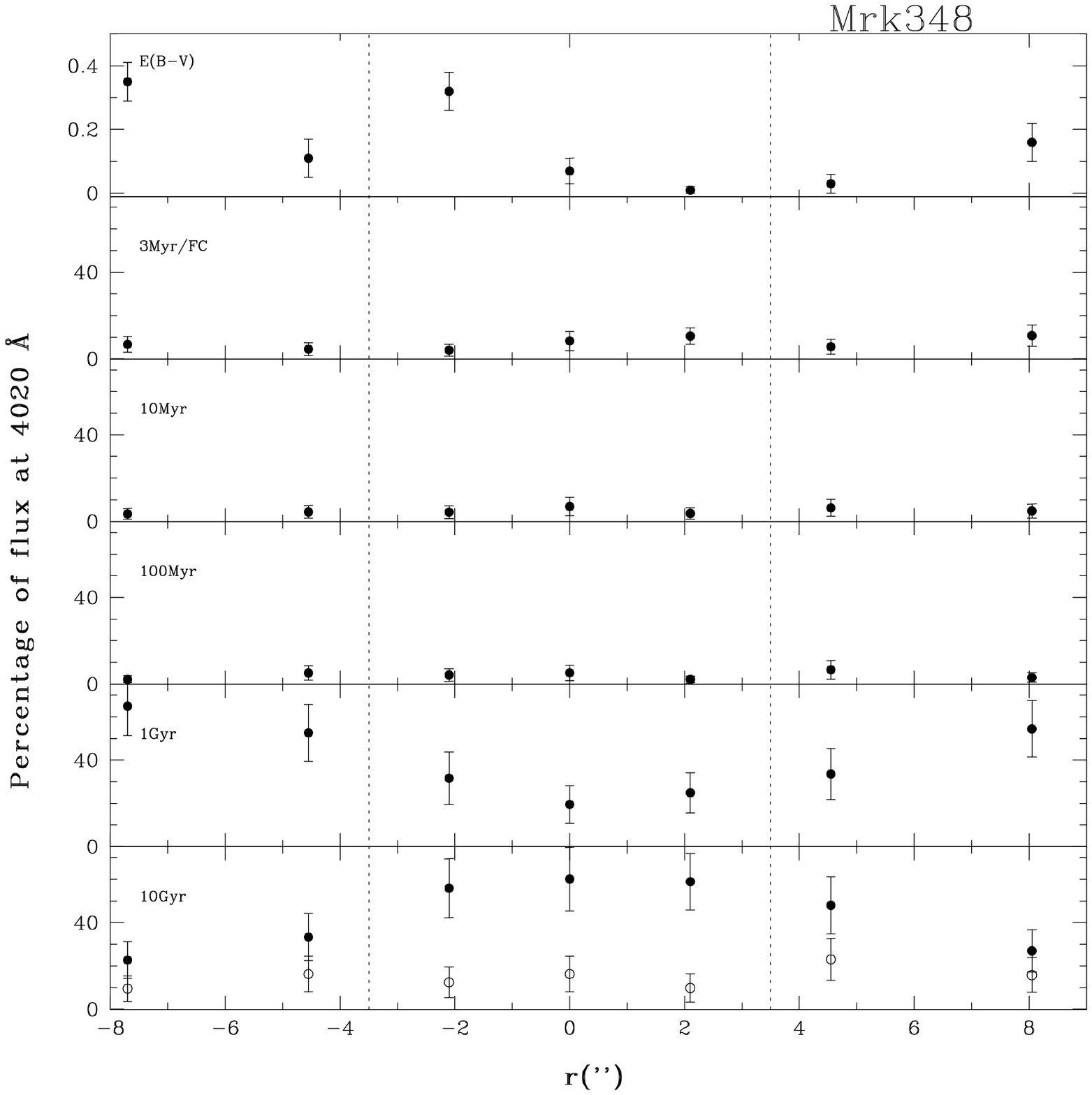}
\includegraphics{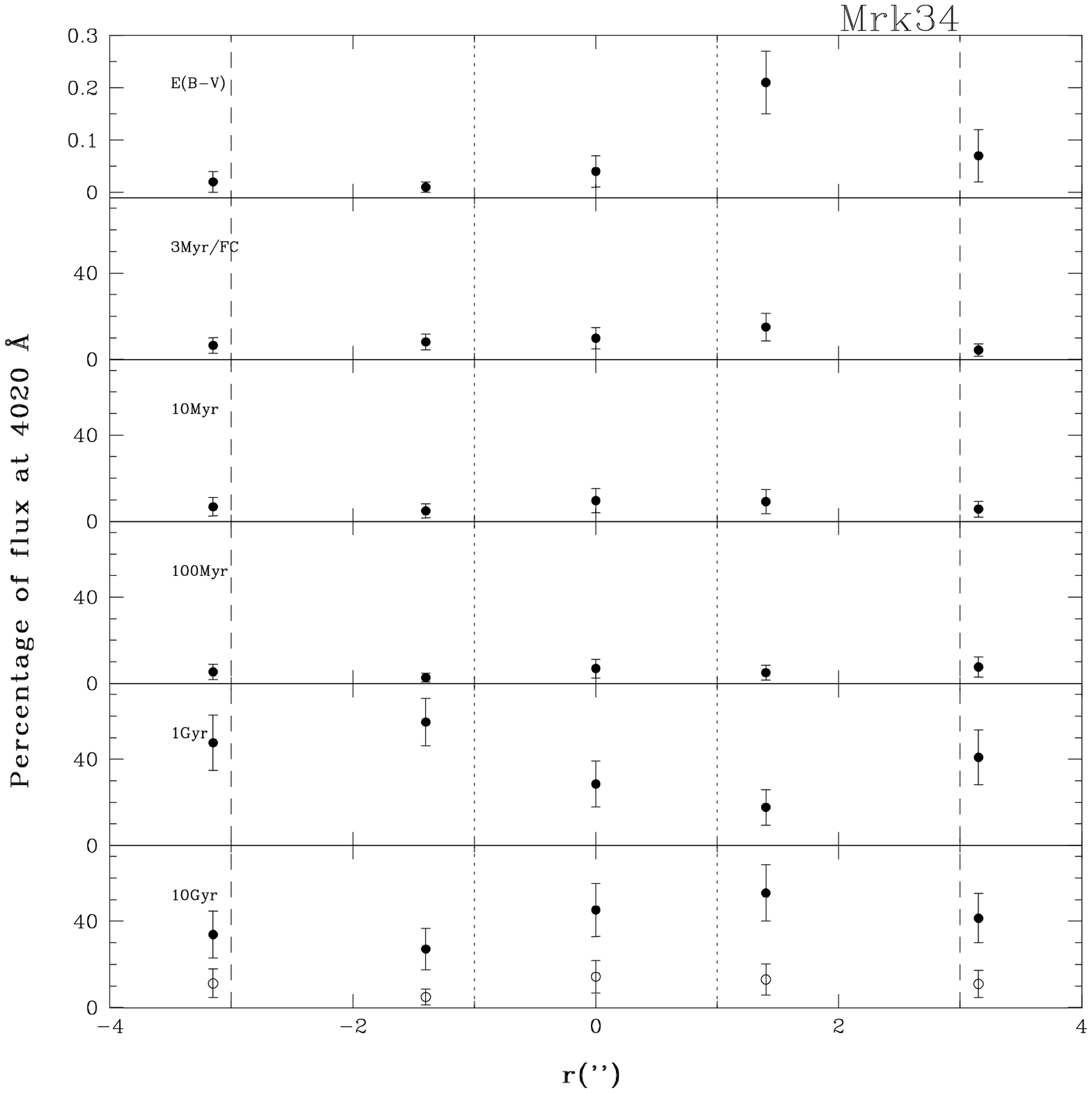}
\end{figure*}

In Figs. \ref{sin1} and \ref{sin2} we show the spatial variation of the spectral synthesis
results for galaxies of this group. The extranuclear old metal rich component contribution
is similar to that of the nucleus  in Mrk 3, Mrk 34 and NGC2110, while in the other galaxies 
this component decreases outwards. The 1\,Gyr  component increases outwards in all galaxies.
The younger components decrease outwards in the galaxies for which they are significant
($>$10 per cent) at the nucleus.

\begin{figure*}
\vspace{12cm}
\caption{Observed and synthetic spectra at the nucleus and outside for one galaxy with dominant
10\,Gyr  metal rich component at the nucleus. The observed spectra were smoothed to match the
lower resolution of the synthetic spectra.}
\label{n5929sin}
\includegraphics{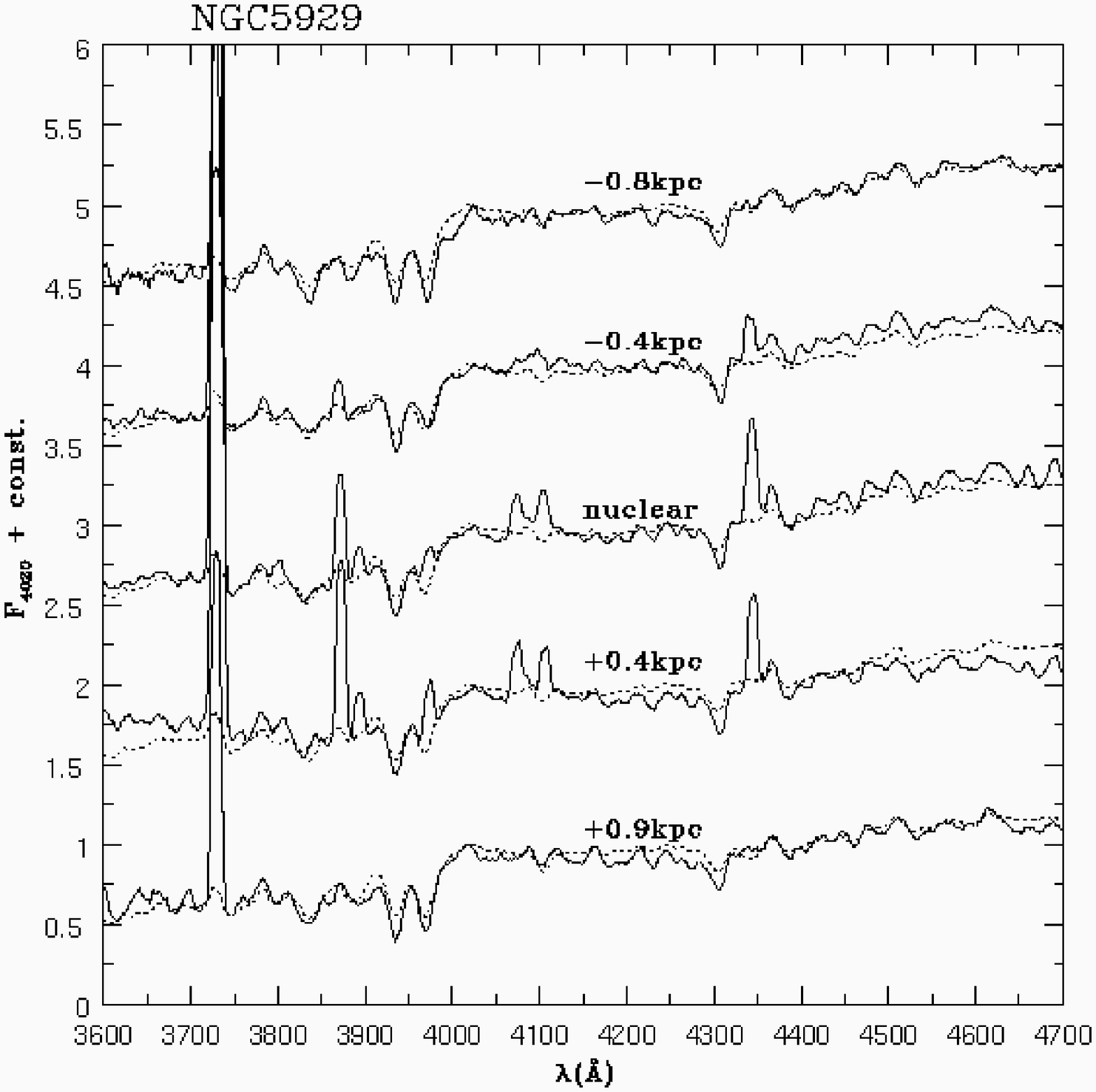}
\end{figure*}

In Fig. \ref{n5929sin} we compare the observed spectrum (solid-line) with the synthetic one
(dashed-line) for one galaxy of this group (NGC5929) at the nuclear and off-nuclear regions.
The synthetic spectra were constructed using the star cluster templates
(Bica \& Alloin 1986, 1987) combined in the proportions given by the synthesis. The synthetic 
spectra reproduce well the observed ones, considering the lower spectral resolution of the 
available templates ($\approx$ 10-20\AA) as compared to the observations ($\approx$ 3\AA).

\subsubsection{Dominant 1\,Gyr stellar population}
The galaxies of this group are Mrk78, NGC1386, IC3639, Mrk1073, NGC1068, Mrk1 and Mrk1066, sorted
according to the importance of the nuclear 1\,Gyr component. In Mrk78 this component contributes 
with about 50 per cent of the total flux at 4020\AA\ and in Mrk1066 with about 20 per cent. All
of them have a significant nuclear contribution of the 10\,Gyr  component, and all but NGC1386
also have significant contribution of the components younger than 1\,Gyr.

\begin{figure*}
\vspace{19cm}
\caption{Synthesis results: galaxies with dominant 1\,Gyr component at the nucleus.}
\label{sin3}
\includegraphics{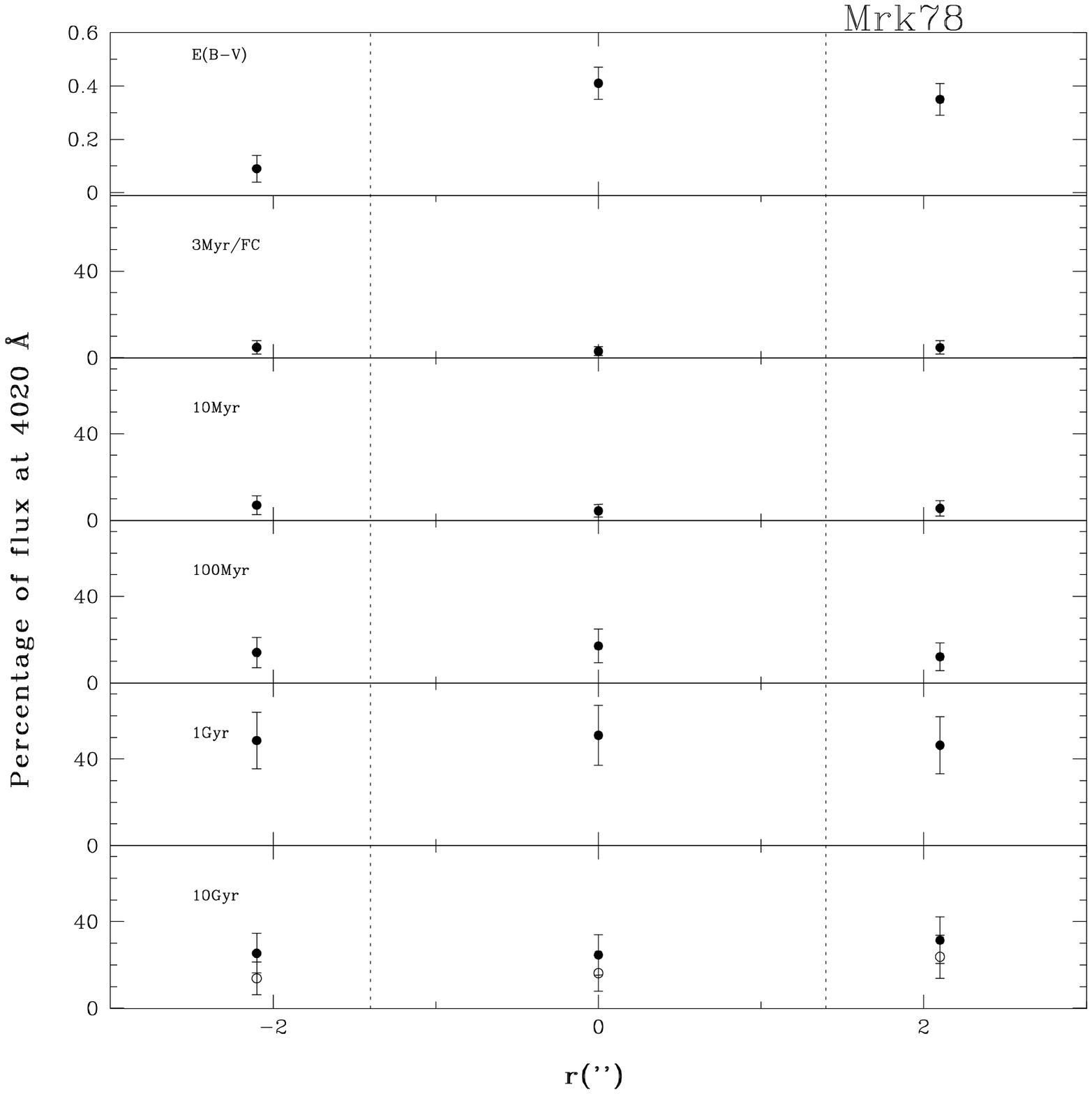}
\includegraphics{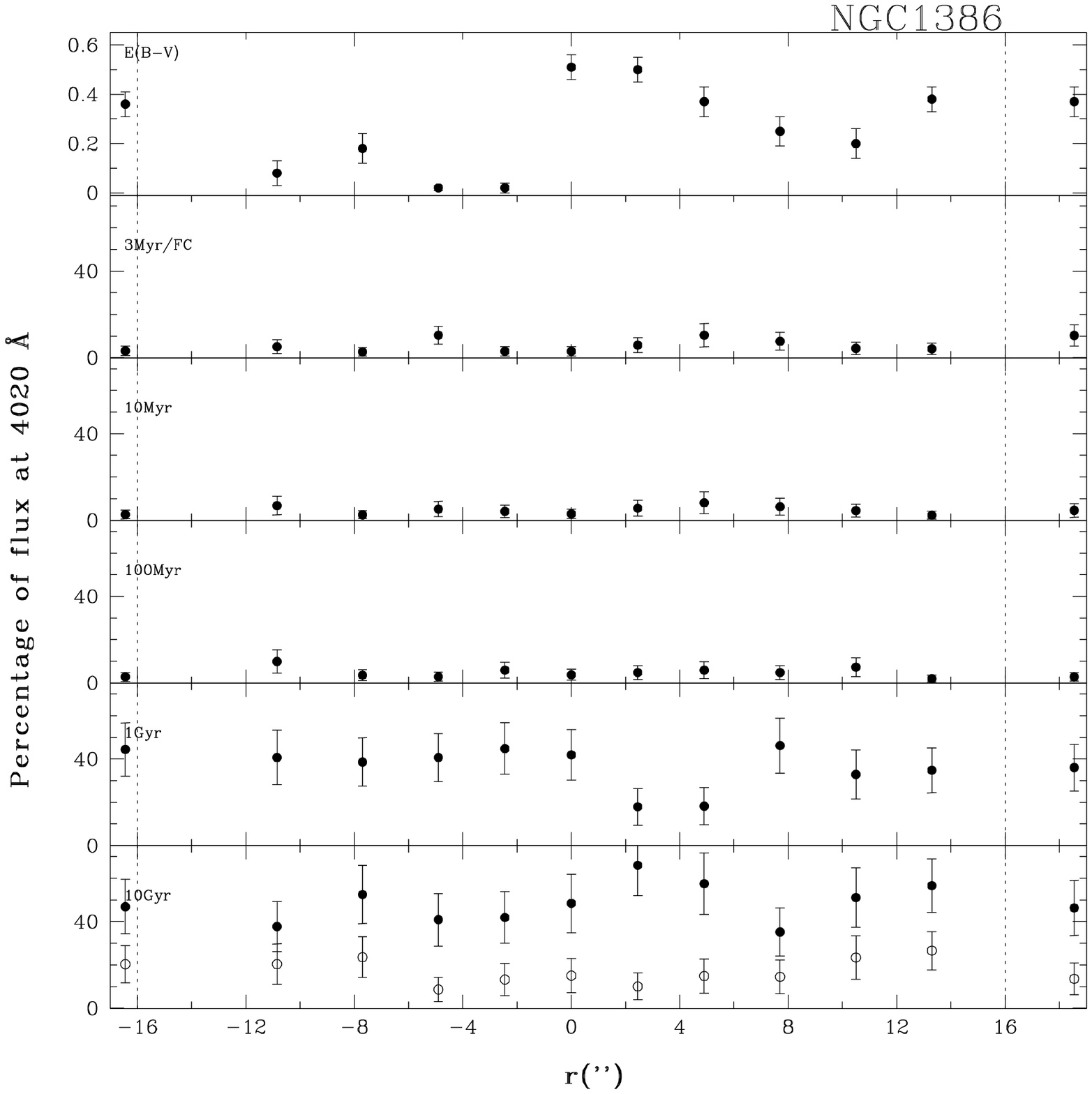}
\includegraphics{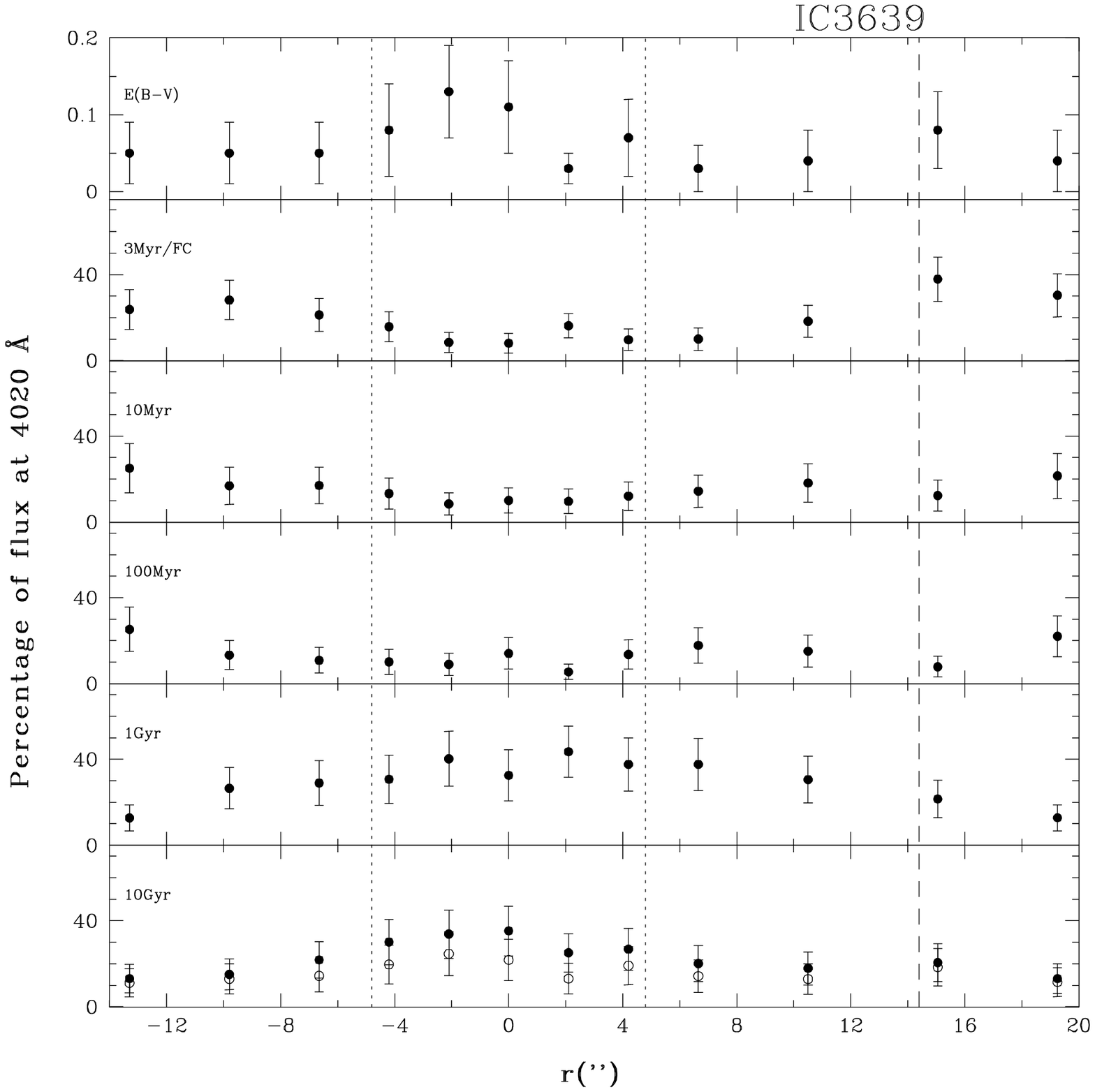}
\includegraphics{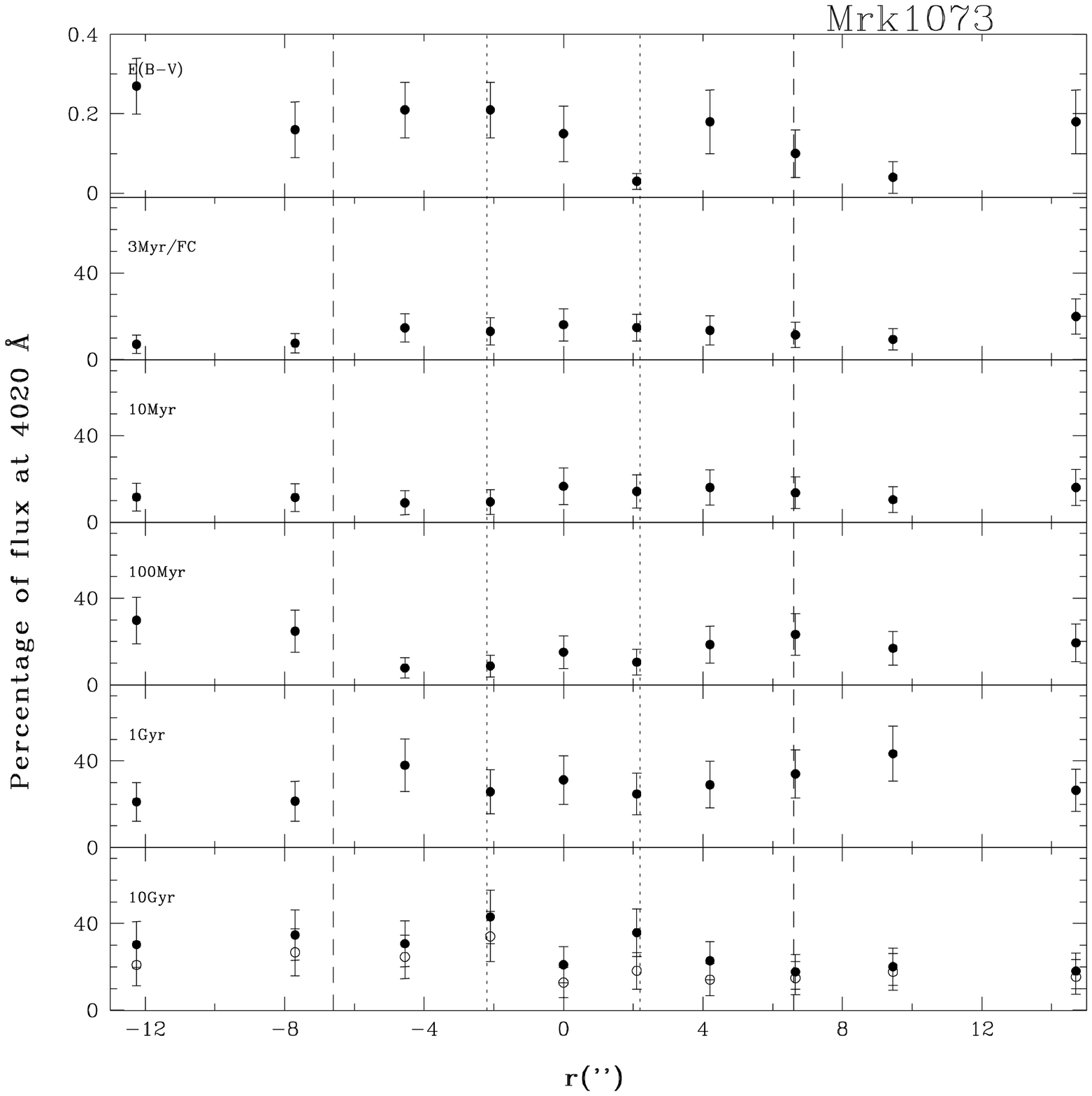}
\end{figure*}

\begin{figure*}
\vspace{19cm}
\caption{Synthesis results: galaxies with dominant 1\,Gyr component at the nucleus.}
\label{sin4}
\includegraphics{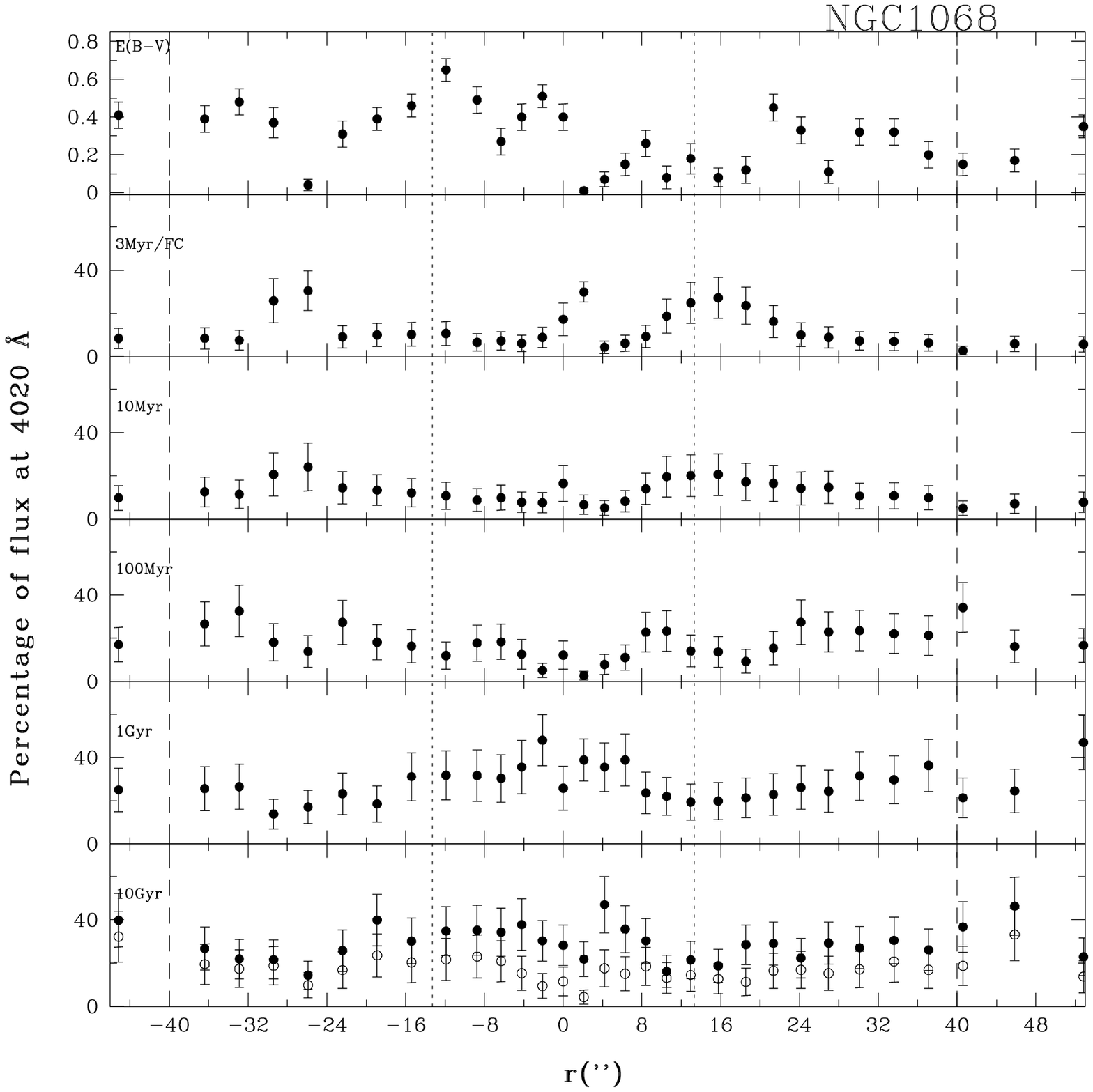}
\includegraphics{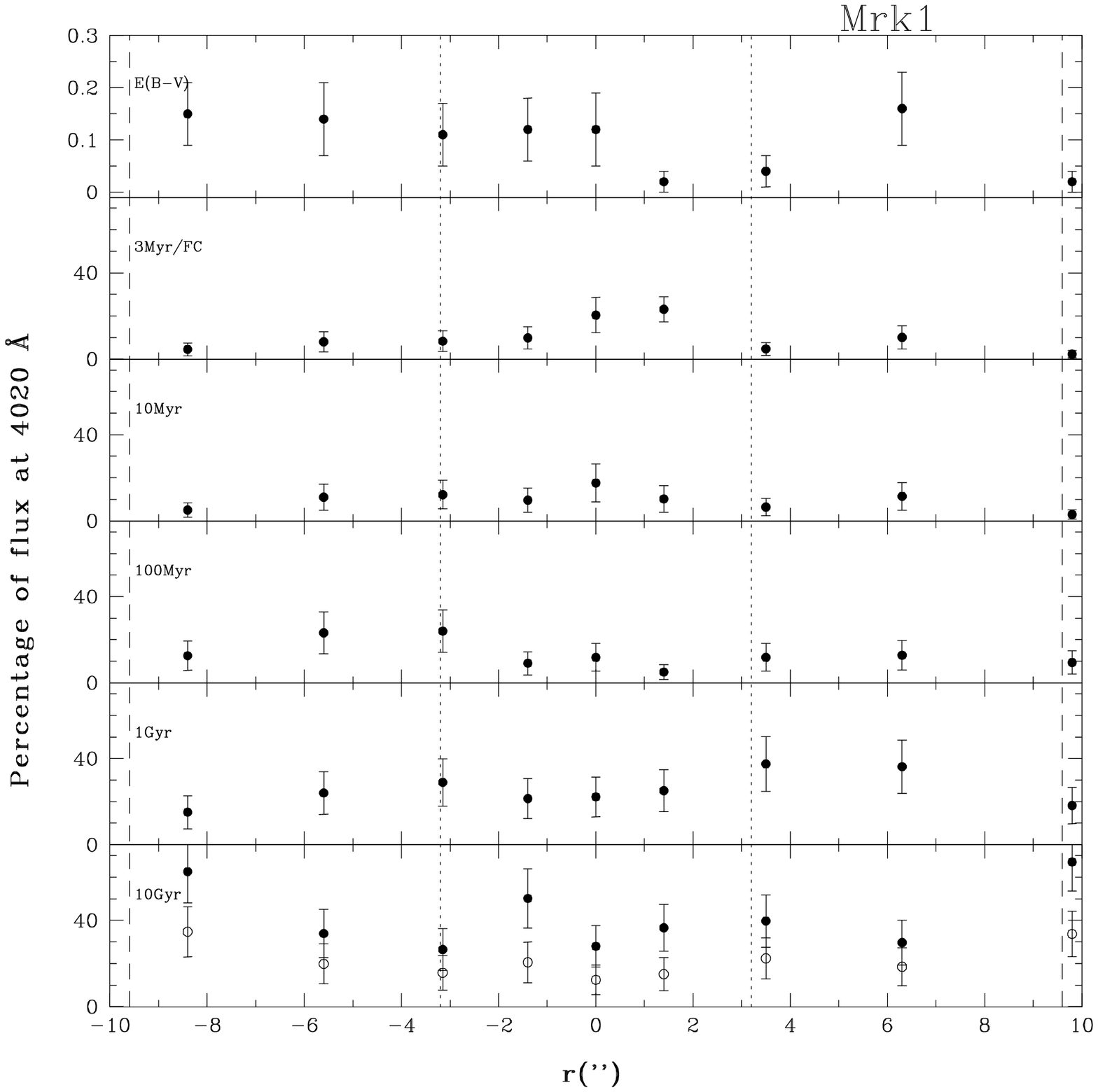}
\includegraphics{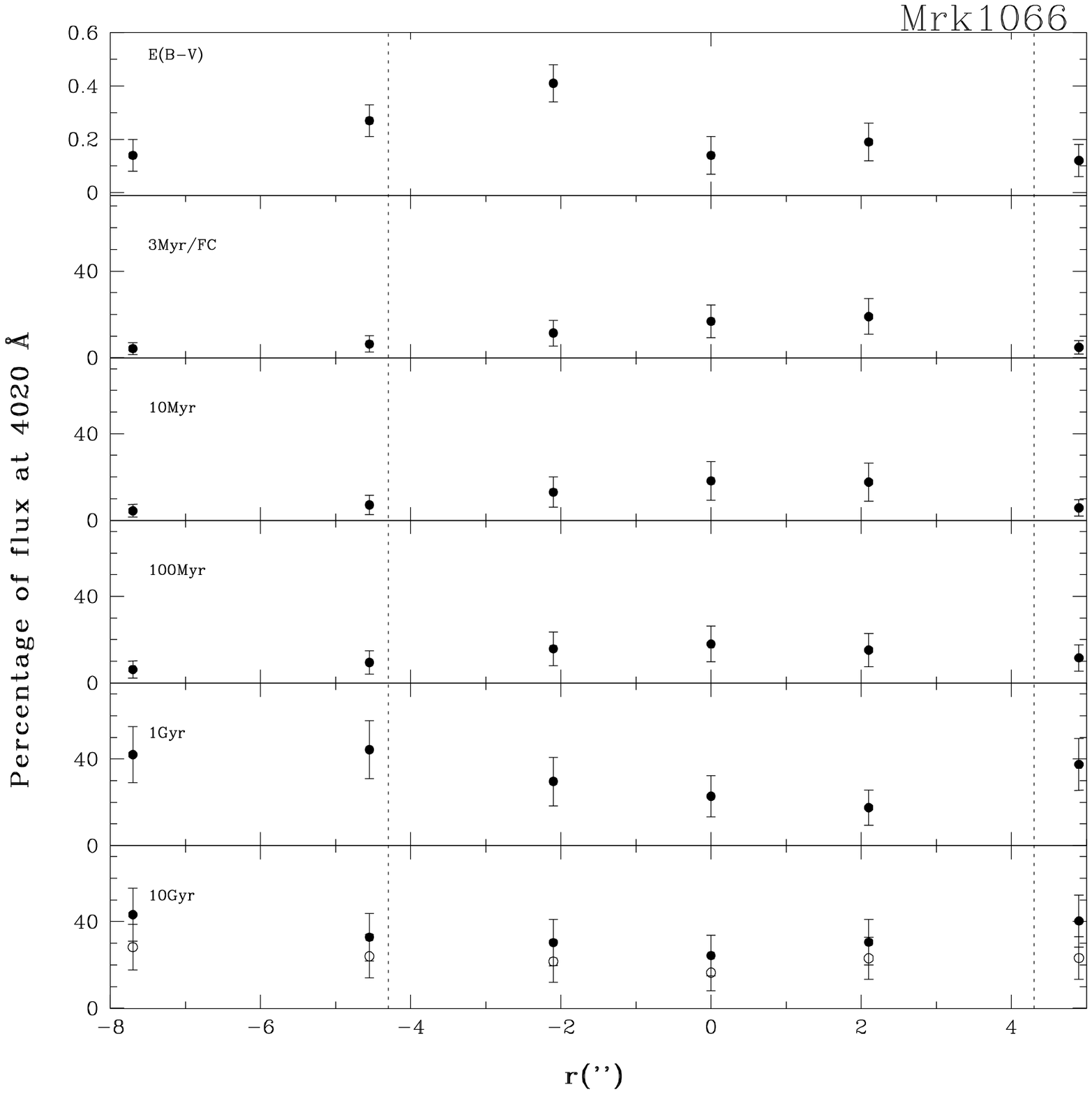}
\end{figure*}

The synthesis results as a function of distance from the nucleus for these galaxies
are shown in Figs. \ref{sin3} and \ref{sin4}. Throughout the central 2\,kpc the galaxies 
Mrk78 and NGC1386 have contributions of the 10\,Gyr, 1\,Gyr and 100\,Myr components similar 
to those at the nucleus. NGC1386 also has significant contributions of the components
100\,Myr old and younger outside the nucleus (at 300 and 660 pc), which is different from the
nuclear one.

In Mrk1066 the components of 10\,Gyr and 1\,Gyr increase and those younger than 1\,Gyr
decrease outwards, while the population in IC3639 has a completely opposite behaviour.

The population outside the nucleus in Mrk1 has contributions of the 1\,Gyr and 100\,Myr
components similar to those at the nucleus and the contribution of the 10\,Gyr  component
increases while that of the components younger than 100\,Myr decreases. In Mrk1073 the 
components of 10\,Gyr and 1\,Gyr have similar contributions to those at the nucleus and
the contribution of the 100\,Myr component increases while that of the components younger
than 100\,Myr decreases outwards.

\begin{figure*}
\vspace{12cm}
\caption{As in Fig.\ref{n5929sin} for one galaxy with dominant 1\,Gyr  component at the 
nucleus.}
\label{m1073sin}
\includegraphics{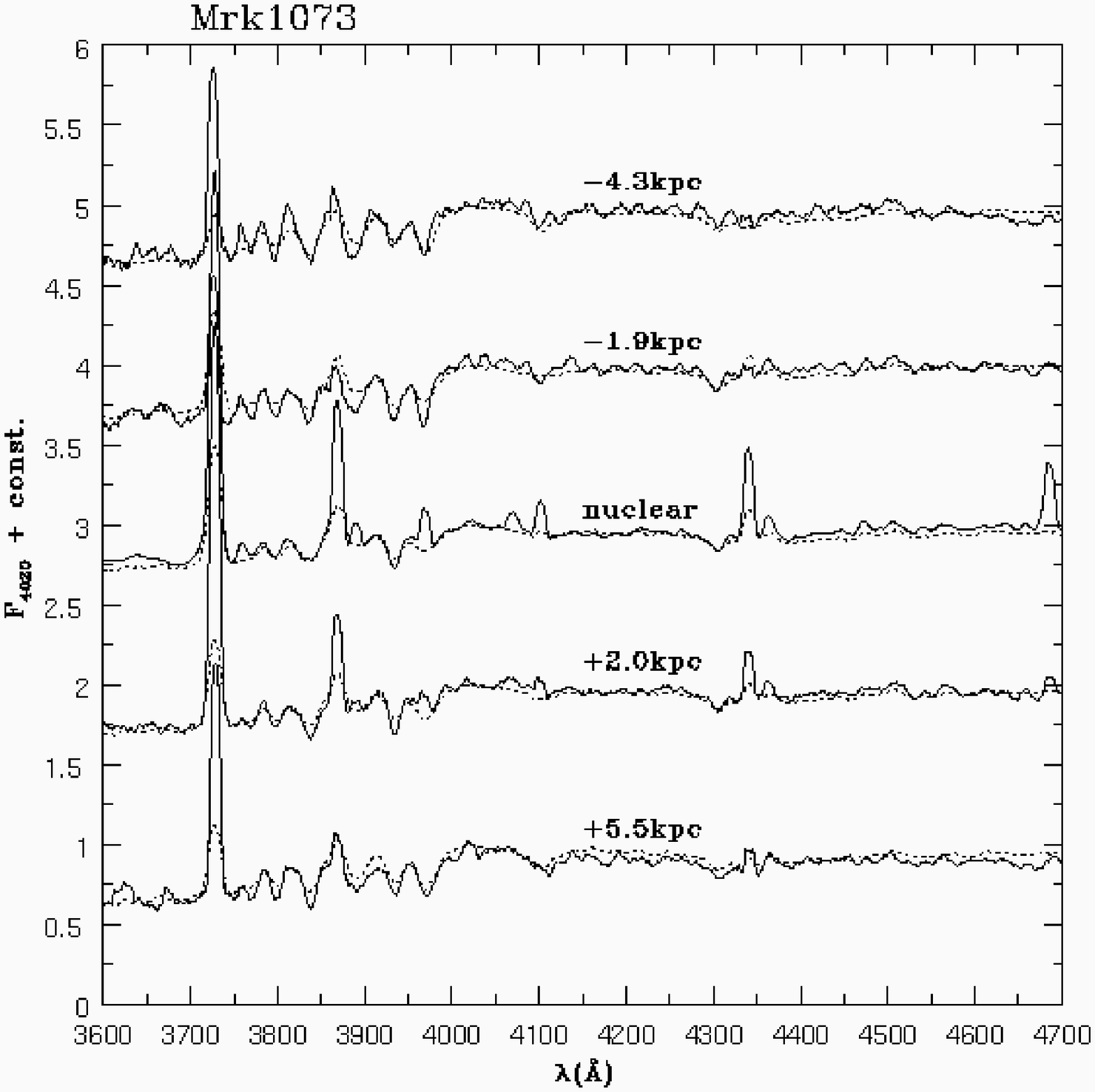}
\end{figure*}

In Fig. \ref{m1073sin} we compare the observed spectrum (solid-line) with the synthetic one
(dashed-line) for one galaxy of this group (Mrk1073) at the nuclear and off-nuclear regions.
All spectra present the high-order Balmer lines in absorption, characteristic of stellar
populations younger than 1\,Gyr.

\noindent {\it 4.2.2.1 NGC1068}

At the nucleus of this galaxy all components are significant but we would like to
highlight the importance of the 3\,Myr/FC component. It contributes with about
17 per cent of the total light at 4020\AA. This result is consistent with the contribution of
polarized light from the hidden Seyfert 1 nucleus (e.g. Antonucci et al. 1994).

Outside the nucleus, all age components continue to be important, contributing with more than
10 per cent to the flux at 4020\AA\ and indicating the presence of recent star-formation
throughout the observed region. In particular at knot J (about 750\,pc from the nucleus, NW
-- positive direction) and at knot C (about 2\,kpc from the nucleus, SE -- negative direction)
the 3Myr/FC component is as important as in the nucleus, contributing about 20 and 25 per cent
respectively. This result confirms that these are active star-forming regions. From the synthesis, 
knot C is younger than knot J. At 3\,kpc from the the nucleus the 3\,Myr/FC component is not
significant.

\begin{figure*}
\vspace{12cm}
\caption{As in Fig.\ref{n5929sin} for NGC1068 and two knots of star formation.}
\label{n1068sin}
\includegraphics{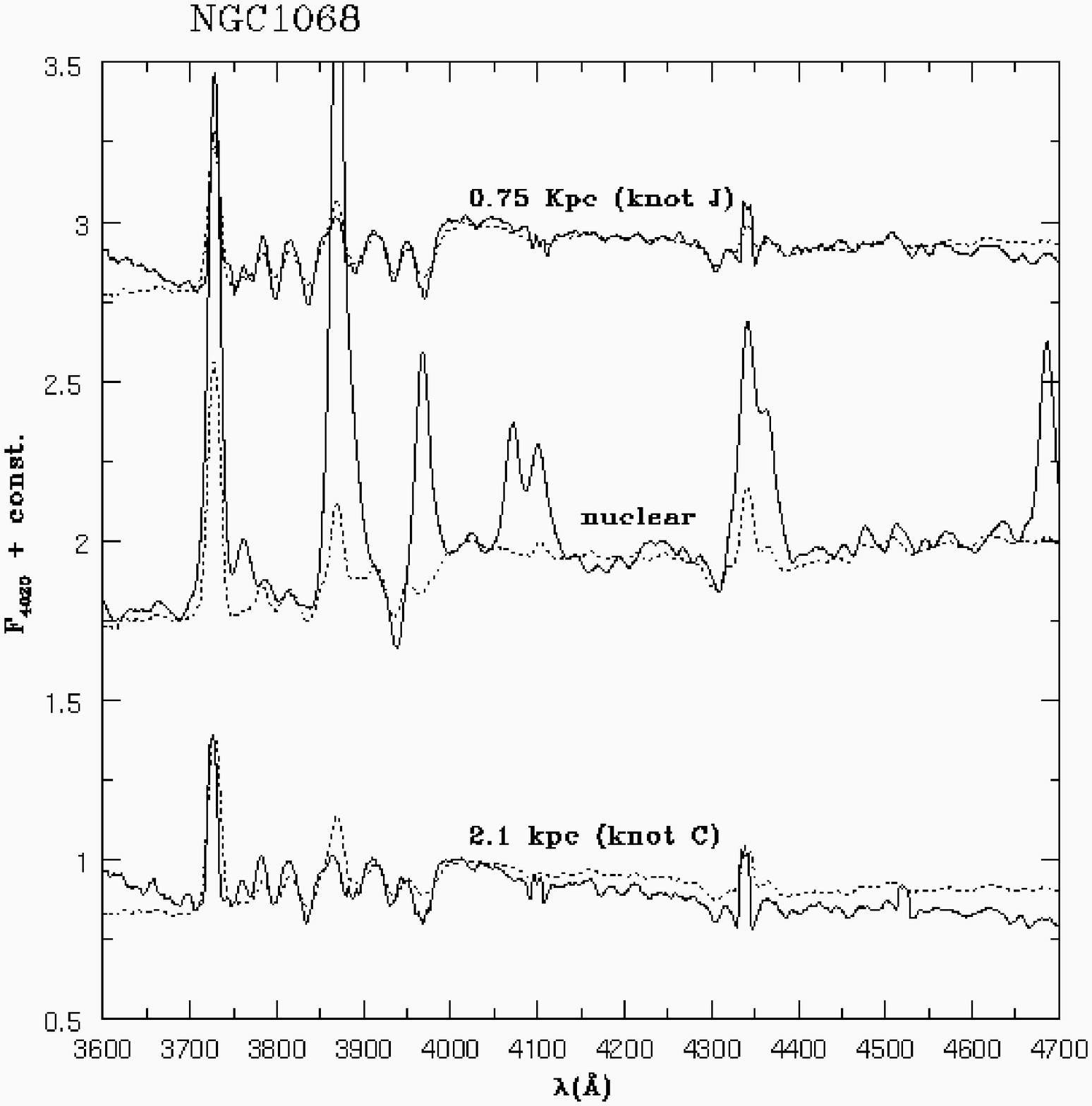}
\end{figure*}

In Fig. \ref{n1068sin} we compare the observed spectrum (solid-line) with the synthetic one
(dashed-line) for the nucleus and knots C and J.

\subsubsection{Dominant 100\,Myr  stellar population}
In this group are Mrk273, Mrk533 and NGC7130, for which the 100\,Myr  component contributes
about 30 per cent to the nuclear flux at 4020 \AA. The 10\,Gyr  component contributes with
about 20 per cent. The other components also have significant contributions at the nucleus
of these galaxies, except 3Myr/FC in Mrk273.

\begin{figure*}
\vspace{19cm}
\caption{Synthesis results: galaxies with dominant 100\,Myr component at the nucleus.}
\label{sin5}
\includegraphics{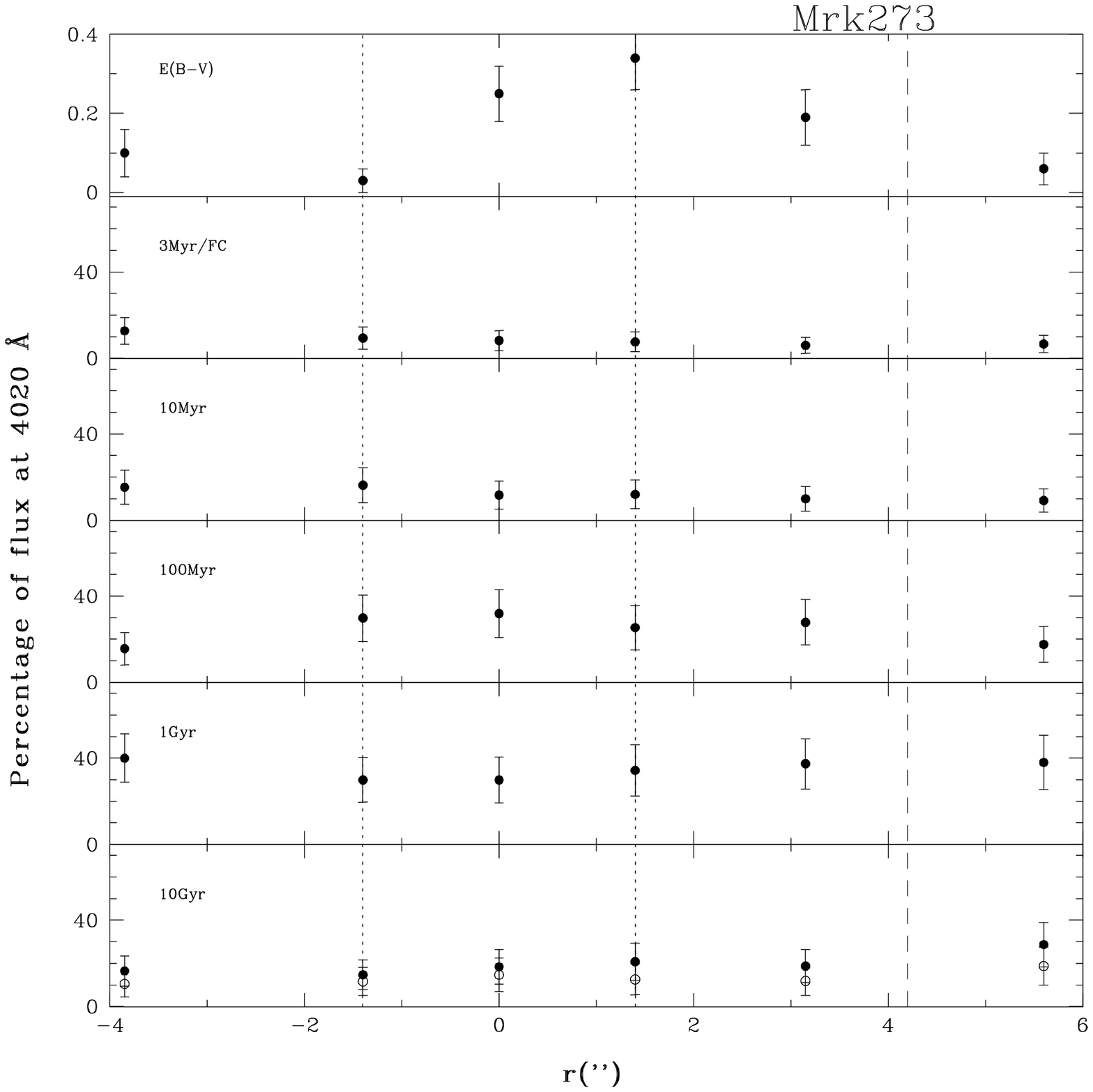}
\includegraphics{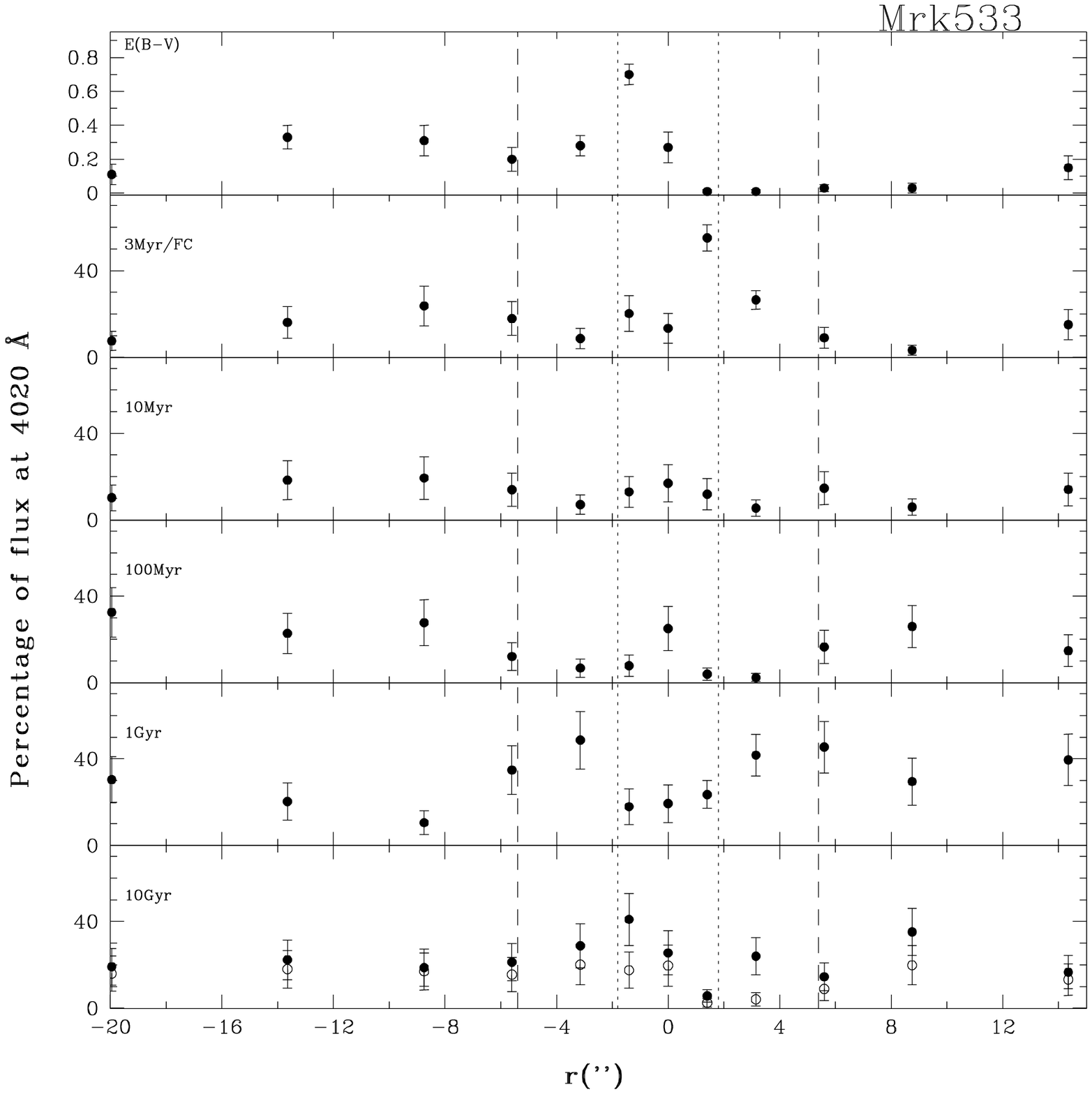}
\includegraphics{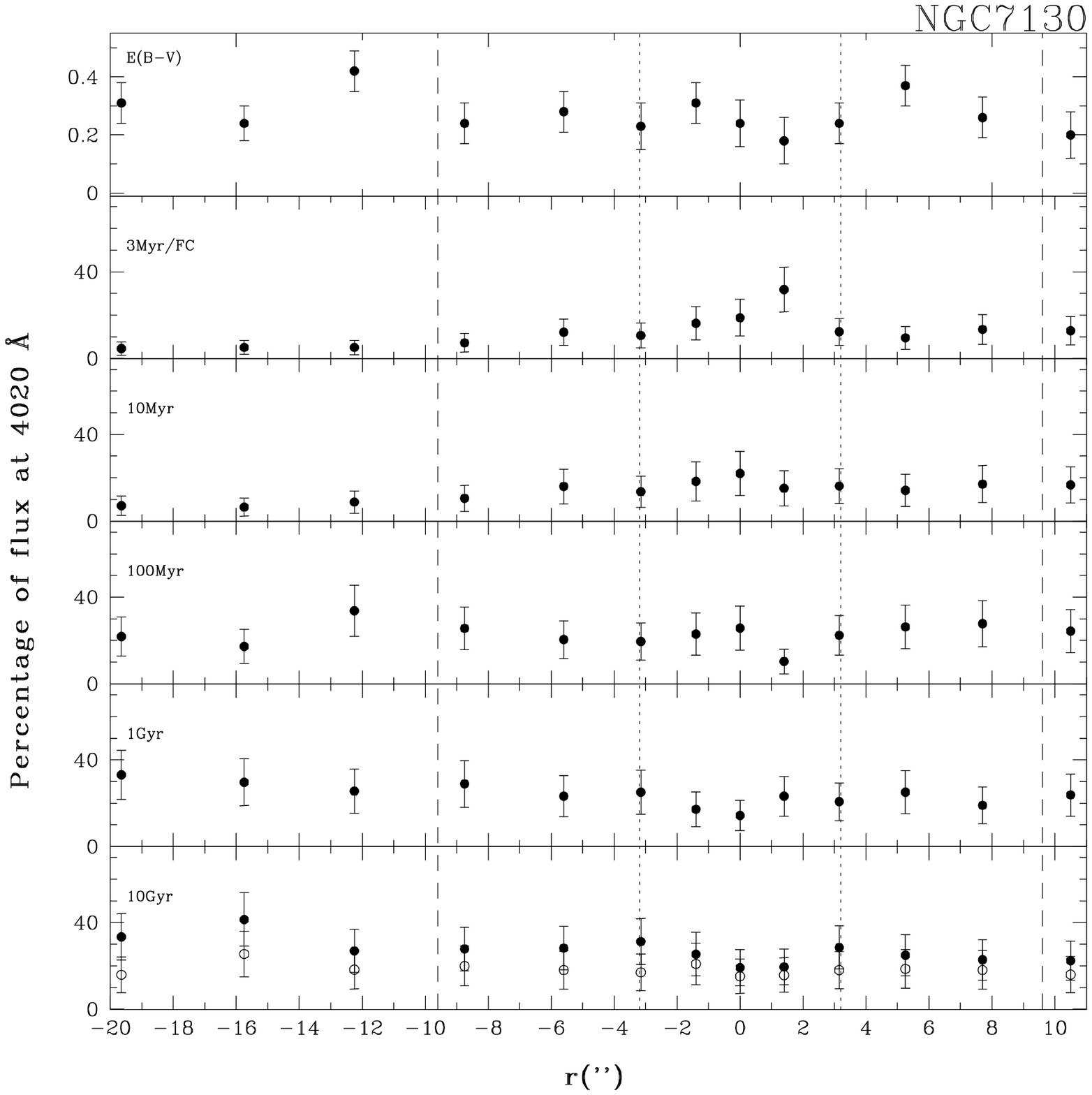}
\end{figure*}

The results of the synthesis for galaxies of this group are shown in Fig.\ref{sin5}.
Outside the nucleus of NGC7130 the contribution of components with 10\,Gyr and 1\,Gyr
increases while that of the components younger than 100\,Myr decreases.

The population of Mrk533 shows large variations in the 10\,Gyr, 100\,Myr and 3\,Myr/FC
components in the central 2\,kpc. In the central 6\,kpc the global tendencies are that
the component with 1\,Gyr increases and the 100\,Myr component decreases outwards, while
the 10\,Gyr component and the components younger than 100\,Myr have contributions similar
to those at the nucleus.

In Mrk273 the contributions do not vary significantly across the central 6\,kpc.

\begin{figure*}
\vspace{12cm}
\caption{As in Fig.\ref{n5929sin} for one galaxy with dominant 100\,Myr  component at the
nucleus.}
\label{m273sin}
\includegraphics{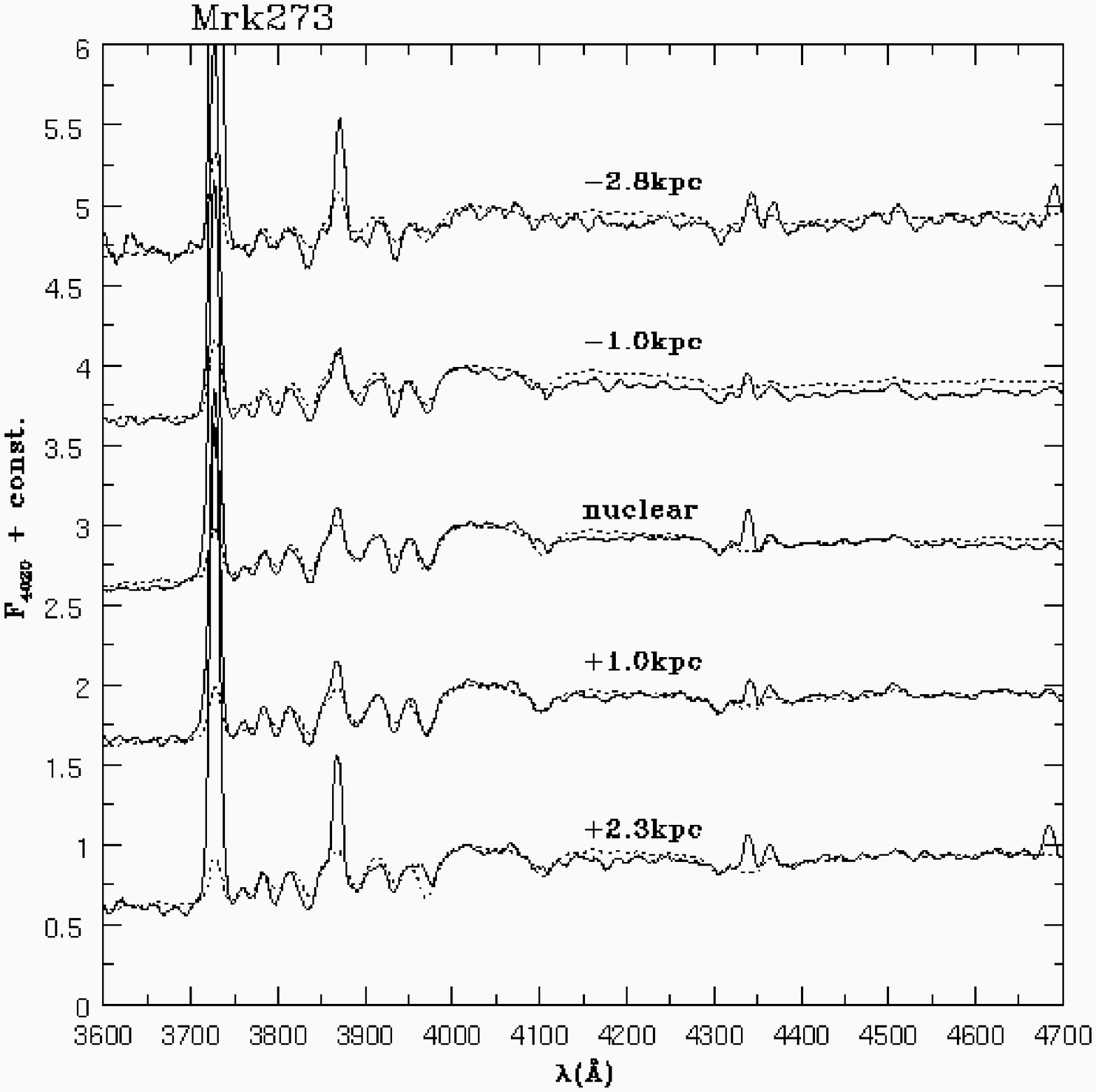}
\end{figure*}

In Fig. \ref{m273sin} we compare the observed spectrum (solid-line) with the synthetic one
(dashed-line) for one galaxy of this group (Mrk273) at the nuclear and off-nuclear regions.
All spectra present the high-order Balmer lines in absorption, characteristic of stellar
populations younger than 1\,Gyr.

\subsubsection{Dominant 3\,Myr/FC component}
Here are Mrk463E, Mrk477 and NGC5135. The 3\,Myr/FC component ranges from 40 per cent to
30 per cent contribution to the nuclear flux at 4020 \AA . At the nucleus Mrk 463E and Mrk477
also have significant flux contributions of the 10\,Gyr, 1\,Gyr and 10\,Myr populations,
about 15 per cent each. Besides those, the 100\,Myr component contributes about 15 per cent at
the nucleus of the NGC5135.

\begin{figure*}
\vspace{19cm}
\caption{Synthesis results: galaxies with dominant 3\,Myr/FC component at the nucleus.}
\label{sin6}
\includegraphics{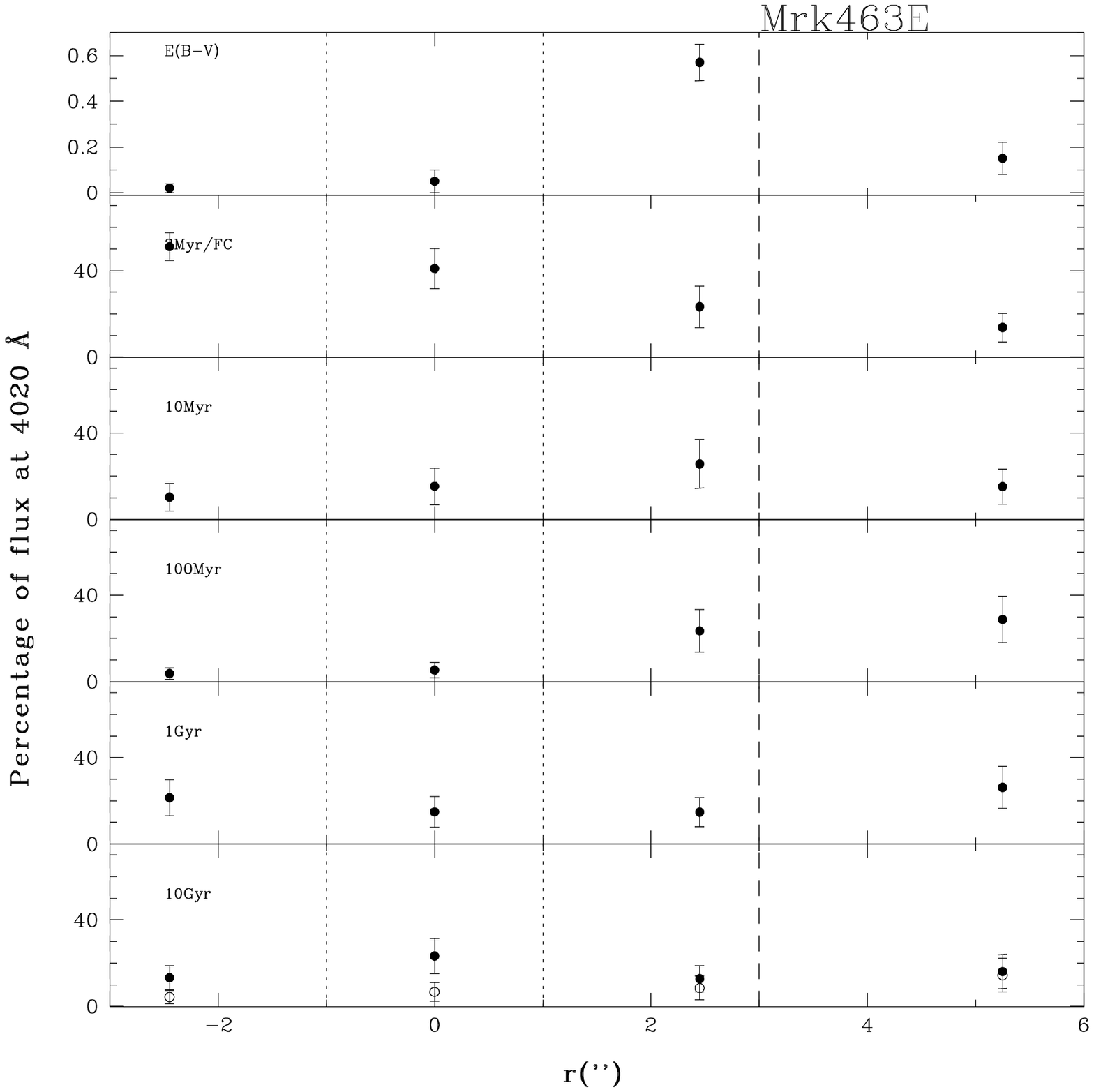}
\includegraphics{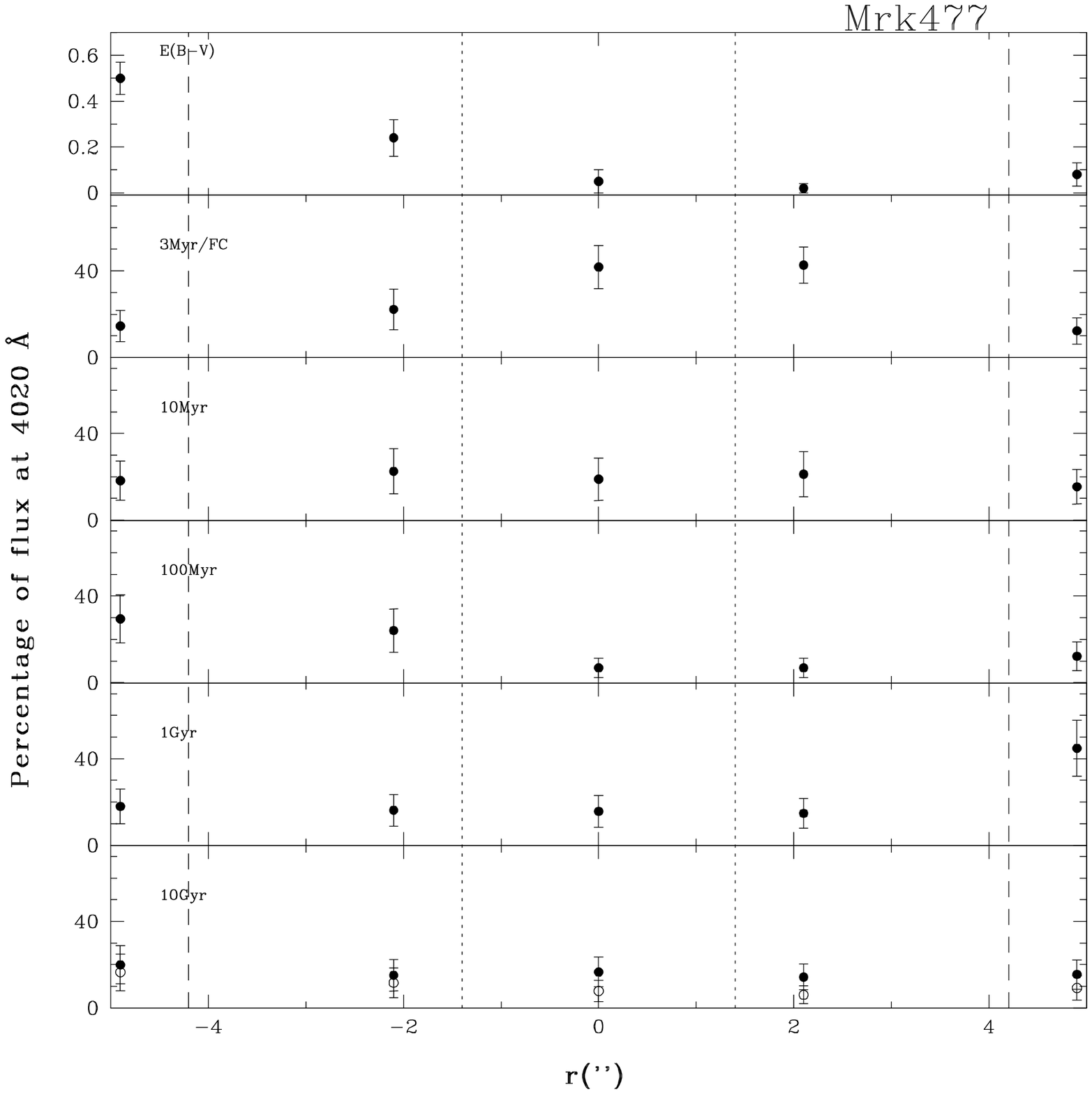}
\includegraphics{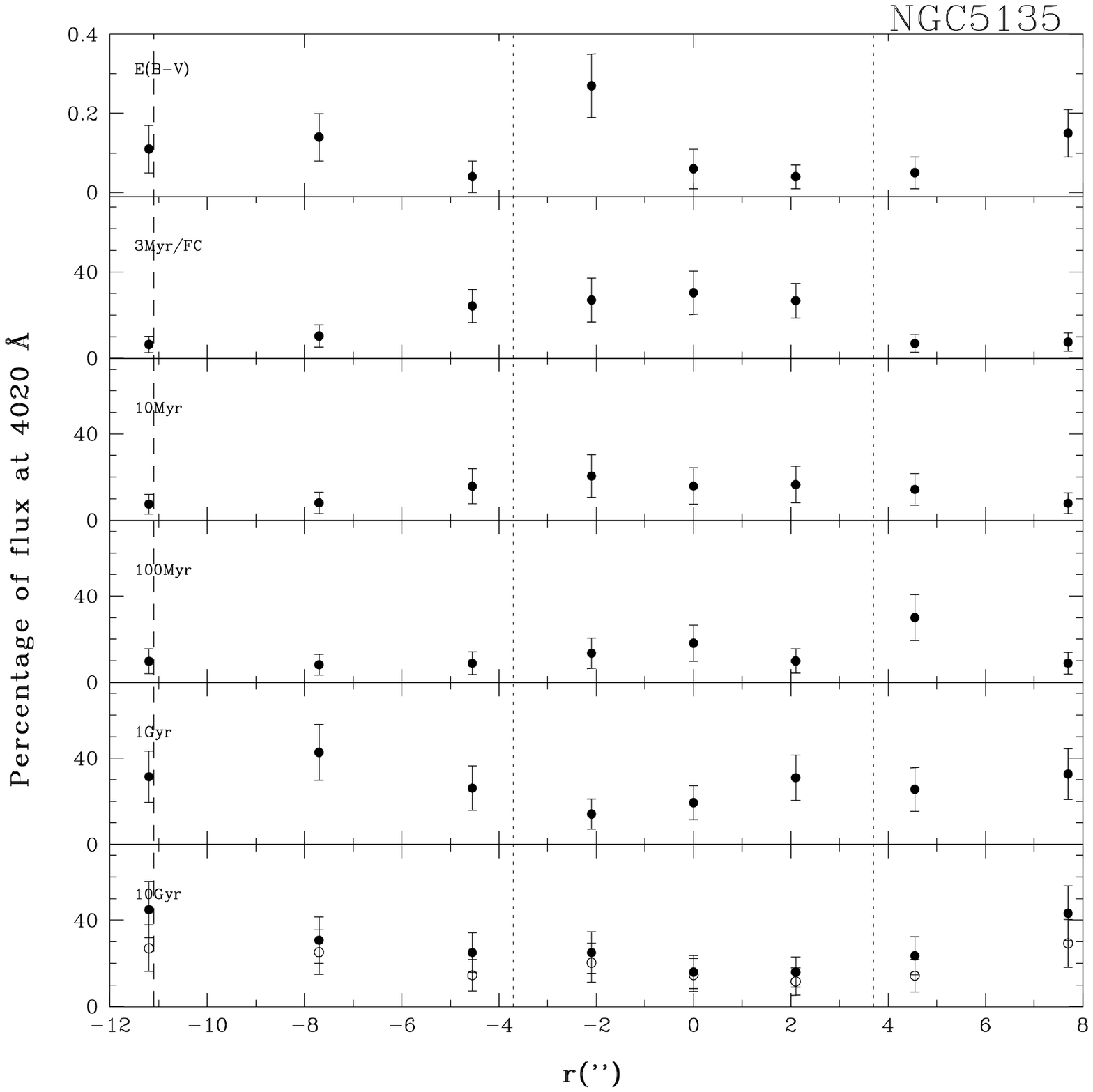}
\end{figure*}

In Fig. \ref{sin6} we show the synthesis results for these galaxies as a function of distance
from the nucleus. In Mrk477 the contribution of the 3\,Myr/FC component decreases and that of
the 100\,Myr component increases outwards. The other components have contributions similar to
those at the nucleus.

Outside the nucleus of Mrk463E the behaviour is not symmetric. At one side (positive
direction) the 3\,Myr/FC component decreases and the 10\,Myr and 100\,Myr
components increase outwards. At the other side these components have the opposite behaviour.

The stellar population of the galaxy Mrk463E has been studied in R01 although with different
spatial sampling, spectral range for the observations and normalization wavelength for the
synthesis. It is intersting to compare the results of the present work with those of R01 in order
to verify the robustness of our method. This comparison shows a general agreement in the sense
that the stellar population is dominated by the young components ($\le$100\,Myr old). Differences
are found in the relative contribution of the 100 and 10\,Myr components. Investigating the
reason for this difference we conclude it is due to the fact that this galaxy has very small W
values and the synthesis is strongly constrained by the continuum, which is affected by
calibration uncertainties. Nevertheless, within the scape of the present paper, these
differences do not alter our main conclusions.

In NGC5135 the contribution of the components of 10\,Gyr and 1\,Gyr increases and that of the
components younger than 1\,Gyr decreases outwards.

\begin{figure*}
\vspace{12cm}
\caption{As in Fig.\ref{n5929sin} for one galaxy with dominant 3\,Myr/FC component at the
nucleus.}
\label{m477sin}
\includegraphics{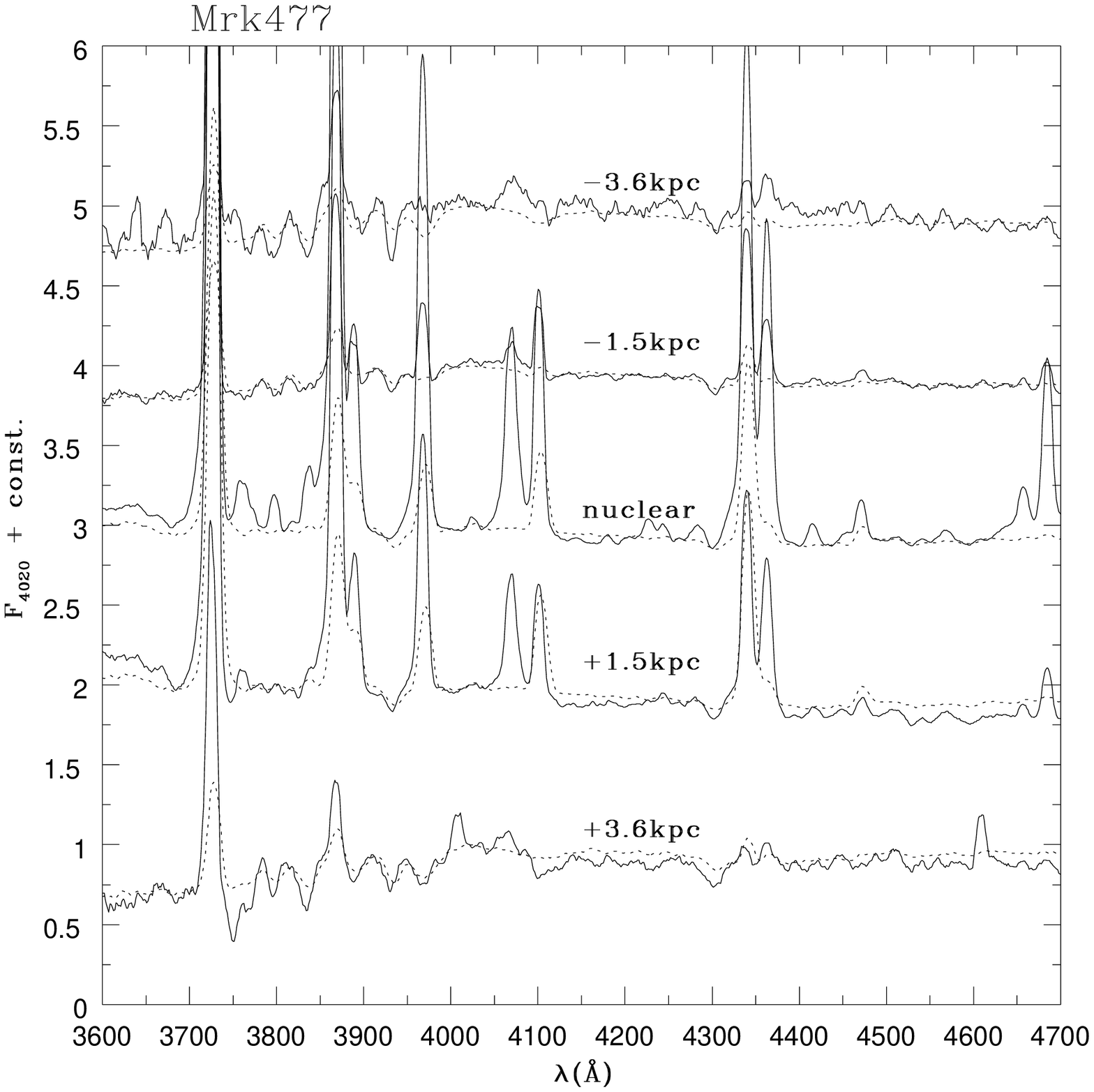}
\end{figure*}

In Fig. \ref{m477sin} we compare the observed spectrum (solid-line) with the synthetic one
(dashed-line) for one galaxy of this group (Mrk477) at the nuclear and off-nuclear regions.
The nuclear spectrum is dominated by gas emission of the AGN narrow line region and the
star-forming regions. Outside the emission decreases and we can see the high-order Balmer
lines in absorption, characteristic of stellar populations younger than 1\,Gyr.

We summarize the synthesis results discussed above in Table \ref{results} which shows the
contribution of four age bins: 10 Gyr, 1\,Gyr, 100Myr+10Myr and 3Myr/FC to the nuclear and
extranuclear spectra (at 1\,kpc and 3\,kpc from the nucleus) of the sample. The Seyfert 2
galaxies are grouped according to the Hubble type (from top to bottom) in the top part of
table: the first group contains the S0 galaxies, the second the Sa-Sb's, the
third the Sbc's and the fourth the galaxies with uncertain morphological type.
In the bottom part of table are those for the non-Seyfert galaxies. For these galaxies, due to
the smaller signal-to-noise ratio of the 1.5m observations, we could only reach distances at
the galaxies somewhat smaller than 1\,kpc. In order to allow a comparison with the Seyfert
data we decided to show the results for the synthesis at 1\,kpc from the nucleus, obtained
by extrapolation from the extranuclear spectra. We did the extrapolation based on the 
hypothesis that the stellar population varies smoothly with the distance from the nucleus, 
like in the S0 non-Seyfert galaxies studied in R01. This seems to be a reasonable assumption,
if one compares Fig.\ref{sin_normal} with Fig.\ref{sin_normaltodos}, which show similar behaviour
of the stellar population gradients for the S0 and later Hubble types. The extrapolated
values are shown between parenthesis in Table \ref{results}.

E(B-V)$_i$ values are found throughout the observed region in the range $0<$E(B-V)$_i<0.7$.
The large E(B-V)$_i$ variations present in almost all Seyfert 2 galaxies are likely due to
a non-uniform dust distribution (Malkan et al. 1998).

\begin{table*}
\caption{Contribution of four age bins to the total flux at 4020\AA, for the sample
galaxies. In the first group (top) are the S0 Sy2 galaxies, in the second the Sa-Sb
Sy2's, in the third the Sbc Sy2's and in the fourth the
galaxies with uncertain morphological type. In the second part of the table are the
non-Seyfert galaxies grouped by morphological type. The numbers on parenthesis were
extrapolated from the extranuclear spectra.}
\label{results}
\begin{center}
\begin{tabular}{lcccccccccccc} \hline
Name                        & 10\,Gyr & & & 1\,Gyr & & & 100$+$10Myr& & & 3Myr/FC & & \\
                            & Nuclear & 1\,kpc & 3\,kpc & Nuclear & 1\,kpc & 3\,kpc & Nuclear
& 1\,kpc & 3\,kpc & Nuclear & 1\,kpc & 3\,kpc \\ \hline
{\it S0 Seyfert 2} \\
Mrk3            & 50 & 56 & -- & 20 & 37 & -- & 10 &  5 & -- & 20 &  2 & -- \\
Mrk348          & 60 & 49 & -- & 20 & 35 & -- & 12 & 10 & -- &  8 &  6 & -- \\
Mrk573          & 65 & 60 & -- & 21 & 29 & -- & 10 &  7 & -- &  4 &  4 & -- \\
Mrk1066         & 24 & 33 & -- & 23 & 40 & -- & 36 & 17 & -- & 17 &  6 & -- \\
NGC1386         & 48 & 50 & -- & 42 & 38 & -- &  7 &  4 & -- &  3 &  8 & -- \\
NGC2110         & 53 & 50 & -- & 32 & 43 & -- & 10 &  4 & -- &  5 &  3 & -- \\
\\
{\it Sa-Sb Seyfert 2} \\
Mrk1073         & 21 & 39 & 26 & 31 & 25 & 30 & 32 & 22 & 36 & 16 & 14 &  8 \\
NGC1068         & 29 & 25 & 43 & 26 & 26 & 26 & 28 & 31 & 24 & 17 & 18 &  7 \\
NGC5135         & 16 & 24 & 45 & 20 & 29 & 31 & 34 & 31 & 18 & 30 & 16 &  6 \\
NGC5929         & 61 & 53 & -- & 27 & 34 & -- &  8 &  9 & -- &  4 &  4 & -- \\
NGC7130         & 19 & 29 & 25 & 14 & 23 & 26 & 48 & 36 & 38 & 19 & 12 & 11 \\
\\
{\it Sbc Seyfert 2} \\
Mrk533          & 26 & 23 & 18 & 19 & 22 & 40 & 41 & 18 & 28 & 14 & 37 & 14 \\
IC3639          & 36 & 26 & 17 & 31 & 30 & 23 & 24 & 30 & 36 &  9 & 14 & 24 \\
\\
{\it S Seyfert 2} \\
Mrk1            & 28 & 33 & 64 & 22 & 34 & 17 & 30 & 27 & 15 & 20 &  6 &  4 \\
Mrk34           & 45 & 40 & 37 & 29 & 37 & 45 & 16 & 12 & 12 & 10 & 11 &  6 \\
Mrk78           & 24 & 27 & -- & 51 & 48 & -- & 21 & 20 & -- &  4 &  5 & -- \\
Mrk273          & 19 & 18 & 23 & 29 & 33 & 38 & 44 & 41 & 32 &  8 &  8 &  7 \\
Mrk463E         & 24 & 19 & 15 & 15 & 17 & 20 & 20 & 28 & 46 & 41 & 36 & 19 \\
Mrk477          & 17 & 15 & 18 & 16 & 16 & 29 & 26 & 36 & 36 & 41 & 33 & 17 \\
NGC7212         & 57 & 55 & 48 & 16 & 18 & 40 & 12 &  9 &  8 & 15 & 18 &  4 \\ \hline
{\it Non-Seyfert Sa-Sb} \\
NGC1367         & 61 & (40) & -- & 33 & (55) & -- &  4 & (4) & -- &  2 & (1) & -- \\
NGC1425         & 62 & (25) & -- & 34 & (75) & -- &  3 & (0) & -- &  1 & (0) & -- \\
NGC3358         & 61 & (70) & -- & 31 & (25) & -- &  5 & (3) & -- &  3 & (2) & -- \\
\\
{\it Non-Seyfert Sbc} \\
NGC3054         & 66 & (30) & -- & 29 & (70) & -- &  3 & (0) & -- &  2 & (0) & -- \\
NGC3223         & 69 & (55) & -- & 26 & (45) & -- &  4 & (0) & -- &  1 & (0) & -- \\
\\
{\it Non-Seyfert Sc} \\
NGC1232         & 56 & (38) & -- & 40 & (62) & -- &  4 & (0) & -- &  1 & (0) & -- \\
NGC1637         & 35 & --   & -- & 30 & --   & -- & 27 & --  & -- &  8 & --  & -- \\ \hline
\end{tabular}
\end{center}
\end{table*}

\section{Discussion}

From the synthesis results, the main characteristics of the stellar population of the present
Seyfert 2 sample are:

a) At the nucleus 75 per cent of the sample (15 galaxies) have contribution of ages $\leq$
100\,Myr (or/and FC) larger than 20 per cent. This result is in agreement with those of
Schmitt et al. (1999), Storchi-Bergmann et al. (2000) and Gonz\'alez Delgado et al. (2001),
who found that in $\approx$ 40 per cent of nearby Seyfert 2 galaxies there are clear
signatures of young stars and in another 30 per cent there is a blue continuum which can be
either due to a FC or a stellar population younger than 10\,Myr. 55 per cent (11 galaxies),
have significant ($>$ 10 per cent) 3\,Myr/FC component at the nucleus.

b) At 1kpc from the nucleus, 70 per cent of the sample have contribution
of ages $\leq$ 100\,Myr (or/and FC) larger than 20 per cent and 50 per cent have a significant
3\,Myr/FC component.

c) At 3 kpc, 45 per cent of the sample have contribution of ages $\leq$ 100\,Myr (or/and FC)
larger than 20 per cent and 25 per cent have a significant 3\,Myr/FC population.

d) The contribution of the 1\,Gyr component increases outwards from the nucleus in 13 galaxies
while the old metal rich component decreases in 7 galaxies. This behaviour was also found
in S0 non-Seyfert galaxies studied by R01.

e) The contribution of old metal poor component is significant in 17 galaxies of the
sample.

\begin{figure*}
\vspace{9.5cm}
\caption{Results of the synthesis performed with all base components (filled dots) and
without the metal poor components of 10 and 1\,Gyr old (crosses) for two cases (see the
discussion in the text section 5). Panels as in Fig.\ref{sin_normal}.}
\label{sinmod}
\includegraphics{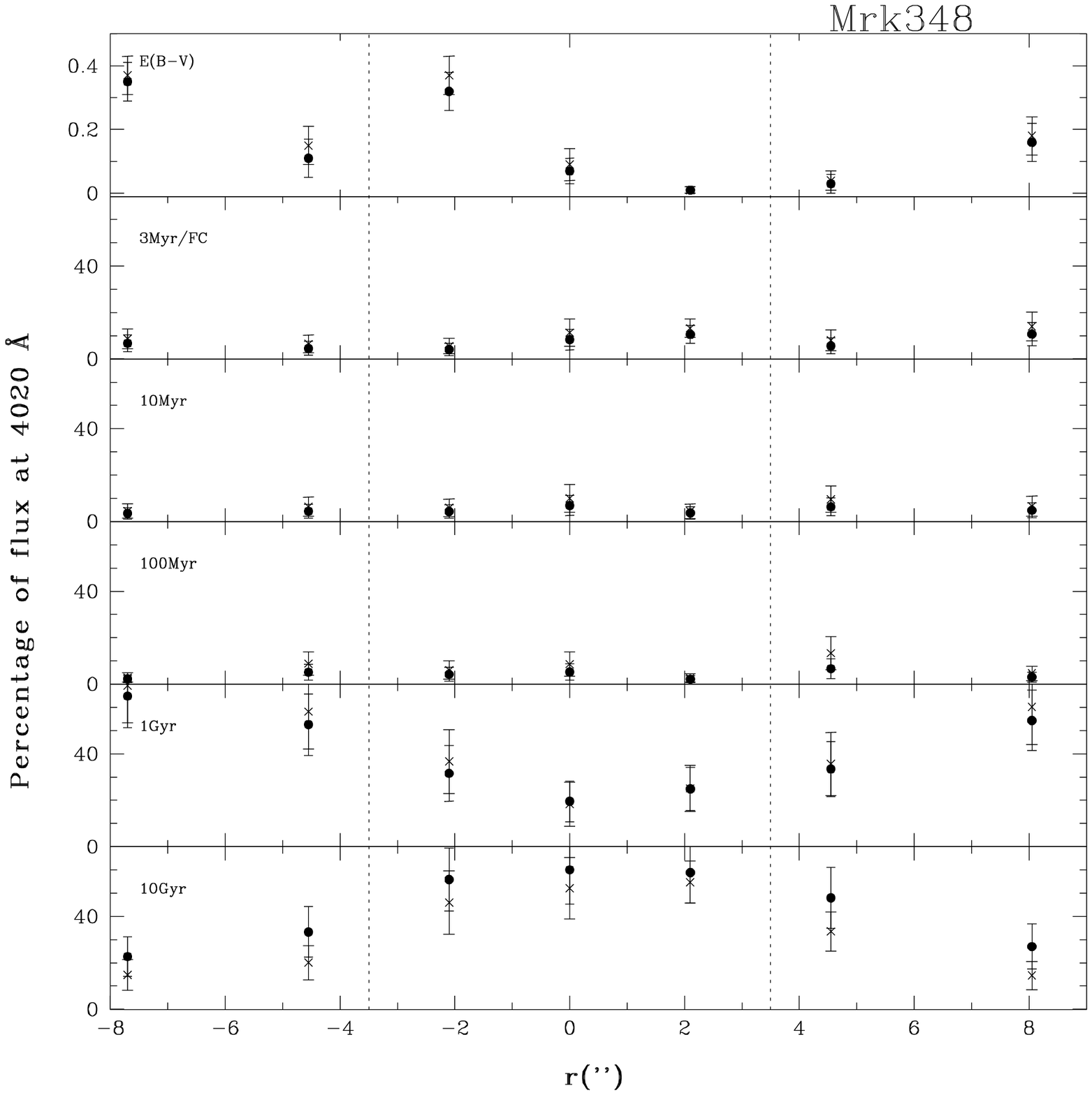}
\includegraphics{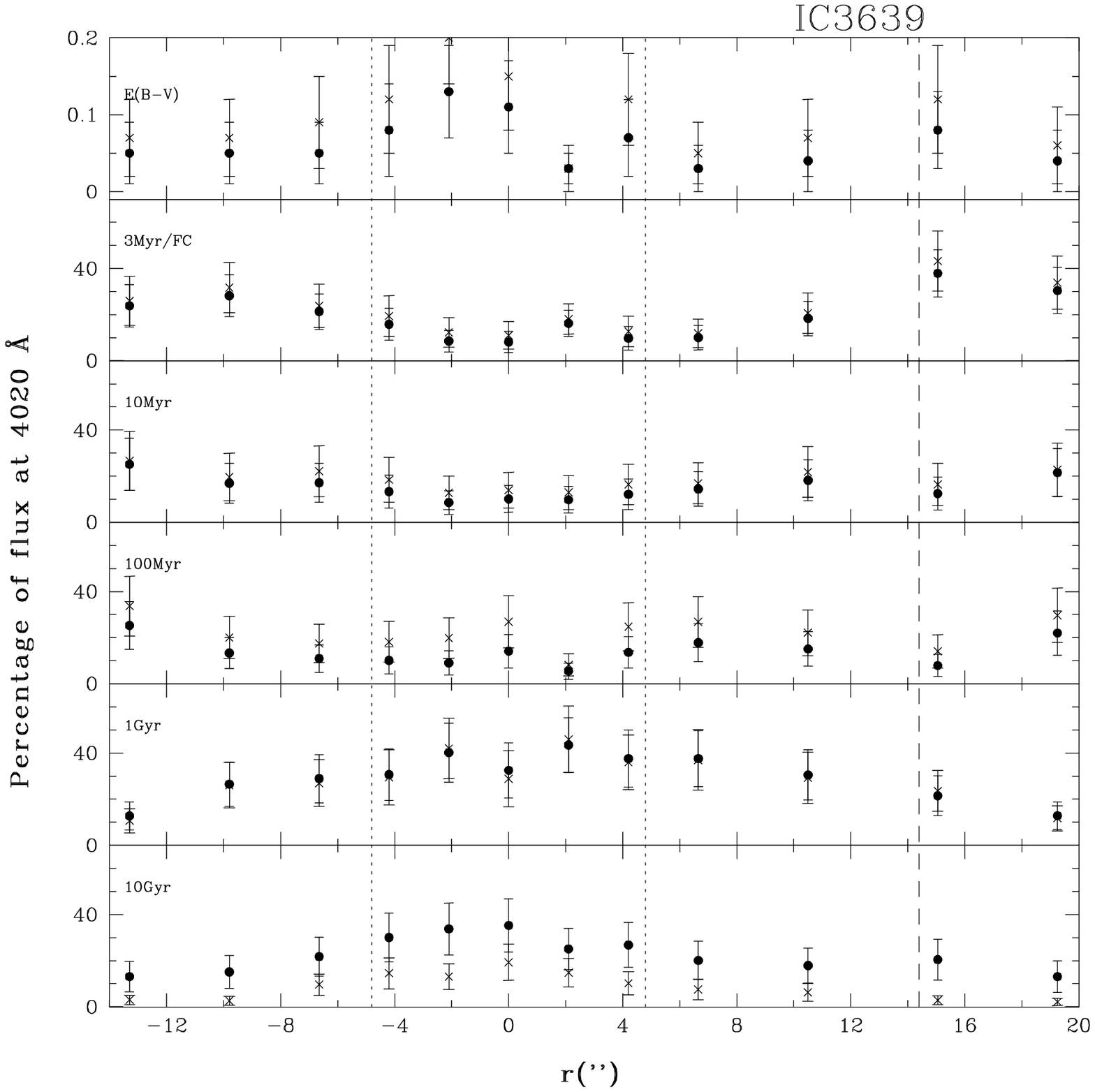}
\end{figure*}

The latter result was also found in R01 and in that work we hypothesized that this
contribution, larger than that found in non-Seyfert S0 galaxies, could be an effect of degeneracy
between the old metal poor component and the 3\,Myr/FC component. In order to verify if
this effect is important we repeated the spectral synthesis excluding the metal poor
components. We show in Fig.\ref{sinmod} the results of the synthesis without these components
(crosses) as compared with the synthesis using all components (filled dots), for two extreme
cases: Mrk348 and IC3639.

In Mrk348, in the new synthesis, the missing contribution of the metal poor components is
shared among all other younger components and the only significant change is in the 10\,Gyr
population which decreases in about 15 per cent of the total flux at 4020\AA. In the other
extreme case of IC3639, while the old population decreases about 20 per cent, the 100\,Myr
component increases about 10 per cent in the new synthesis. The remaining 10 per cent is
shared among the other components.

We thus conclude that such possible degeneracy effects are not large in the synthesis.
In addition, although the 10\,Gyr  metal poor component presents a small contribution in
S0's, B88 shows that this component is larger in later-type galaxies,
and we would thus need to compare the results for the Seyferts with those of non-Seyfert
galaxies of the same Hubble type, considering that,  besides 7 S0/Sa's, the present sample
comprises 4 Sab/Sb's and 2 Sbc's, and 7 galaxies with uncertain morphology.

\subsection{Comparison with synthesis results for non-active galaxies}

A comparison of the Seyfert 2 stellar population properties and their radial variations with
those of non-active galaxies of the same Hubble type is necessary in order to
search for systematic differences in the Seyfert sample, which could be related to the AGN
phenomenon.

In Table \ref{results} we summarize the individual synthesis results for the Seyfert 2 and non-Seyfert
galaxies and in Table \ref{average} we show the corresponding average results for each Hubble type,
including also those for the three non-Seyfert S0 galaxies from R01 and the reference sample of
B88 (in this latter case we show the integrated results from apertures of
$\approx$ 1\,kpc$\times$1\,kpc).

Presently, we have three S0 non-Seyfert galaxies studied with the same method applied
to the present Seyfert 2 sample, large spatial coverage (up to 3\,kpc from the nucleus) and
typical sampling regions at the galaxy of 200\,pc.
For Hubble types later than S0, we have the seven non-Seyfert galaxies studied in this work, with
a more restrict spatial coverage and a similar typical sampling. In order to compare the
results for these Sa to Sc galaxies with those for the active ones, we had to extrapolate the stellar
population synthesis to 1\,kpc using the nuclear and available extranuclear spectra.

Because the number of non-Seyfert galaxies is reduced we also use as reference the study of B88. We have
performed the spectral synthesis of the templates using the same Ws and Cs used for our sample.
The synthesis results were then averaged in order to obtain the ``typical'' population of each
Hubble type Sa, Sb and Sc. As the dimensions of the regions studied by B88 are much larger than
those corresponding to our nuclear extractions, we need to take into account both the nuclear
and extranuclear synthesis results from our Seyfert 2 sample in the comparison with the
non-Seyfert galaxies.

Comparing the synthesis results for the Seyferts with those for non-Seyferts of the same
Hubble type we realize that, at the nucleus, the contribution of the 10\,Gyr  component is smaller
in Seyfert 2 galaxies than in non-Seyfert galaxies in eleven (of the thirteen, which have Hubble types
from S0 to Sbc) of them. The component with 1\,Gyr is similar in Seyfert 2 and non-Seyfert galaxies in
four objects, larger in five Sy2 and smaller in four. In terms of average values (for each Hubble
type) the 1\,Gyr contribution is larger for S0 Sy2, and comparable to that of
non-Seyferts for Sa to Sbc Hubble type. The components younger than 1\,Gyr and FC are
larger in Sy2 than in non-Seyferts in eleven galaxies.

At 1\,kpc from the nucleus we found for S0 galaxies the contribution of the 10\,Gyr
component is always smaller in Sy2 than in non-Seyferts, while the contribution of the 1\,Gyr
and younger components are normally larger. In the later Hubble type galaxies we have
usually a smaller contribution of both the 10\,Gyr  and 1\,Gyr components and a larger
contribution of the youngest components in Sy2's than in non-Seyferts.

There are 7 galaxies with uncertain morphology: Mrk1, Mrk34, Mrk78, Mrk273, Mrk463E,
Mrk477 and NGC7212, all but Mrk34 and Mrk78 in clear interaction. At the nucleus all of
them have significant contribution of components younger than 1\,Gyr, while at 3\,kpc
from the nucleus only for two of them this does not happen.
We do not have a reference study, similar to the one we have performed here,
for the stellar population in non-active interacting galaxies, but it is
a well known result that these galaxies frequently show recent episodes
of star-formation, triggered by the interaction (Sanders \& Mirabel 1996 and references
therein). Thus the results we have obtained for these Seyfert 2 galaxies do not seem peculiar.

Regarding to E(B-V)$_i$ values, non-Seyferts S0 galaxies present small internal reddening throghout
the central 6\,kpc  while Seyfert galaxies with same Hubble type have normally larger values
and large variation. For later-type galaxies, we did not find significant differences between
Seyfert and non-Seyfert galaxies. Both present large E(B-V)$_i$ variations (from 0 to 0.7),
likely due to a non-uniform dust distribution.

\begin{table*}
\caption{Average contribution of four age bins to the total flux at 4020\AA, for different
Hubble types. The numbers on parenthesis were extrapolated from the extranuclear spectra.}
\label{average}
\begin{center}
\begin{tabular}{lcccccccccccc} \hline
Hubble type                 & 10\,Gyr & & & 1\,Gyr & & & 100$+$10Myr& & & 3Myr/FC & & \\
                            & Nuclear & 1\,kpc & 3\,kpc & Nuclear & 1\,kpc & 3\,kpc & Nuclear
& 1\,kpc & 3\,kpc & Nuclear & 1\,kpc & 3\,kpc \\ \hline
S0 Seyfert 2                & 50 & 50 & -- & 26 & 37 & -- & 14 &  8 & -- & 10 &  5 &  -- \\
Non-Seyfert S0              & 86 & 77 & 50 &  6 & 15 & 36 &  9 &  8 & 12 &  0 &  0 &  1 \\
\\
Sa-Sb Seyfert 2             & 29 & 34 & 35 & 24 & 27 & 28 & 30 & 26 & 29 & 17 & 13 &  8 \\
Non-Seyfert Sa-Sb           & 61 & (45) & -- & 33 & (52) & -- & 4 & (2) & -- & 2 & (1) & -- \\
Bica' Sa$^a$                & 56 & -- & -- & 37 & -- & -- &  4 & -- & -- &  2 & -- & --\\
Bica' Sb$^a$                & 53 & -- & -- & 35 & -- & -- &  7 & -- & -- &  5 & -- & -- \\
\\
Sbc Seyfert 2               & 31 & 25 & 17 & 25 & 26 & 32 & 33 & 24 & 32 & 12 & 26 & 19 \\
Non-Seyfert Sbc             & 68 & (43) & -- & 27 & (57) & -- & 3 & (0) & -- & 2 & (0) & -- \\
Non-Seyfert Sc              & 46 & -- & -- & 35 & -- & -- & 15 & -- & -- &  4 & -- & -- \\
Bica' Sc$^a$                & 27 & -- & -- & 27 & -- & -- & 26 & -- & -- & 20 & -- & -- \\
\\
S Seyfert 2                 & 31 & 30 & 34 & 25 & 29 & 32 & 24 & 24 & 24 & 20 & 16 & 10 \\ \hline
\end{tabular}

$^a$ Results for non-Seyfert galaxies from Bica (1988), in a region of 1\,kpc $\times$ 1\,kpc.
\end{center}
\end{table*}

\section{Summary and Conclusions}

In this paper we have analysed the stellar population of a sample
of 20 Seyfert 2 galaxies as a function of distance from the nucleus,
comparing the contribution of different age components
with those for non-Seyfert galaxies of the same Hubble type.

The main conclusions can be summarized as follows.

 1) The radial distribution of the Ca{\sevensize II}K and G-band Ws
show smaller values at the nucleus than that at $\approx$ 1\,kpc
from it for 11 of the 20 Seyfert 2 galaxies of the
sample, suggesting dilution by a blue continuum
from a FC or young stars; two present larger W values
at the nucleus than outside, similar to non-Seyfert S0 galaxies, and in the remaining
there is no obvious dilution, nor a systematic variation with the distance similar
to that observed for the S0 galaxy.

2) Stellar population synthesis shows that, while
at the nucleus, 75 per cent of the galaxies present
contribution $\geq$ 20 per cent of ages $\le$100\,Myr (or/and FC),
this proportion decreases to 45 per cent at 3\,kpc.
In particular, 55 per cent of the galaxies have
contribution $>$ 10 per cent of the 3\,Myr/FC component
at the nucleus, but only 25 per cent of them have
this component at 3\,kpc.

3) Our results point to a systematic difference between the stellar population of
Seyfert galaxies and those of non-Seyfert galaxies for the Hubble types from S0 to Sbc.
At the nucleus and up to 1\,kpc from it the contribution of ages younger than 1\,Gyr is
in most cases larger in the Seyferts than in non-Seyfert galaxies. Regarding the 1\,Gyr
component, at the nucleus its contribution is for the Seyferts, on average, larger for
S0's and similar to that of non-Seyfert for later type galaxies. Outside the nucleus
it is again larger for the S0 Sy2s but smaller for the later type Seyfert when compared
to the non-Seyferts. The 10\,Gyr component
shows a larger contribution in non-Seyferts than in Seyferts. Both the Seyferts and
non-Seyferts show a decrease of the contribution of the 10\,Gyr component and an increase of
the 1\,Gyr component with distance from the nucleus, although the gradient seems to be
steeper for the non-Seyfert galaxies.

In the evolutionary scenario proposed by Storchi-Bergmann et al. (2001) and Cid
Fernandes et al. (2001), the results above indicate that the interactions which
trigger the AGN and circumnuclear bursts of star-formation, also trigger bursts along
the body of the galaxy, at least within the inner 1\,kpc.

\section*{Acknowledgments}
D.R., T.S.B. and R.C.F. acknowledge support from the Brazilian Institutions CNPq, CAPES and FAPERGS.
The National Radio Astronomy Observatory is a facility of the National Science Foundation operated
under cooperative agreement by Associated Universities, Inc. We thank an anonymous referee
for valuable suggestions which helped to improve the paper.

\label{lastpage}

\end{document}